\documentclass{aa}  

\usepackage{graphicx}
\usepackage{gensymb}
\usepackage{verbatim}
\usepackage{siunitx}
\usepackage{amssymb}
\usepackage{amsmath}
\usepackage{setspace}
\usepackage[table,xcdraw]{xcolor}
\usepackage{tabularx}
\usepackage[normalem]{ulem} % for strikethrough using \sout
\usepackage{txfonts}
\usepackage[colorlinks,allcolors=blue]{hyperref}
\newcommand{\add}[1]{{{#1}}}
\newcommand{\remove}[1]{\textcolor{red}{\sout{}}}

\begin{document}

   \title{The SEDIGISM survey: Molecular cloud morphology}

   \subtitle{I. Classification and star formation}
   \author{K. R. Neralwar\inst{1}\thanks{Member of the International Max Planck Research School (IMPRS) for Astronomy and Astrophysics at the Universities of Bonn and Cologne.},
          D. Colombo\inst{1},
          A. Duarte-Cabral\inst{2},
          J.\,S.\,Urquhart\inst{3},
          M. Mattern\inst{4},
          F. Wyrowski\inst{1},
          K. M. Menten\inst{1},
          P. Barnes\inst{5,6},
          \'A. S\'anchez-Monge\inst{7},
          H. Beuther\inst{8},
          A. J. Rigby\inst{2},
          P. Mazumdar\inst{1},
          D. Eden\inst{9},
          T. Csengeri\inst{10},
          C.L. Dobbs\inst{11},
          V. S. Veena\inst{7},
          S. Neupane\inst{1},
          T. Henning\inst{8},
          F. Schuller\inst{1,12},
          S. Leurini\inst{13},
          M. Wienen\inst{1},
          A. Y. Yang\inst{1},
          S.\,E. Ragan\inst{2},
          S. Medina\inst{1},
          Q. Nguyen-Luong\inst{14}
          }%          

   \institute{Max-Planck-Institut f\"ur Radioastronomie, Auf dem H\"ugel 69, 53121 Bonn, Germany \\ email: \texttt{kneralwar@mpifr-bonn.mpg.de}
   \and
   School of Physics \& Astronomy, Cardiff University, Queen’s building, The parade, Cardiff CF24 3AA, UK
   \and
   Centre for Astrophysics and Planetary Science, University of Kent, Canterbury, CT2\,7NH, UK
   \and
   Laboratoire d’Astrophysique (AIM), CEA, CNRS, Universit{\'e} Paris-Saclay, Universit{\'e} Paris Diderot, Sorbonne Paris Cit{\'e}, 91191 Gif-sur-Yvette, France
   \and
   Space Science Institute, 4765 Walnut St Suite B, Boulder, CO 80301, USA
   \and
   School of Science and Technology, University of New England, Armidale NSW 2351, Australia
   \and
   I.\ Physikalisches Institut, Universität zu K\"oln, Z\"ulpicher Strasse 77, 50937 Cologne, Germany
   \and
   Max-Planck-Institut f\"ur Astronomie, K\"onigstuhl 17, D-69117 Heidelberg, Germany
   \and
   Astrophysics Research Institute, Liverpool John Moores University, IC2, Liverpool Science Park, 146 Brownlow Hill, Liverpool, L3 5RF, UK
   \and
   Laboratoire d’astrophysique de Bordeaux, CNRS, Univ. Bordeaux, B18N, allée Geoffroy Saint-Hilaire, F-33615 Pessac, France
   \and
   School of Physics and Astronomy, University of Exeter, Stocker Road, Exeter EX4 4QL, UK
   \and
   Leibniz-Institut für Astrophysik Potsdam (AIP), An der Sternwarte 16, D-14482 Potsdam, Germany
   \and
   INAF – Osservatorio Astronomico di Cagliari, Via della Scienza 5, I-09047 Selargius (CA), Italy
   \and
   McMaster University, 1 James St N, Hamilton, ON L8P 1A2, Canada
}

  \date{Received XXX; accepted XXX}

% \abstract{}{}{}{}{} 
% 5 {} token are mandatory
  \abstract 
  {We present one of the very first extensive classifications of a large sample of molecular clouds based on their morphology. This is achieved using a recently published catalogue of 10663 clouds obtained from the first data release of the SEDIGISM survey. The clouds are classified into four different morphologies by visual inspection and using an automated algorithm -- $J$ plots. The visual inspection also serves as a test for the $J$ plots algorithm, as this is the first time it has been used on molecular gas. Generally, it has been found that the structure of molecular clouds is highly filamentary and our observations indeed verify that most of our molecular clouds are elongated structures. Based on our visual classification of the 10663 SEDIGISM clouds, 15\% are ring-like, 57\% are elongated, 15\% are concentrated and 10\% are clumpy clouds. The remaining clouds do not belong to any of these morphology classes and are termed unclassified. We compare the SEDIGISM molecular clouds with structures identified through other surveys, i.e. ATLASGAL elongated structures and the bubbles from Milky Way Project (MWP). We find that many of the ATLASGAL and MWP structures are velocity coherent. ATLASGAL elongated structures overlap with $\approx$ 21\% of the SEDIGISM elongated structures (elongated and clumpy clouds) and MWP bubbles overlap with $\approx$ 25\% of the SEDIGISM ring-like clouds. We also analyse the star-formation associated with different cloud morphologies using two different techniques. The first technique examines star formation efficiency (SFE) and the dense gas fraction (DGF), based on SEDIGISM clouds and ATLASGAL clumps data. The second technique uses the high-mass star formation (HMSF) threshold for molecular clouds. The results indicate that clouds with ring-like and clumpy morphologies show a higher degree of star formation.}

   \titlerunning{SEDIGISM: molecular cloud morphology I}
   \authorrunning{K. R. Neralwar, D. Colombo et al.}
   \maketitle
%
%-------------------------------------------------------------------

\section{Introduction}\label{sec: intro}

Molecular clouds are often approximated as self-gravitating spheres \citep{crops, benedettini_2021}; especially for the calculation of the cloud size (radius). Nevertheless, modern surveys have shown that molecular gas is organised in a more complex fashion. In particular, molecular gas appears permeated by filamentary structures \citep{andre_2010, nessie_1, higal_MW, ATLASGAL_filaments, ana_2017, filaments_catherine_zucker, arzoumanian_2019, suri_2019, zhang_2019, abe_2021, priestley_2021, Yuan_2021,colombo2021}. Thus, an analysis of the connection between molecular clouds and structures like filaments and bubble can improve our understanding of the various phenomena in the interstellar medium (ISM), e.g. star formation.

The presence of filaments in molecular clouds as sites of formation of pre-stellar and proto-stellar cores has been a popular topic of discussion in star formation research \citep{andre_2014, ATLASGAL_filaments, arzoumanian_2017,filament_intro_2_nessie_strucuture, zhang_2019, bonne_2020}. The existence of filaments in star-forming regions has been evident 
through decades of observations \citep{globular_filament_ref_nessie_Stru, GMF_fil_1, marsh_2016, suri_2019, schisano_2020}, accompanied by numerical simulations \citep{fil_sim_5, fil_sim_1, fil_sim_6, fil_sim_4, fil_sim_3, fil_sim_2}. The size of filaments range from sub-parsec scales \citep[mostly seen in nearby regions;][]{andre_2010, arzoumanian_2019} %(solar neighbourhood)
to hundreds of parsecs \citep{schisano_2014, Herschel_fil_1, filament_intro_2_nessie_strucuture, schisano_2020, wang_2020, lin_2020}.
These large-scale filaments are often associated with galactic spiral arms \citep{Herschel_fil_1} and are seen as chains of smaller filaments, e.g. the active star formation sites in the filaments are visible in infrared observations whereas the quiescent parts are IR dark \citep[infrared dark clouds;][]{peretto_2010}. 

Herschel images \citep[Hi-GAL survey;][]{higal_MW, zavagno_2010, schisano_2020} have revolutionised the study of filaments in the Milky Way, exhibiting their abundance in the Galaxy (within molecular clouds) and introducing constraints on their formation and evolution. 
The formation of filaments in galaxies is often attributed to the shock waves permeating the ISM \citep[][\& references therein]{arzoumanian_2018}. Cloud collisions also result in filament formation, and these filaments fragment into smaller components due to turbulence and gravitational instabilities \citep{balfour_2015, bonne_2020_b, liow_2020, dobbs_2021, clarke_2020, fukui_2021}. These dense (super-critical) filaments often \remove{gravitationally}
fragment into star-forming cores thus leading to star formation \citep{andre_2010, konyves_2015, arzoumanian_2017_b}.

%\subsection{Bubbles}
The molecular gas in clouds often gets dispersed and expelled due to stellar radiation and winds resulting in the formation of ring-like objects called bubbles or shells. Infrared bubbles are often associated with triggered star formation and usually encompass an HII region \citep{deharveng_2010,schneider_2020}. Whether the ring-like appearance of bubbles is a consequence of its 3D structure or a projection effect, is debatable \citep{churchwell_2006, Beaumont_2010, pabst_2020}. The expansion of HII regions %due to excitation by massive stars
drives a shock wave into the molecular clouds sweeping up the gas \citep{francis_1998, bialy_2021}. It further causes the entrapment of neutral material between the ionised and shock fronts giving rise to dense rings of molecular gas. Stellar winds from massive stars create X-ray dominated regions, aiding the formation of these bubbles and the cold material in the shell, that may become gravitationally unstable and host star formation \citep{zavagno_2006, deharveng_2009}. Low-mass stars can readily be formed due to small-scale instabilities like the Jeans instability, whereas large-scale gravitational instabilities can eventually lead to high-mass star formation \citep{ habing_1972, elmegreen_1977, krumholz_2006}. The mass of a star is also influenced by its environment \citep{rosen_2020}.

Many infrared detections of bubbles came from images by the Infrared Space Observatory and the Midcourse Space Experiment, which culminated in high-resolution surveys of the Milky Way like Galactic Legacy Infrared Mid-Plane Survey Extraordinaire \citep[GLIMPSE;][]{churchwell_2006} and Herschel infrared Galactic Plane Survey \citep[Hi-GAL; ][]{higal_MW, Hi-Gal_bubbles}. These have been used to understand the mechanisms behind bubble formation while associating them with objects like supernovae, planetary nebulae, open clusters, Wolf-Rayet stars, and OB stars. \citet{churchwell_2007} postulates that most of the bubbles are produced by stars with strong winds, O stars and B stars. OB stars often have HII regions associated with them whereas late-B stars produce small bubbles by exciting polycyclic aromatic hydrocarbon (PAH) bands without forming HII regions. Simulations including ionisation find that bubbles or shells readily form within molecular clouds \citep[e.g., ][]{dale_2005, geen_2015, ali_2019, li_2019, bending_2020, fukushima_2021, grudic_2021}. Typically these simulations start with spherical molecular clouds, and for more massive clouds or at earlier times the feedback may not be sufficient to break out of the clouds, and a complete ring occurs containing an HII region. Whereas for lower mass clouds, the feedback is often sufficient to break out of and in some cases disperse the cloud. Generally numerical simulations find that photoionisation appears to dominate compared to stellar winds \citep{dale_2014, haid_2018, haid_2019, grudic_2021, geen_2021, ali_2022}.

As particular gas morphologies appear to be connected to a specific set of physical phenomena, a study of the molecular cloud morphology can aid the study of star formation and other complex processes like stellar feedback, that drive turbulence in the ISM, result in disruption of molecular clouds, and lead to the formation of structures (e.g. bubbles). 
Recently, the identification of molecular clouds from large-scale surveys have been carried out efficiently using automated methods (e.g. dendrogram analysis \citep{dendrograms}, $\sc{scimes}$ \citep{SCIMES}, CPROPS \citep{crops, rosolowsky_2021}). % and deep learning algorithms (e.g. CASI-3D \citep{Xu_2020}).
A robust large-scale cloud catalogue is presented by \cite{ana_paper} (hereafter called DC21), that has identified molecular clouds from the SEDIGISM survey using the $\sc{scimes}$ algorithm. We employ the $J$ plot algorithm \citep{jaffa_2018} and visual classification technique to classify the SEDIGISM molecular clouds into various morphologies, and this forms the core of this work. We thus try to understand whether different cloud morphologies host different types of star formation. %These morpholgically classified clouds are also compared to their dense infrared counterparts from from continuum surveys.% to check how they fare in comparison to their dense infrared counterparts.

In this paper, we present the morphological classification for the 10663 clouds identified in the SEDIGISM survey by DC21. In Sec. \ref{sec: data} we describe the three datasets/surveys -- SEDIGISM, ATLASGAL and MWP -- used throughout the paper. Sec. \ref{sec: methods} describes the two methods used to classify the clouds into different morphologies (i.e. $J$ plots and by-eye classification). In Sec. \ref{sec: reliability of j plots}, we discuss the pros and cons of the $J$ plot classification, by comparing it with visual classification. %We also test the two morphological classifications by comparing the aspect ratios for the clouds obtained using different techniques. 
Our results are presented and discussed in Sec. \ref{sec: results and discussions}. In Sec. \ref{sec: counterparts} we compare the SEDIGISM clouds to the filamentary structures from the ATLASGAL survey and dust bubbles from MWP. We thus reveal the SEDIGISM clouds overlapping with these structures and find possible coherent ATLASGAL and MWP structures. In Sec. \ref{sec: star formation prop} we use two different methods to study the star formation associated with different morphologies. Finally, we summarise our findings in Sec. \ref{sec: summary}.

%--------------------------------------------------------------------

\section{Data}\label{sec: data}

We use catalogues from three surveys in this paper -- molecular clouds from Structure, Excitation and Dynamics of the Inner Galactic InterStellar Medium (SEDIGISM), elongated (filamentary) structures from APEX Telescope Large Area Survey of the Galaxy (ATLASGAL) and bubbles derived from the infrared data in the Milky Way Project (MWP).

\subsection{SEDIGISM}

%The major dataset is the catalogue of 10663 from SEDIGISM survey (presented in DC21). 
The Structure, Excitation and Dynamics of the Inner Galactic InterStellar Medium (SEDIGISM) survey covers a region of 84 deg$^2$ between $-60 \degree \leq l \leq +18 \degree$ and  $|b| \leq 0.5 \degree$ ($b$ varies in some regions) using various molecular tracers; in particular, the $J = 2$--$1$ transitions of $^{13}$CO and C$^{18}$O. These observations were conducted using the 12m Atacama Pathfinder EXperiment (APEX, \citealt{gusten_2006}) during 2013--2017. A complete description of the SEDIGISM\footnote{\url{https://sedigism.mpifr-bonn.mpg.de/index.html}} survey is presented in \citet{schuller_2017} and \citet{SEDIGISM_1}.

The survey data (complete contiguous dataset) is presented in the form of 77 datacubes of approximately $2 \degree \times 1 \degree$ with velocities from -200 to 200 $\mathrm{km \; s}^{-1}$ and pixel size of \ang{;;9.5}, and these are centred on all integer Galactic longitudes. The DR1 includes $^{13}$CO observations, with a FWHM beam size of \ang{;;28} and typical 1$\sigma$ sensitivity of 0.8--1.0 K per $0.25 \; \mathrm{km \; s}^{-1}$.
A catalogue of 10663 molecular clouds (full sample) has been identified from the contiguous dataset of SEDIGISM survey DR1 ($^{13}$CO) and is presented in DC21. The molecular clouds have been extracted using the Spectral Clustering for Interstellar Molecular Emission Segmentation $\sc{scimes}$ algorithm (v.0.3.2) \citep{SCIMES, scimes_2}. These clouds from the catalogue are hereafter referred to as SEDIGISM clouds.
%DC21 defined a sub-sample of 6664 clouds -- science sample, which are well-resolved sources (cloud area $> 3 \sigma_{\mathrm{beam}}$) with reliable distance estimates and that do not lie on the edge of data cubes. 
Furthermore, we use SEDIGISM clouds to define two new cloud sub-samples for our morphological analysis (see Sec. \ref{sec: vc and mr sample}).

\subsection{ATLASGAL}

APEX Telescope Large Area Survey of the Galaxy \citep[ATLASGAL,][]{ATLASGAL_survey} is an unbiased survey of the inner Galaxy aimed at studying sites of star formation. It observes the dust continuum emission at $\SI{870}{\mu m}$, in the region $280\degree < l < 60\degree$ and $|b| < 1.5\degree$ ($b$ varies between $-2 \degree$ and $ 1 \degree$ for $l < 300 \degree$). The observations were carried out using the APEX LABOCA instrument \citep{siringo_2009} at a typical noise level of $50-70 \; \mathrm{mJy \; beam^{-1}}$ and a beam size of $\ang{;;19.2}$. The survey has identified more than 10000 dense clumps \citep{contrearas_2013, Csengeri_2014, urquhart_2014, Urquhart_2018} of masses $\sim 500 \; \mathrm{M}_\odot$ and sizes $\sim 0.5 \; \mathrm{pc}$. \citet{james_paper} has compared these clumps to the SEDIGISM clouds and obtained star formation efficiencies (SFE) and dense gas fractions (DGF) for these clouds. The SFE and DGF were obtained using cloud masses, integrated clump masses and their bolometric luminosities. We explore the differences in SFE and DGF for various cloud morphologies in Sec. \ref{sec: sfe dgf prop}.

\citet{ATLASGAL_filaments} has identified spatially coherent filamentary structures from ATLASGAL data using the Discrete Persistent Structures Extractor (DisPerSE) algorithm. DisPerSE is a source extractor algorithm based on discrete Morse theory which identifies topological features from 2D and 3D datasets, thus extracting the necessary skeletons. A catalogue of 1812 structures was obtained using this method, and these were subsequently visually inspected and classified into five types: marginally resolved clumps, resolved elongated structures, filaments, networks of filaments and complexes. We use the three elongated-type structures (i.e. filaments, networks of filaments and resolved elongated structures) and compare them with SEDIGISM clouds (Sec. \ref{sec: counterparts}). Hereafter we refer to these three structures collectively as ATLASGAL elongated structures (AG-El) whereas the filaments are individually refereed to as ATLASGAL filaments(AG-Fil).

\subsection{MWP}

Milky Way Project \citep[MWP;][]{simpson_2012, jayasingghe_2019} is a citizen science project aimed at identification of bubbles and bow shocks from the infrared images obtained using Spitzer Space Telescope Galactic plane surveys. It was launched in 2010 and provided the users with $\SI{4.5}{\mu m}$ and $\SI{8}{\mu m}$ images from GLIMPSE \citep{benjamin_2003, churchwell_2009} and $\SI{24}{\mu m}$ images from the MIPSGAL survey \citep{carey_2009}. The DR1 \citep{simpson_2012} was released in 2012 and has produced a catalogue of over 5000 bubbles.

The second data release of MWP culminates the visual analysis of $31000+$ citizen scientists during 2012--2017. It has identified Galactic structures by inspection of images from GLIMPSE, MIPSGAL, Spitzer Mapping of the Outer Galaxy \citep[SMOG; ][]{carey_2008} and Cygnux-X \citep{hora_2009} surveys. The project observes $0\degree < l < 65\degree$, $295\degree < l < 360\degree$ and $|b| < 1\degree$ and a few additional regions.
The bubbles were identified using an ellipse drawing tool to mark the location and shape of bubbles.
The identification quality was tested by comparing the bubbles to observations carried out by trained experts and using machine-learning algorithms.  
We use the 2600 bubbles \citep{jayasingghe_2019} identified from the DR2 to compare with the SEDIGISM clouds in Sec. \ref{sec: counterparts}.
%--------------------------------------------------------------------

\section{Methodology%: Cloud classification
}\label{sec: methods}

%This section describes the methods used to classify the SEDIGISM clouds (full sample; 10663 clouds) into different morphologies. 
We use the SEDIGISM data cubes with 3D masks (produced by DC21) to generate integrated intensity maps (Fig. \ref{fig: ring-like cloud image}--\ref{fig: NA cloud image}) of individual clouds. These are 2D (integrated intensity masked) images obtained by integrating the intensity in a 3D cube along the velocity axis. % (over all velocities).
These images are used to classify the clouds into different morphologies using two methods. The first method -- $J$ plot -- is an automated algorithm whereas the second method -- by-eye classification -- implies classification of clouds carried out visually \add{by the lead author}.
The cloud classifications are provided as a catalogue, that is further described in App. \ref{app: cloud catalogue}.

\subsection{$J$ plots}

$J$ plots \citep{jaffa_2018} is a method to classify and quantify a pixelated structure into different morphologies using its moment of inertia, i.e., the degree of elongation and the degree of concentration. The classification procedure involves calculation of the principal moments of inertia ($I_1 \, \& \, I_2$) for each cloud, using the surface density and area covered (in pixels). $I_1$ and $I_2$ are the principal moments of inertia along the two principal axes of the structure, such that the first principal axis is associated with the smaller principal moment, thus $I_1 \leq I_2$. These moments are then compared with the principal moment for a uniform surface density disk $\left(I_0 = \mathrm{\frac{AM}{4 \pi}} \right)$ of same area (A) and mass (M) and hence converted into ``$J$'' moments $J_1$ and $J_2$ as

\begin{equation}
    J_i = \frac{I_0 - I_i}{I_0 + I_i}\; , \quad
    i = 1, 2 \; .
\end{equation}

$J$ plots makes use of the connection between moment of inertia and mass concentration. For example, an increase in the concentration of mass towards the centre results in the decrease of the moment of inertia of the structure. This is used to identify centrally concentrated disks. The original algorithm \citep[described in][]{jaffa_2018} uses the dendrograms to segment images and generate hierarchical structures directly from the raw data. However, in this work, the input to the algorithm are the 2D cloud images (e.g. Fig. \ref{fig: ring-like cloud image}) generated via $\sc{scimes}$ and its only purpose is morphological classification. The structures are classified into three types based on their $J$ moments:

\begin{enumerate}
    \item Centrally concentrated disk (core) : $J_1 > 0 , \; J_2 > 0$.
    \item Elongated ellipse (filament) : $J_1 > 0  , \; J_2 < 0$.
    \item Ring (limb-brightened bubble) : $J_1 < 0 , \; J_2 < 0$.
\end{enumerate}

\add{The original implementation of $J$ plots was based on dust continuum surface-density maps} \citep{jaffa_2018}. \add{However, we apply the $J$ plots algorithm on CO integrated-intensity images, and it % Thus, our implementation of the $J$ plots compares the pixel weights of the structures while classifying instead of the surface densities. Moreover, the principal moments are obtained using the total structure weight rather than the cloud mass. It 
leads to a slight change in the terminology. Due to the non-trivial derivation of distances in the Milky Way, it is difficult to directly obtain the surface density and mass of the structures from the integrated-intensity images and thus we use the pixel weights and total structure weights in their place, respectively. However, these differences do not pose any changes in the overall classification scheme.} \add{The morphological classes resulting from the $J$-plot classification} \remove{These classifications} are hereafter collectively referred to as $J$-structures, and individually as $J$-core, $J$-filament and $J$-bubble (e.g. Fig. \ref{fig: j plot vs by-eye ring}).

\subsection{By-eye classification}\label{sec: by eye classification}
%--------------------------------------------------------------------

\begin{figure*}
    \centering
    \includegraphics[width = 0.45\textwidth, keepaspectratio]{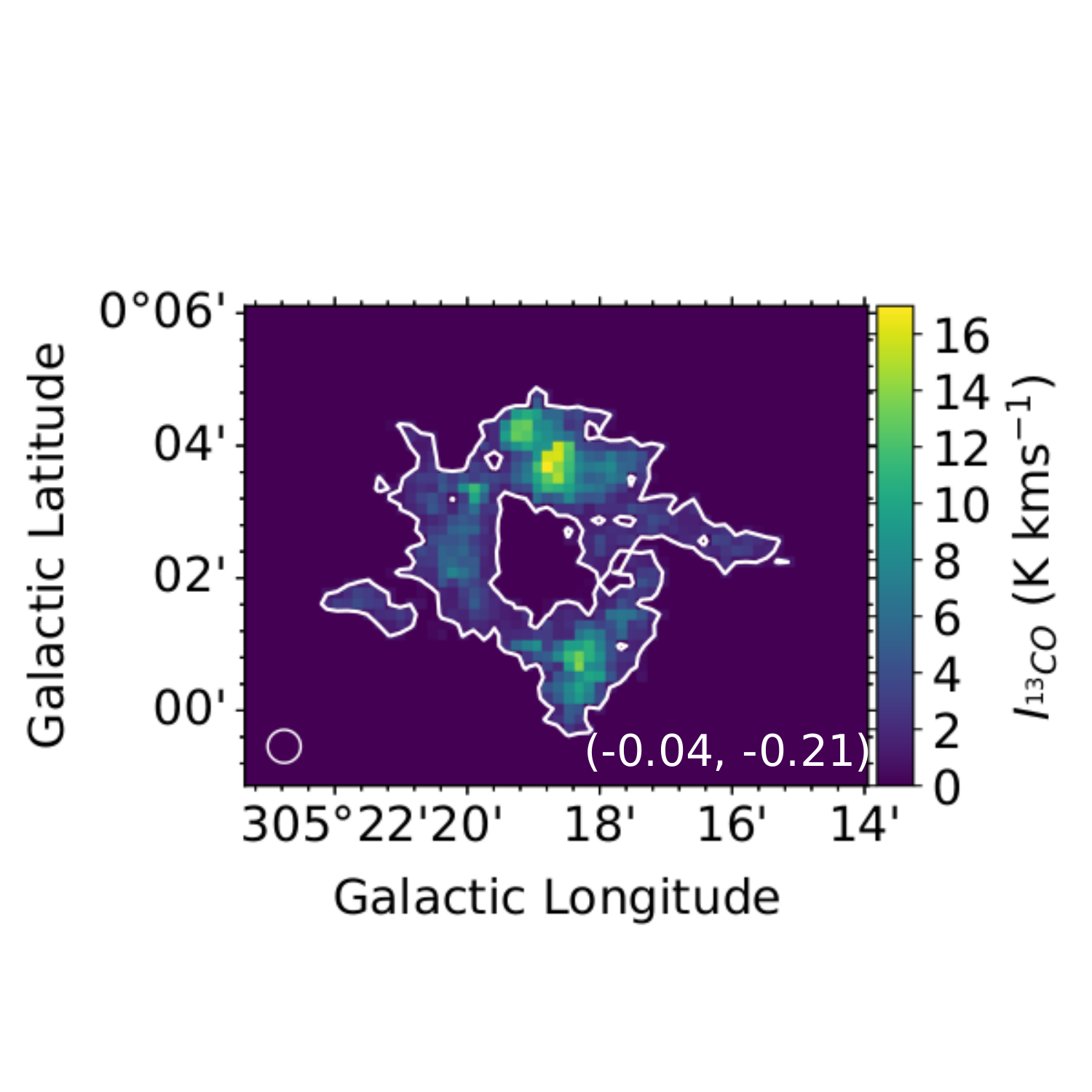}
    \includegraphics[width = 0.45\textwidth, keepaspectratio]{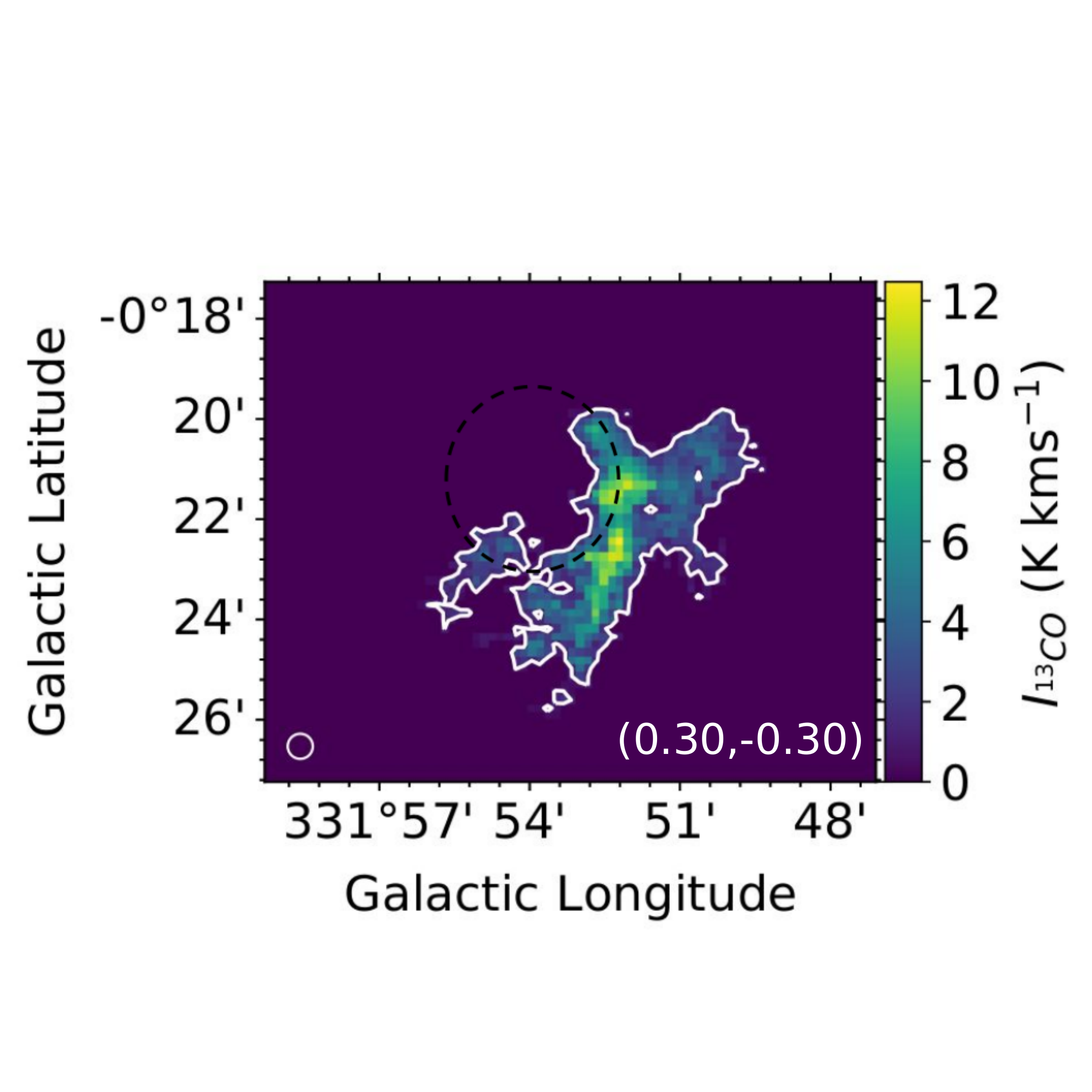}
    \caption{Examples of two ring-like clouds as per the visual classification method. \add{Left: $J$-bubble; cloud id: 238. Right: $J$-filament; cloud id: 3148. The cloud ids can be used to identify the cloud structure from the main catalogue (Table \ref{table: cloud catalog}).} The images are integrated intensity (moment 0) maps of the $^{13}$CO (2 -- 1) transition. \add{The numbers in the right bottom corner of the images represent the $J_1$ and the $J_2$ moments respectively. The colour bar represents the $^{13}$CO integrated intensity in K $\mathrm{km \; s}^{-1}$. The white contour represents the cloud edge.} The white circle at the bottom left of the figure represents the telescope beam size. The black ellipse on the second image represents an artistic impression of the visualised ring.}
    \label{fig: ring-like cloud image}
\end{figure*}

\begin{figure*}
    \centering
    \includegraphics[width = 0.45\textwidth, keepaspectratio]{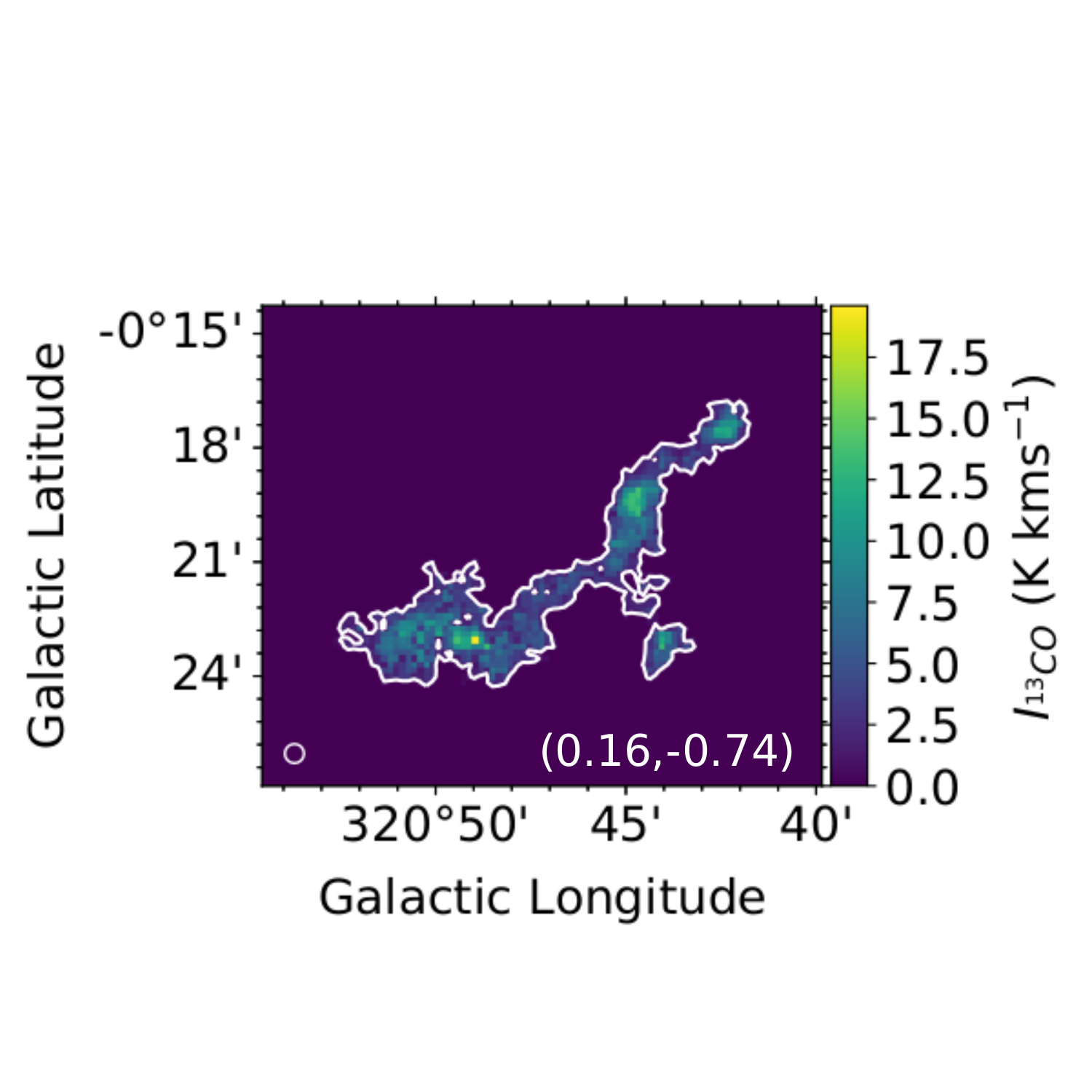}
    \includegraphics[width = 0.45\textwidth, keepaspectratio]{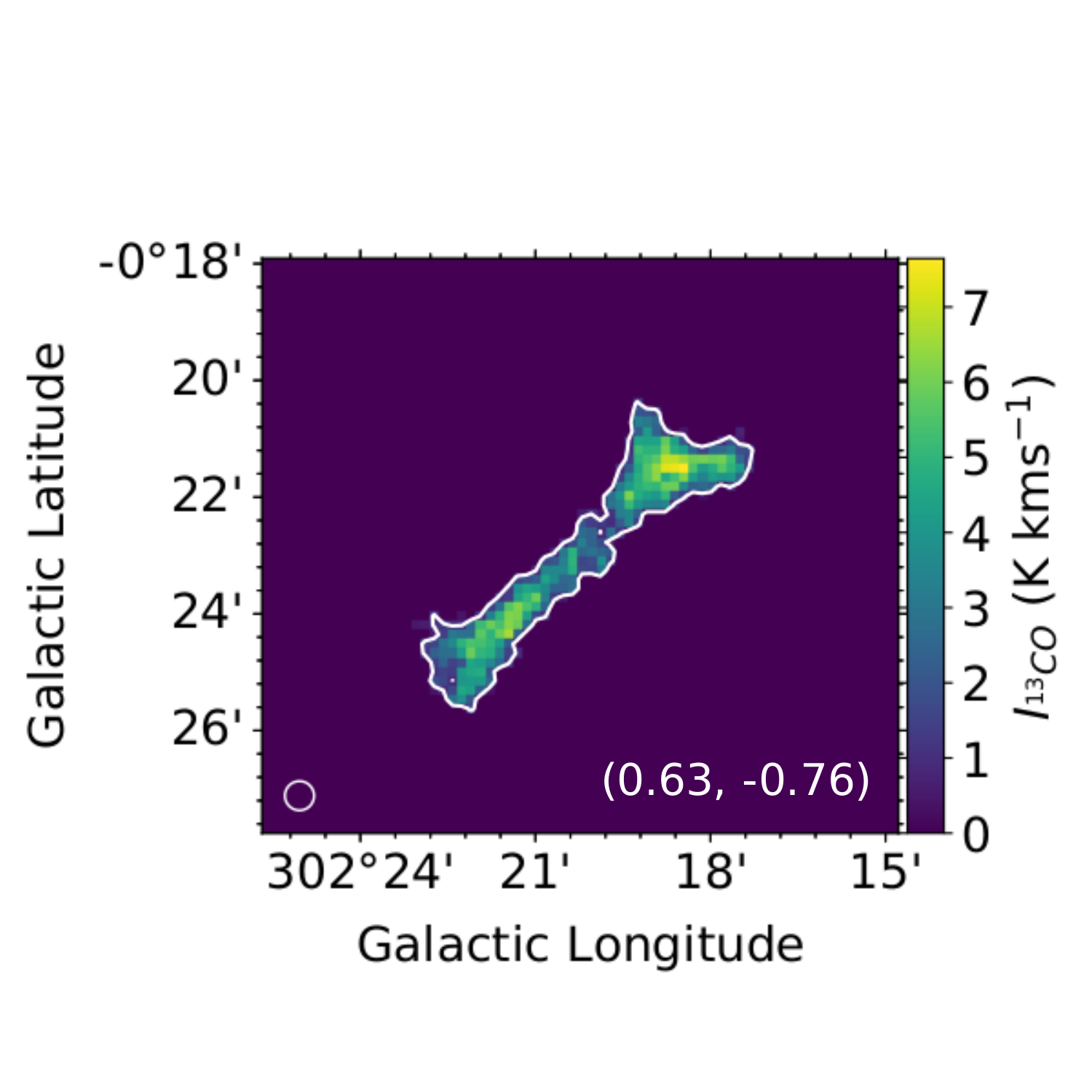}
    \caption{Examples of two elongated clouds as per the visual classification method. \add{Left: $J$-filament ; cloud id: 1726. Right: $J$-filament ; cloud id: 130.} The conventions follow Fig. \ref{fig: ring-like cloud image}.}
    \label{fig: elongated cloud image}
\end{figure*}

\begin{figure*}
    \centering
    \includegraphics[width = 0.45\textwidth, keepaspectratio]{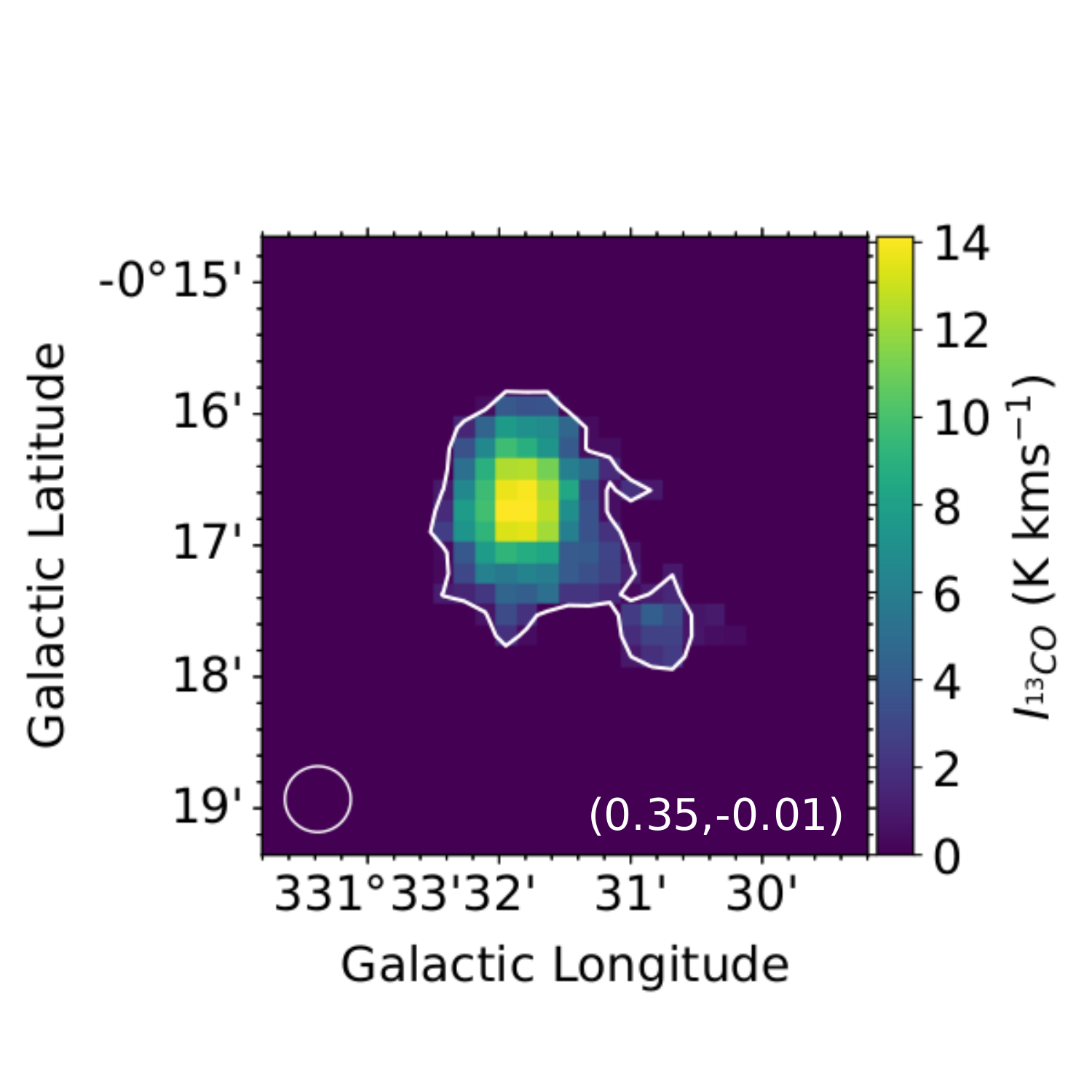}
    \includegraphics[width = 0.45\textwidth, keepaspectratio]{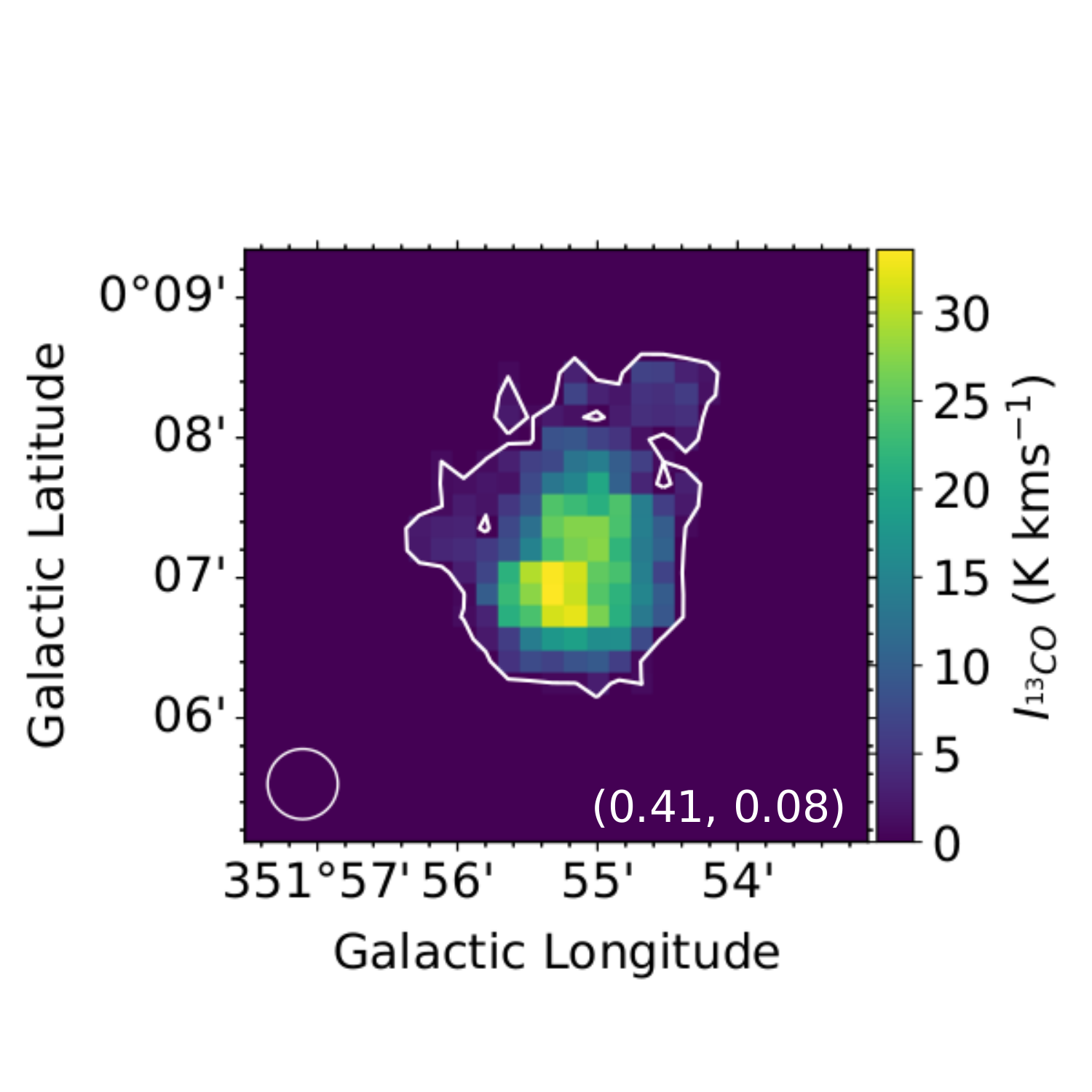}
    \caption{Examples of two concentrated clouds as per the visual classification method. \add{Left: $J$-filament ; cloud id: 3133. Right: $J$-core ; cloud id: 6567.} The conventions follow Fig. \ref{fig: ring-like cloud image}.}
    \label{fig: concentrated cloud image}
\end{figure*}

\begin{figure*}
    \centering
    \includegraphics[width = 0.45\textwidth, keepaspectratio]{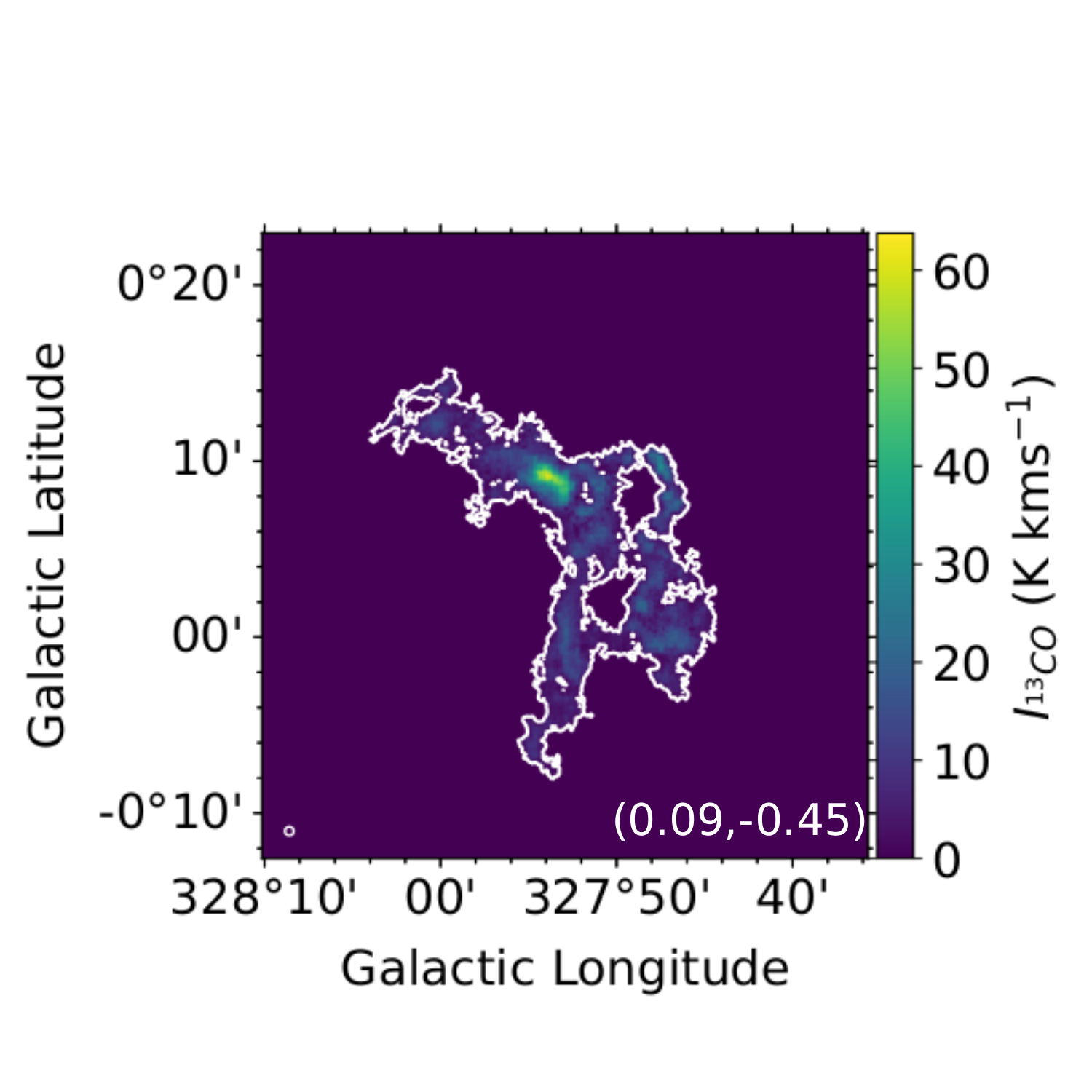}
    \includegraphics[width = 0.45\textwidth, keepaspectratio]{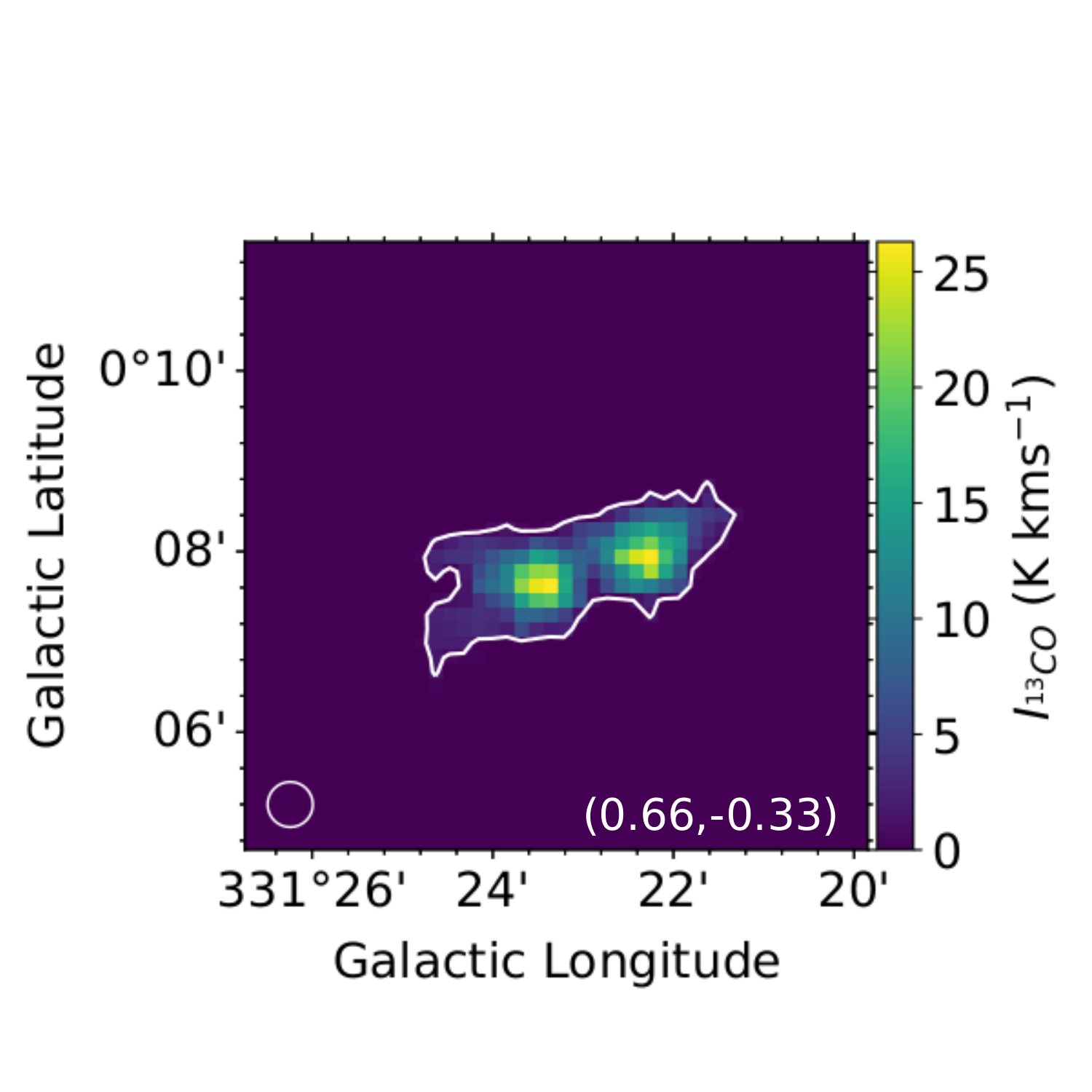}
    \caption{Examples of two clumpy clouds as per the visual classification method. \add{Left: $J$-filament ; cloud id: 2500. Right: $J$-filament ; cloud id: 3010.} The conventions follow Fig. \ref{fig: ring-like cloud image}.}
    \label{fig: clumpy cloud image}
\end{figure*}

We performed a visual inspection of the integrated intensity maps of clouds (e.g. Fig. \ref{fig: ring-like cloud image}) and classified them into four different morphological classes. Three of the classes -- ring-like, elongated and concentrated cloud -- are inspired from $J$ plots. We defined the fourth class of clouds, referred to as clumpy clouds, by combining smaller sub-classes \footnote{We initially had six visual sub-classes which were combined to get the current four classes. Complete and partial bubbles were combined to get the ring-like clouds. Elongated clouds with single and multiple denser clumps/regions were combined to get clumpy clouds. \add{A detailed description is provided in App. \ref{App: ks test}.}} on the basis of the two sample Kolmogorov–Smirnov (KS) test \add{and the two-sided Mann–Whitney U (MWU) test}. \remove{The KS test is a non-parametric test with a null hypothesis that two samples are drawn from the same distribution.} \add{The KS and MWU tests are non-parametric tests. The KS test has a null hypothesis that the two samples are drawn from the same distribution. The MWU test has a null hypothesis that neither distribution has statistical dominance over the other. We discuss the two test in details in App. \ref{App: ks test}. }The p-values presented \remove{throughout} \add{in} this work are obtained using the \add{two sample} KS test \add{and the two-sided MWU test}.
The visual classification provides a method to independently verify the $J$ plot classification. The morphologies are discussed below: 

\begin{enumerate}
    \item Ring-like clouds: clouds that resemble a ring or a bubble (Fig. \ref{fig: ring-like cloud image}). Comparable to $J$-bubble.
    \item Elongated clouds: clouds which are elongated in nature (Fig. \ref{fig: elongated cloud image}) and maintain a visibly uniform structure. Comparable to $J$-filament.
    \item Concentrated clouds: clouds which have most of their integrated intensity densely packed in a compact region (Fig. \ref{fig: concentrated cloud image}). These clouds have a spherical geometry. Comparable to $J$-core.
    \item Clumpy clouds: elongated clouds with presence of one or more smaller (visible) denser regions/clumps (Fig. \ref{fig: clumpy cloud image}). %The clouds with a large elongated structure accompanied by a small dense region fall into this category. These clouds are different from the concentrated clouds which present a spherical geometry. 
    Comparable to $J$-filament.
\end{enumerate}

\begin{figure*}
    \centering
    \includegraphics[width = 0.45\textwidth, keepaspectratio]{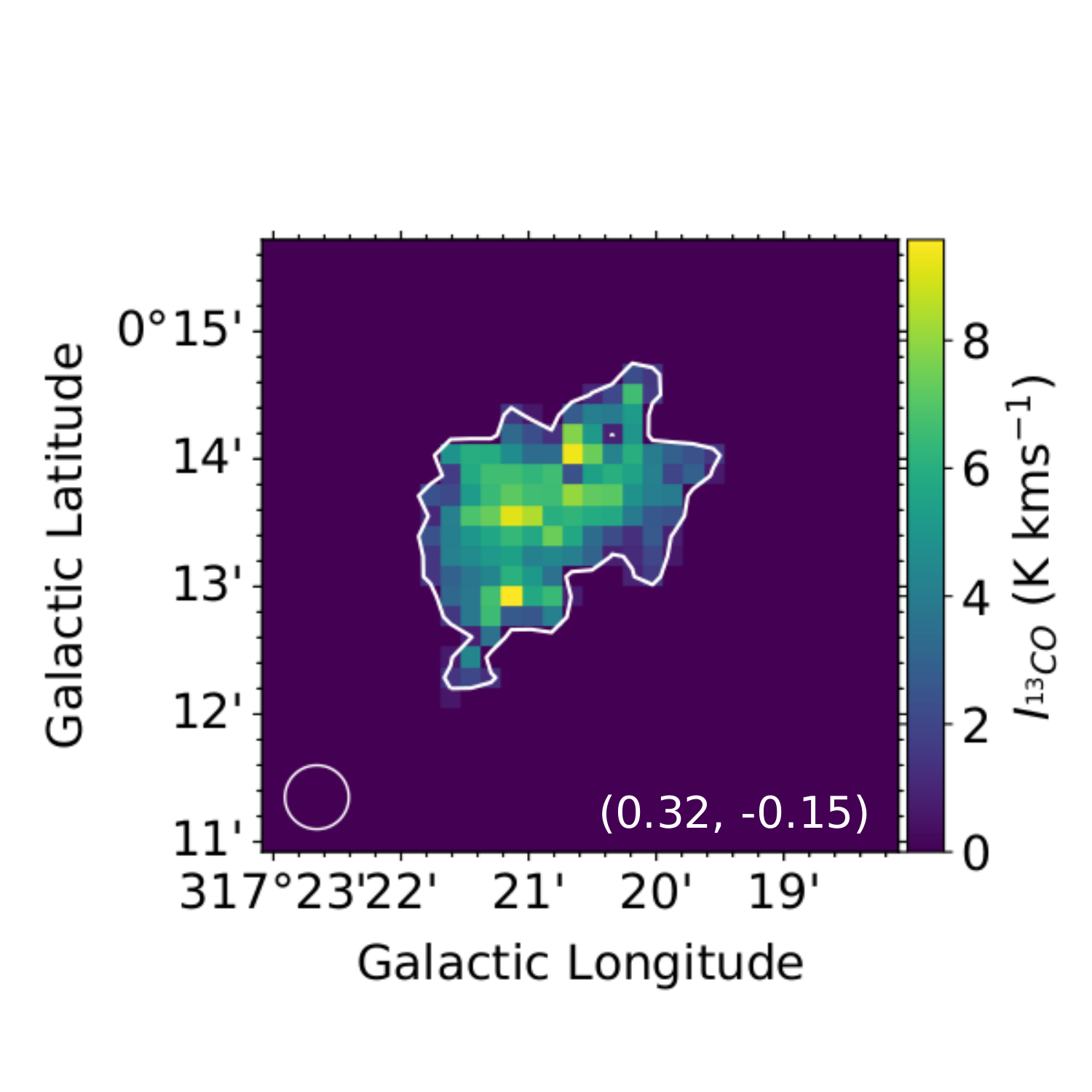}
    \includegraphics[width = 0.45\textwidth, keepaspectratio]{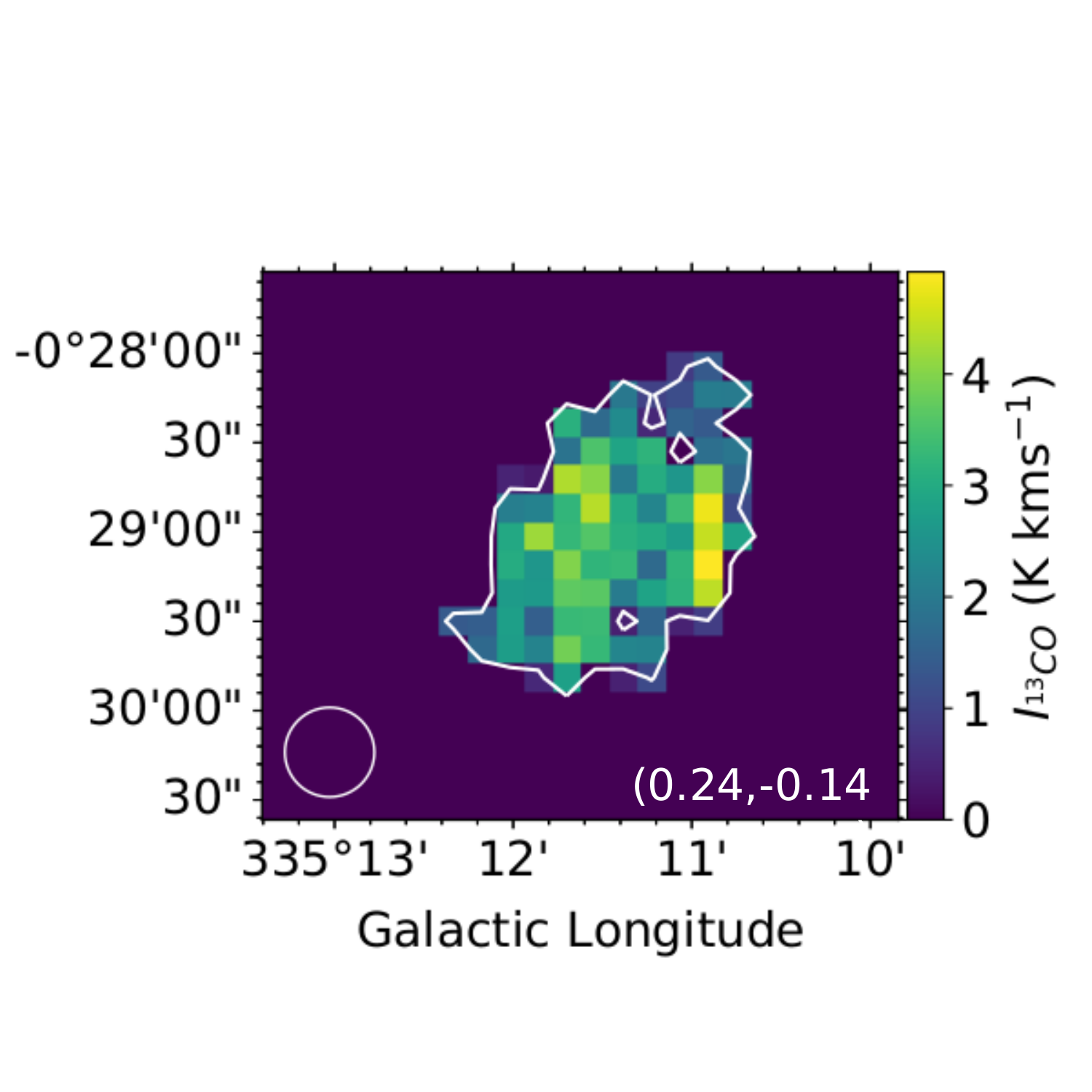}
    \caption{Examples of two unclassified clouds. \add{Left: $J$-filament ; cloud id: 1445. Right: $J$-filament ; cloud id: 3976.} The conventions follow Fig. \ref{fig: ring-like cloud image}.}
    \label{fig: NA cloud image}
\end{figure*}

There are 298 clouds that do not fit the description of any of the above morphologies (Fig. \ref{fig: NA cloud image}). They are typically diffuse in nature and close to the resolution element of the survey. These clouds are termed as unclassified, and we exclude them in our analysis (refer Sec. \ref{sec: vc and mr sample}).

\subsection{Limitations of \add{the classification methods} \remove{$J$ plots}}

$J$ plots is aimed at producing an automated classification of filaments, cores and bubbles by analysing their surface-density emission.
The algorithm is currently limited to detection (using dendrograms) and analysis of 2D structures from images. \citet{jaffa_2018} has applied $J$ plots on the region around RCW 120 using data from the Hi-GAL survey and successfully identified the previously known bubble and other ring-like structures. It also confirmed that in the third quadrant of the $J$ plots ($J_1 < 0 \, \& \, J_2 < 0$), the distance from the origin reflects the thickness of the ring (bubble). Similarly, it has also identified and quantified filaments from the smoothed-particle hydrodynamics (SPH) simulation of \citet{Clarke_2017}.

We use the velocity coherent SEDIGISM clouds identified in 3D PPV space and provide them to $J$ plots after integrating along the velocity axis. % for the identification of their morphology. The 3D datacubes are converted to 2D by integrating along the velocity axis before the use of $J$ plots, such that 
Therefore, the morphological analysis is done in 2D space. 
%This is the first time that $J$ plots has been applied on 3D molecular line data. This raises concerns on its reliability to classify clouds identified in a 3D space.
$J$ plot assumes strict limits of principal (and $J$) moments for all structures, e.g., for all bubbles, both the $J$ moments are negative. 
This may lead to an incorrect classification of some structures. For example, the interaction of an OB star with the ambient ISM can lead to deformities in the circular shape of a bubble \citep{jayasingghe_2019}. It could lead to one $J$ moment being positive for the structure, resulting in its identification as a filament instead of a bubble. Hence, elliptic bubbles might be incorrectly classified by $J$ plots.
We also see in Table \ref{tab: quantitative full sample} that $J$ plot classifies 87\% of the clouds as filaments while visual classification identifies only 67\% of the clouds as elongated structures (elongated \& clumpy clouds).
%We try to solve these issues by defining the following two samples.
\add{However, our visual (by-eye) classification isn't completely unambiguous. For example, some clouds display both clumpy structure and partial rings. We also see clouds with small curved branches along otherwise long filaments. To minimise the uncertainties in cloud classification, we present the morphologically reliable (MR) sample in Sec. \ref{sec: vc and mr sample}.}

As our analysis is performed in 2D space, we only see a projected image of the molecular clouds and this can lead to an incorrect morphological classification. For example, a filament lying completely in the line of sight of the telescope can appear to be concentrated in a small region, thus leading to it classification as a core. However, due to the large size of our sample, the projection effects are unlikely to affect the general results.

We also test if the morphological classification is affected by the noise in the data. The SEDIGISM cloud catalogue (DC21) contains the S/N ratio for each cloud. We choose the 1000 clouds with the $|J_1|$ values closest to zero, and choose the 100 most noisy clouds (lowest S/N) from them. Similarly, we choose the 100 most noisy clouds closest to $|J_2| = 0$. These 200 clouds are introduced with random noise ($\sim$ average noise in the SEDIGISM data cube to which the cloud belongs) and their morphologies are identified with $J$ plots. We observe that only three clouds\footnote{These clouds look identical under the visual classification irrespective of the noise.} show a change in the morphology (from core to filament), which indicates that the $J$ plots classification is mostly robust against the changes in noise.  %Moreover, our usage of $J$ plots is only limited to morphological classification of the clouds identified by DC21.

\begin{figure}
    \centering
    \includegraphics[width = 0.5 \textwidth, keepaspectratio]{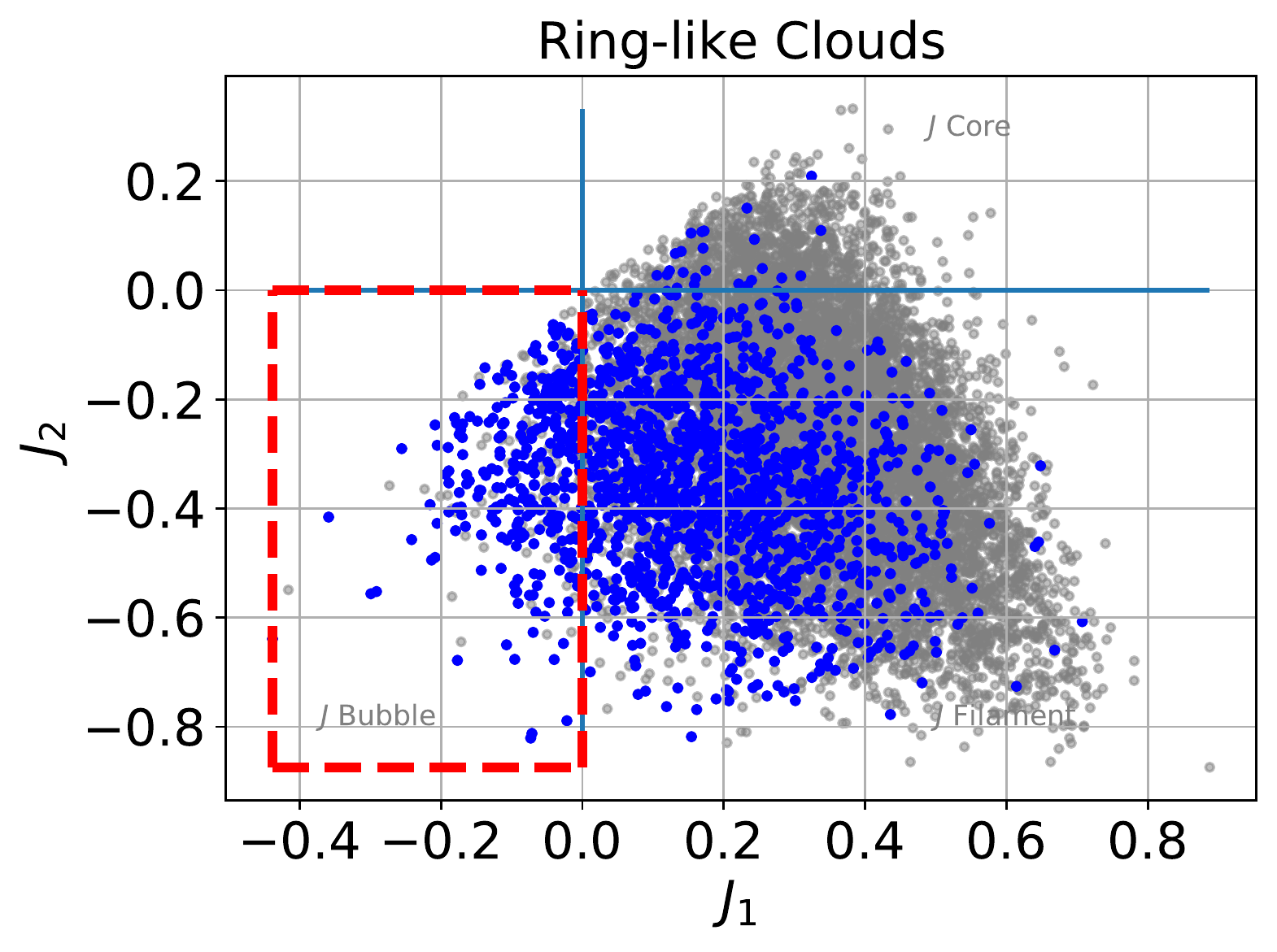}
    \caption{$J$ plot for visual class: ring-like clouds. \add{The blue dots represent the ring-like clouds whereas the grey dots represent all the clouds in VC sample. The blue dots lying inside the red dashed rectangle represent the morphologically reliable (MR) sample.} \remove{The red dashed rectangle represents the morphologically reliable (MR) sample.}}
    \label{fig: j plot vs by-eye ring}
\end{figure}

\begin{figure}
    \centering
    \includegraphics[width = 0.5 \textwidth, keepaspectratio]{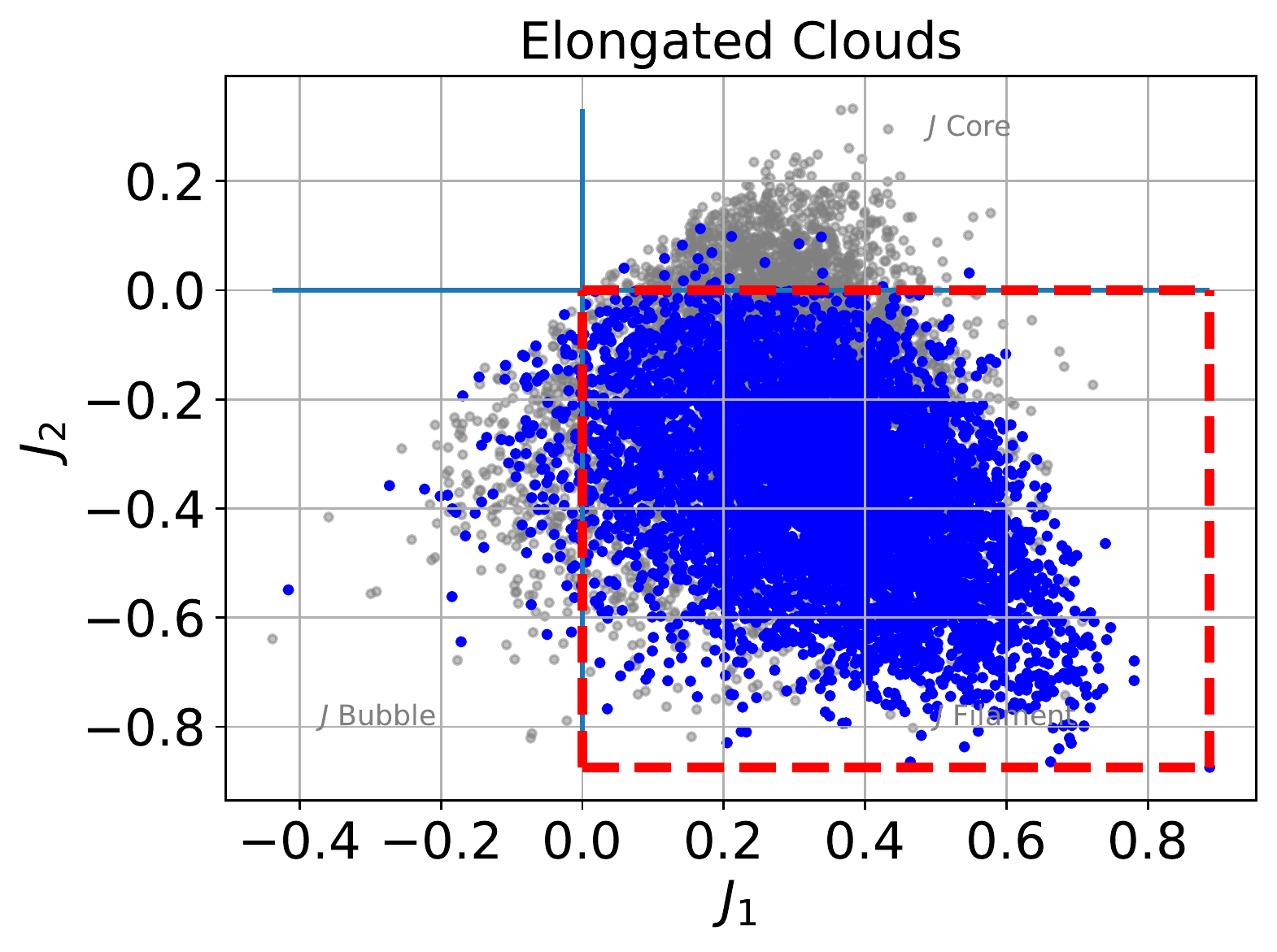}
    \caption{$J$ plot for visual class: elongated clouds. \add{The symbols and conventions follow Fig. \ref{fig: j plot vs by-eye ring}.} \remove{The red dashed rectangle represents the MR sample.}}
    \label{fig: j plot vs by-eye elon}
\end{figure}

\begin{figure}[h]
    \centering
    \includegraphics[width = 0.5 \textwidth, keepaspectratio]{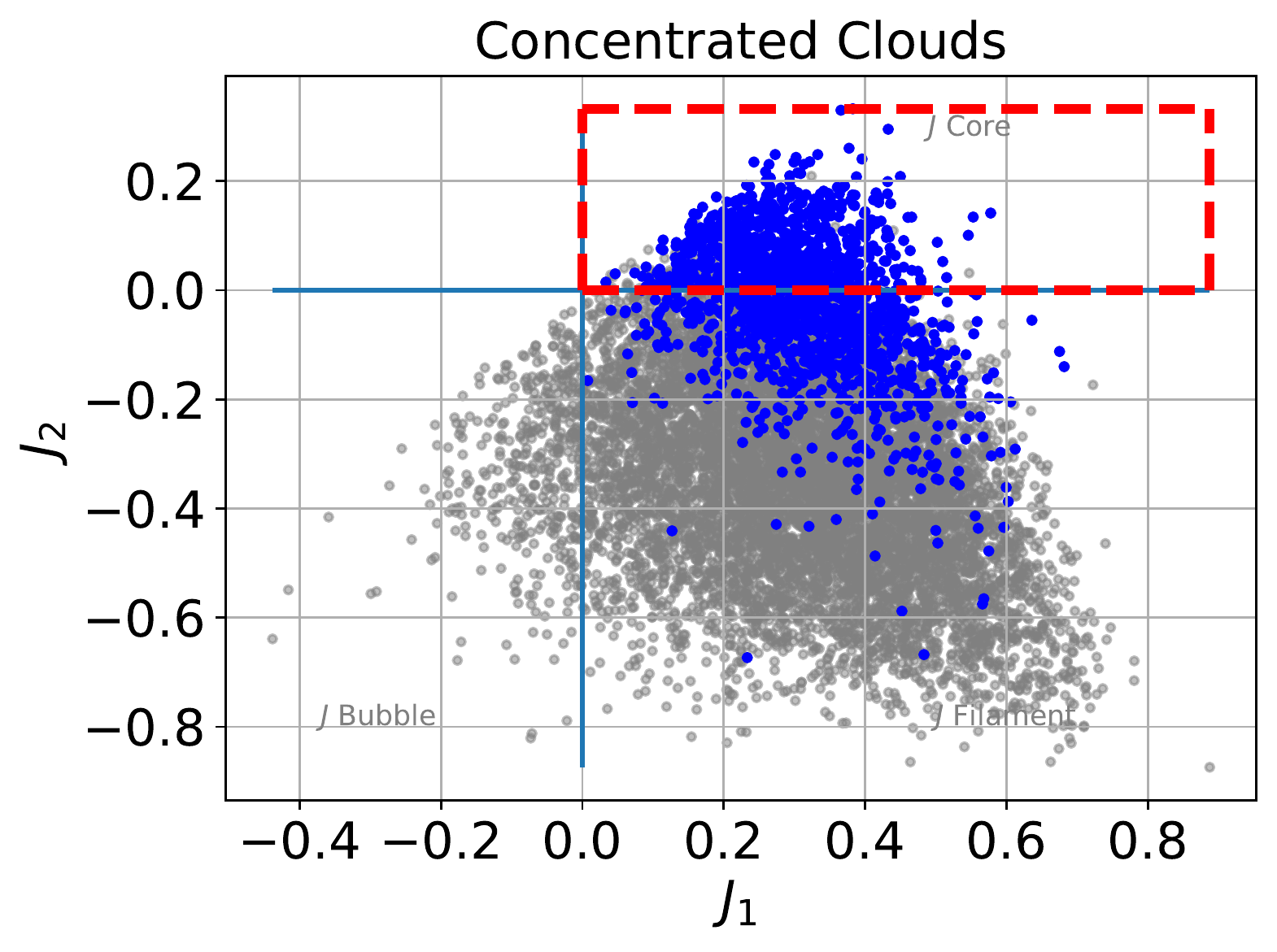}
    \caption{$J$ plot for visual class: concentrated clouds. \add{The symbols and conventions follow Fig. \ref{fig: j plot vs by-eye ring}.} \remove{The red dashed rectangle represents the MR sample.}}
    \label{fig: j plot vs by-eye conc}
\end{figure}

\begin{figure}[h]
    \centering
    \includegraphics[width = 0.5 \textwidth, keepaspectratio]{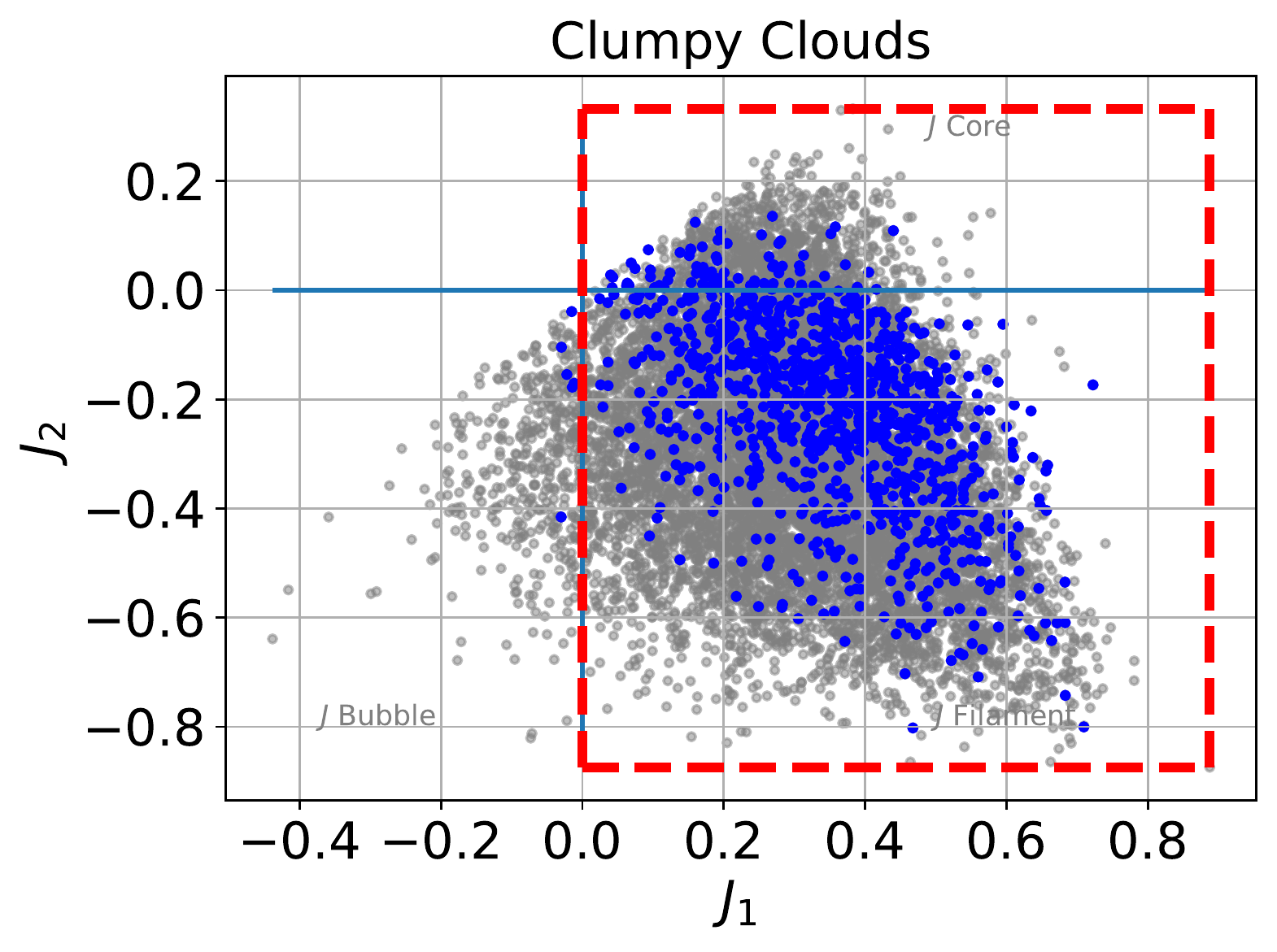}
    \caption{$J$ plot for visual class: clumpy clouds. \add{The symbols and conventions follow Fig. \ref{fig: j plot vs by-eye ring}.} \remove{The red dashed rectangle represents the MR sample.}}
    \label{fig: j plot vs by-eye clumpy}
\end{figure}
%%%%%%%%%%%%%%%%%%%%%%%%%%%%%%%%%%%%%%%%%%%%%%%%%%%%%%%%%%%%%%%%%%
%%%%%%%%%%%%%%%%%%%%%%%%%%%%%%%%%%%%%%%%%%%%%%%%%%%%%%%%%%%%%%%%%%

%\section{Reliability of $J$ plots}
\section{Morphological classification}\label{sec: reliability of j plots}

\subsection{Visually classified (VC) and morphologically reliable (MR) samples}\label{sec: vc and mr sample}

We use the SEDIGISM cloud sample to create two samples to study the cloud morphology. These are the visually classified (VC) and morphologically reliable (MR) samples. The VC sample is obtained from the by-eye classification of full sample by discarding the unclassified clouds (Sec. \ref{sec: methods}). The MR sample is a sub-sample of the VC sample, containing only the clouds for which the $J$ plots and by-eye classification morphologies are in agreement (i.e. the clouds' morphology is consistent for the two methods; refer Fig. \ref{fig: j plot vs by-eye ring} -- \ref{fig: j plot vs by-eye clumpy}). Therefore a cloud belongs to the MR sample only if it satisfies one of these conditions:

\begin{itemize}
    \item It is a ring-like cloud classified as bubble by $J$ plot.
    \item It is an elongated cloud classified as filament by $J$ plot.
    \item It is a concentrated cloud classified as core by $J$ plot.
    \item It is a clumpy cloud classified as core or filament by $J$ plot.
\end{itemize}

The clumpy clouds contain only the elongated clouds with dense regions/clumps in them. We include the $J$ cores as they might have elongated structure but recognised as a core due to high central density (Fig. \ref{fig: j plot vs by-eye clumpy}).
%A quantitative description of both the samples for different morphologies is provided in
In Table \ref{tab: quantitative full sample}, we list the total number of clouds assigned to each group. In brief, the MR sample contains clouds with the most reliable morphological classification and should be preferred. However, it excludes a large number of SEDIGISM clouds and has a low sample size, especially for ring-like clouds. A larger sample is the VC sample but it might contain subjective biases. %Hence, the choice of samples is dependent on the type of analysis.
In the parallel paper (Neralwar et. al. in prep), we confirm that the distance distributions of both the samples %w.r.t. distance 
follow each other closely. We also study the SEDIGISM cloud properties using the two samples and their results are in agreement. 

\begin{table}
\caption{Quantitative description of cloud groups. The columns represent the $J$ plot classes whereas rows represent the visual classes. The blue shaded cells represent the VC sample containing 10365 clouds. The grey shaded cells represent the MR sample containing 8086 clouds.}
\label{tab: quantitative full sample}
\centering
\resizebox{\columnwidth}{!}{\begin{tabular}{lllll}
\hline
Morphological class & $J$-bubble & $J$-filament & $J$-core & Total \\ \hline
Ring-like cloud       & \cellcolor[HTML]{C0C0C0}285        & 1306         & 24       & \cellcolor[HTML]{8FABFC}1615  \\
Elongated cloud       & 118        & \cellcolor[HTML]{C0C0C0}5915         & 22       & \cellcolor[HTML]{8FABFC}6055  \\
Concentrated cloud    & 0          & 803          & \cellcolor[HTML]{C0C0C0}822      & \cellcolor[HTML]{8FABFC}1625  \\
Clumpy cloud          & 6          & \cellcolor[HTML]{C0C0C0}984          & \cellcolor[HTML]{C0C0C0}80       & \cellcolor[HTML]{8FABFC}1070  \\
Unclassified          & 1          & 283          & 14       & 298   \\
Total                 & 410        & 9291         & 962      & 10663 \\ \hline
\end{tabular}}

\end{table}

\begin{comment}
\begin{table}[]
\caption{Quantitative description of the clouds in visually classified (VC) and morphologically reliable (MR) samples for different morphologies. The shaded cells represent the MR sample containing 5020 clouds.}
\label{tab: quantitative vc and mr samples}
\centering
\begin{tabular}{lllll}
\hline
Morphological class & $J$-bubble                  & $J$-filament                 & $J$-core                    & Total \\ \hline
Ring-like cloud     & \cellcolor[HTML]{C0C0C0}156 & 791                          & 14                          & 961   \\
Elongated cloud     & 60                          & \cellcolor[HTML]{C0C0C0}3607 & 9                           & 3676  \\
Concentrated cloud  & 0                           & 566                          & \cellcolor[HTML]{C0C0C0}528 & 1094  \\
Clumpy cloud        & 4                           & \cellcolor[HTML]{C0C0C0}688  & \cellcolor[HTML]{C0C0C0}41  & 733   \\
Total               & 220                         & 5652                         & 592                         & 6464  \\ \hline
\end{tabular}

\end{table}
\end{comment}

\subsection{Aspect ratio: Moments technique vs Medial Axis}

\begin{figure}
    \centering
    \includegraphics[width = 0.5 \textwidth, keepaspectratio]{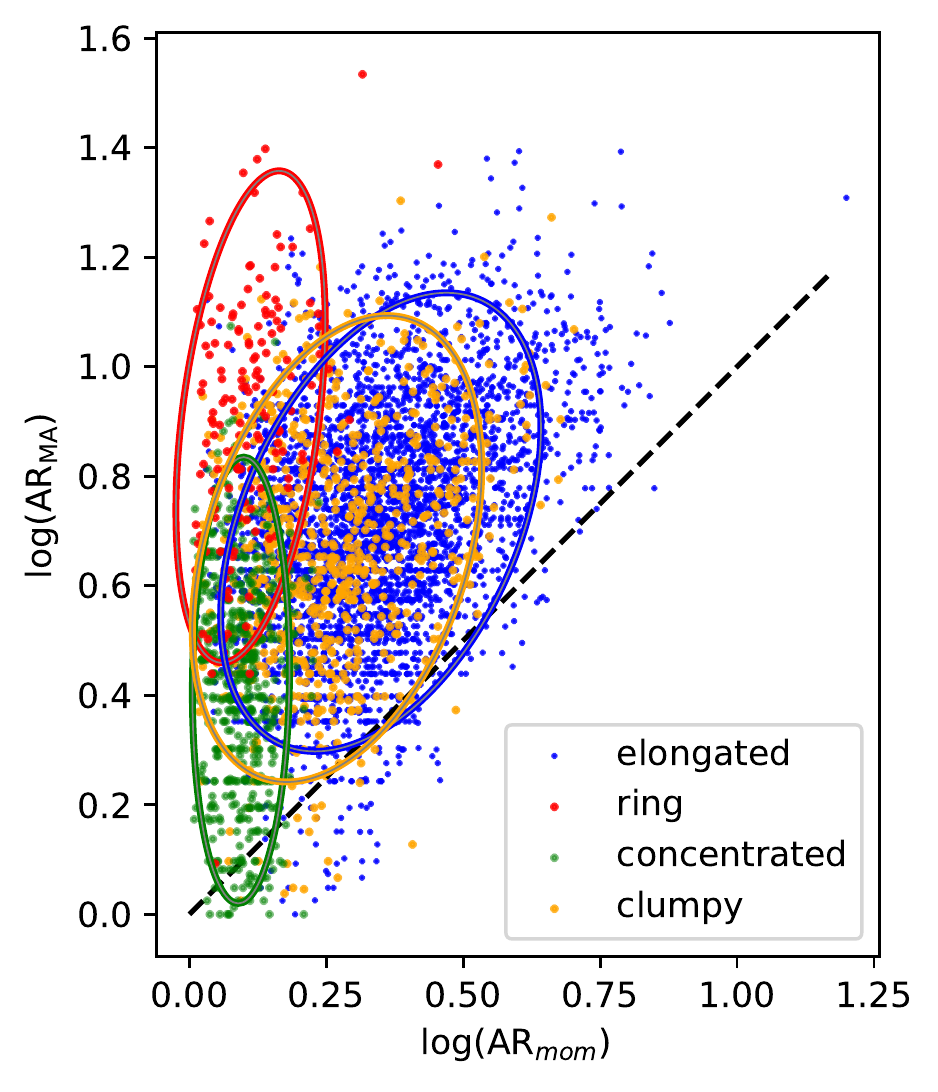}
    \caption{Comparison between aspect ratios obtained using the moment technique (AR$_{\mathrm{mom}}$) and the medial axis technique (AR$_{\mathrm{MA}}$) for different cloud morphologies. The \remove{confidence} ellipses \add{encompass approximately 95\% of the respective data points (2-sigma level).} \remove{represent the 2-sigma level.} \add{The black dashed line represents the 1:1 relation.}}
    \label{fig: asp rat}
\end{figure}

We use our most reliable sample -- MR sample -- to check whether the aspect ratios obtained using two different methods show some trends with respect to different morphologies (Fig. \ref{fig: asp rat}) or whether both methods provide the same information regarding the aspect ratio of the structures.
The first method to obtain the aspect ratio is the moment technique \citep[described in][]{crops}. It uses principal component analysis (PCA) to determine the orientation of the major axis of a cloud. The geometric mean of the second spatial moments gives the rms size of the cloud along the two orthogonal axes. The ratio of this intensity weighted major and minor axes gives the aspect ratio -- AR$_{mom}$. The AR$_{mom}$ values for each cloud have been automatically calculated while building the catalogue for the dendrograms of emission and are provided in the SEDIGISM cloud catalogue (DC21).

The second method to obtain an estimate of the aspect ratio is using a geometrical medial axis (a quantity also computed by DC21, and provided in the cloud catalogue). Such a medial axis is the longest-running spine passing centrally through the entire length of a 2D projected cloud mask. The cloud length is the medial axis length, whereas the cloud (medial axis) width is twice the average distance from this central spine to the cloud edge. The ratio of the medial axis length to medial axis width gives the aspect ratio -- AR$_{MA}$. The medial axis length is free of the assumption that clouds have any particular geometry and hence provides an alternative estimate of the cloud ``shape''.

We compare the two aspect ratios for the MR sample in Fig. \ref{fig: asp rat}, where a rough distinction between the different morphologies is seen.
The two approaches give different aspect ratios for the same cloud depending on the cloud's morphology and potentially also the internal structure (as the moment method takes into account the pixel intensity to calculate the two semi-axis lengths). We might expect the ARs to diverge for complex structures, while to give more comparable measurements for simpler shaped clouds, such as the concentrated objects. For instance, the majority of ring-like clouds do not form a complete ring. The moment technique (PCA) intrinsically perceives these clouds as elliptical structures whereas medial axis runs along their shell. This leads to AR$_{MA}$ being a ratio of the perimeter and thickness of cloud shell whereas AR$_{mom}$ traces a more spherical geometry, relating more to the size of the bubble that created the ring. Hence, we expect ring-like clouds to have a typically low AR$_{mom}$ but a high AR$_{MA}$. This effect might be not so pronounced in the case of other morphologies, leading to concentrated clouds having both aspect ratios with low values, while elongated and clumpy clouds are expected to have high AR$_{mom}$ and AR$_{MA}$. Nevertheless, in almost every case, AR$_{MA}$ appears to be larger than AR$_{mom}$. We conclude that aspect ratio measuring methods can give widely different results depending on the cloud morphology and the complexity, and they need to be used with care. %For example, the AR$_{mom}$ might provide better understanding of the shape of the concentrated clouds whereas AR$_{MA}$ might be a better estimate for the elongated clouds.

\begin{comment}
\begin{table}[]
\caption{Aspect ratios for clouds}
\label{table: cloud aspect ratio}
\begin{tabular}{ccc}
\hline
Cloud        & AR$_{\mathrm{mom}}$ & AR$_{\mathrm{MA}}$ \\ \hline
Ring-like    & Low                  & High                \\ 
Elongated    & High                 & High                \\ 
Concentrated & Low                  & Low                 \\ 
Clumpy       & High                 & High                \\ \hline
\end{tabular}

\end{table}
\end{comment}

%--------------------------------------------------------------------
\section{Results and Discussion}\label{sec: results and discussions}
\subsection{Cloud counterparts from other surveys}\label{sec: counterparts}

In this section, we compare the positions of SEDIGISM clouds and the structures from ATLASGAL and MWP. It helps us to understand how the low density molecular gas follows the dense structures observed through ATLASGAL. The comparison may shed light on the connection between AG-El and elongated clouds. 
MWP traces PAH regions, HII regions and warm dust, which have a higher excitation temperature than the $^{13}$CO emission \citep{mazumdar_2021}. These regions (bubbles) are often a consequence of the stellar feedback from massive stars. We use these bubbles to study whether the molecular gas surrounding these regions is detected as our ring-like clouds. ATLASGAL and MWP are continuum surveys and the corresponding structures lack velocity information. Therefore, we compare the positions of clouds and (ATLASGAL and MWP) structures only in two-dimensional position-position (p-p) space. The positions of pixels for each AG-El are provided in \citet{ATLASGAL_filaments}. The MWP bubble catalogue \citep{jayasingghe_2019} contains centroids for the ellipses (bubbles) and their major and minor axes. These parameters give us the position of the bubble but they do not provide us any information about the completeness of the ring. We use these parameters and pixel positions to compare the position of these structures with SEDIGISM clouds.

The lack of velocity information for continuum structures can be addressed for the structures that uniquely match with a single SEDIGISM cloud. SEDIGISM clouds are obtained using the $\sc{scimes}$ algorithm on the 3D PPV data cubes and therefore are coherent structures. Hence, the ATLASGAL and MWP structures overlapping with only a single SEDIGISM cloud are expected to be coherent as well. The velocity of the matching cloud thus provides an estimate for the velocity of the overlapping continuum structure. However, we are not able to identify all the coherent structures from the ATLASGAL and MWP sample. This is due to $\sc{scimes}$ focusing on identifying clouds with similar properties (e.g. similar sized clouds), which leads to large coherent structures getting identified as separate SEDIGISM clouds.

We use `ATLASGAL elongated structures (AG-El)' to collectively refer to resolved elongated structures, filaments and networks of filaments from \citet{ATLASGAL_filaments}. Additionally, the filaments from \citet{ATLASGAL_filaments} are analysed separately and referred to as `ATLASGAL filaments (AG-Fil)'. The kinematics of these filaments (few excluded) have been previously studied by \cite{mass_vel_scaling_relation} through comparing them with the SEDIGISM data, i.e. without segmentation into clouds \citep{schuller_2017, SEDIGISM_1}.
`MWP bubbles' refer to bubbles from the Milky Way project DR2 \citep{jayasingghe_2019}. A quantitative description of the structures from these two surveys that overlap with the SEDIGISM clouds (full sample, 10663 clouds) is provided in Table \ref{table: overlapping structures}.  
%There are no networks of filaments which overlap with exactly one SEDIGISM cloud, whereas .

\begin{table}[h]
\caption{Number of structures from ATLASGAL and MWP (described in \ref{sec: data}) overlapping with SEDIGISM clouds (full sample; 10663 clouds).
The `Total' column refers to the total structures from the respective catalogue, the `Coverage' column refers to the structures that lie inside the SEDIGISM coverage, the `Overlap' column refers to the structures overlapping with SEDIGISM clouds and the `Single overlap' column notes the ATLASGAL and MWP structures overlapping with exactly one SEDIGISM cloud.}
\label{table: overlapping structures}
\centering
\begin{tabularx}{\columnwidth}{Xcccc}
\hline
Structures         & Total & Coverage & Overlap & Single overlap \\ \hline
AG-El  & 937    &498       &497      & 43               \\
AG-Fil & 517     &271      &270    & 24               \\
MWP  bubbles   & 2600       &1222       &1188       & 201               \\ \hline
\end{tabularx}

\end{table}
% The number of 'Total' Structures with centers in SEDIGISM coverage are ATLASGAL elon - 523, ATLASGAL fil - 283 and MWP bubbles 1942

\begin{comment}
Similarly, a quantitative description of the clouds overlapping with their counterparts is presented in Table \ref{table: overlapping clouds vc} \&
\ref{table: overlapping clouds mr} for the VC and MR samples respectively.  We find a total of %4968 new elongated clouds in the full sample
4914 new elongated clouds in the VC sample and 
4803 new elongated clouds in the MR sample. These are the elongated SEDIGISM clouds that do not overlap with an elongated structure from the ATLASGAL survey. 
We also compare the SEDIGISM clouds to the 2600 bubbles from MWP.
There are %932 ring-like clouds in the full sample 
1010 ring-like clouds in the VC sample and 185 in the MR sample that do not overlap with the MWP bubbles.
\end{comment}

\begin{table}[]
\caption{\add{Number of overlapping clouds from the VC sample. The table presents the full sample of clouds which overlap with the structures presented in Table \ref{table: overlapping structures}. The column AG-El, AG-Fil and MWP represents the number of clouds overlapping with the ATLASGAL elongated structures} \citep{ATLASGAL_filaments}, ATLASGAL filaments \citep{ATLASGAL_filaments} and MWP bubbles \citep{jayasingghe_2019} respectively.}
\label{table: overlapping clouds vc}
\centering
\begin{tabular}{lccc}
\hline
Clouds       & AG-El & AG-Fil & MWP  \\ \hline
Ring-like    & 562      & 318       & 605  \\
Elongated    & 1141     & 562       & 1257 \\
Concentrated & 261      & 108       & 311  \\
Clumpy       & 299      & 150       & 348  \\ \hline
VC sample    & 2263     & 1138      & 2521 \\
Unclassified & 28       & 15        & 52   \\ \hline
Total        & 2291     & 1153      & 2573 \\ \hline
\end{tabular}

\end{table}

\begin{table}[]
\caption{Number of overlapping clouds from the MR sample lying completely inside SEDIGISM coverage (edge = 0; described in DC21). \add{The symbols and conventions follow Table \ref{table: overlapping clouds vc}.} The last column represents the total number of clouds of various morphologies belonging to the MR sample and lying completely inside the SEDIGISM coverage}
\label{table: overlapping clouds mr}
\centering
\begin{tabular}{lcccc}
\hline
Clouds       & AG-El & AG-Fil & MWP  & Total\\ \hline
Ring-like    & 83       & 42        & 87    & 258  \\
Elongated    & 1076     & 525       & 1154  & 5596 \\
Concentrated & 129      & 50        & 143   & 805  \\
Clumpy       & 289      & 141       & 324   & 1002 \\
Total        & 1577     & 758       & 1708  & 7918 \\ \hline
\end{tabular}

\end{table}

The overlap between the ATLASGAL structures, MWP bubbles and the SEDIGISM sample for the whole survey is presented in App. \ref{app: cloud overlap}.
Almost all of the AG-El lying in the SEDIGISM range have a cloud counterpart.  
This is easily understood, since the dust emission from ATLASGAL typically traces the high column density regions within larger molecular clouds. In that sense, the AG-El, are often surrounded by lower density material, which is seen in the $^{13}\mathrm{CO}$ emission. However, only $\approx 21\%$ of the SEDIGISM clouds (full sample) have a high-density filamentary ridge seen as an ATLASGAL elongated structure. This is in agreement with the findings from DC21, in which only 16\% of clouds had an ATLASGAL clump counterpart. 
Out of 10663 clouds from the full sample, only 2291 have an AG-El counterpart, of which 1141 are elongated clouds and 299 are clumpy clouds. 
Similarly, 1153 clouds have AG-Fil counterpart, of which 562 are elongated and 150 are clumpy.
\remove{A quantitative description for the MR sample clouds lying completely inside the SEDIGISM coverage is provided in Table \ref{table: overlapping clouds mr}.}
\add{A quantitative description of the clouds overlapping with their counterparts is presented in Table \ref{table: overlapping clouds vc} \&
\ref{table: overlapping clouds mr} for the VC and the MR samples respectively.}

%AG-El are often seen to have complete overlap with SEDIGISM clouds. 
We find that more than 90\% MWP bubbles have a SEDIGISM molecular cloud counterpart and these bubbles are usually overlap with patches of SEDIGISM clouds. This is an expected behaviour for the molecular gas surrounding an HII region \citep{jianjun_2020, tiwari_2021}, which is often disrupted by the feedback processes responsible for forming the bubbles. 
Also, due to the nature of the MWP identification procedure, the knowledge of the completeness of the ring is not provided, forcing us to consider them as complete ellipses.
Out of 10663 clouds in the SEDIGISM full sample only 2573 (24\%) have a MWP bubble counterpart, of which 605 are ring-like clouds. \add{It accounts to $\approx 23\%$ which is different from the general distribution of ring-like clouds ($\approx 15\%$ of clouds in VC sample are ring like), suggesting that ring-like clouds are disproportionately related to MWP bubbles.} A MWP bubble could be seen overlapping with multiple SEDIGISM clouds simply due to $\sc{scimes}$ identifying an actual coherent cloud as separate structures.
Not all of these clouds might be recognised as ring-like clouds, leading to a lower number of ring-like clouds overlapping with the MWP bubbles. Moreover, some of our ring-like clouds might form as a result of turbulence in the ISM and lack a surrounding HII region, thus not being detected by MWP.

\citet{mass_vel_scaling_relation} has studied the kinematics of 283 ATLASGAL filament candidates \citep{ATLASGAL_filaments} using data from the SEDIGISM survey, in order to figure out which of these are velocity coherent structures. The filament candidates belong to the region of overlap between the two surveys and lack velocity information due to being identified from dust continuum data. The process for obtaining the kinematic information for the filament candidates involve overlaying the filament pixels on the SEDIGISM emission grid and identifying velocity components for emission peaks of averaged spectra for each structure. \citet{mass_vel_scaling_relation} found that 260 filament candidates have accompanying SEDIGISM emission and 180 filaments are fully coherent structures containing a single velocity component.

We identify 271 AG-Fil in the SEDIGISM coverage, out of which 270 have an overlapping SEDIGISM counterpart and 24 show an overlap with a single SEDIGISM molecular cloud. These numbers are different from those reported by \citet[]{mass_vel_scaling_relation} because we exclude the AG-Fil that overlap with the edge of the SEDIGISM coverage and consider the cloud-filament overlap only in 2D (p-p) space.
%270 AG-Fil have an overlapping SEDIGISM counterpart. It includes the AG-Fil even with a partial (tiny) SEDIGISM overlap (in 2D) and this could be the reason for a different number if overlapping structures compared to \citet[260 filament candidates][]{mass_vel_scaling_relation}.
%We find that 24 AG-Fil that show a single overlap (refer Table \ref{table: overlapping structures}). There is a difference in the number of coherent filaments as compared to \citet{ATLASGAL_filaments} 
Moreover, we have used a different algorithm to identify the SEDIGISM structures. The algorithm developed by \citet[]{mass_vel_scaling_relation} identifies all velocity components along the line-of-sight that are correlated with the AG-Fil. It derives the kinematic properties of these velocity components and identifies coherent structures in position-position-velocity space. We characterise an AG-Fil as coherent if it overlaps with a single SEDIGISM cloud (identified by $\sc{scimes}$). $\sc{scimes}$ is oriented towards the identification of molecular clouds with similar properties, from the SEDIGISM survey. Hence, many of the coherent AG-Fil \citep[from][]{mass_vel_scaling_relation} overlap with multiple SEDIGISM clouds and this leads to the different number of coherent filaments between the two analyses.

Differently from \citet{mass_vel_scaling_relation} we also check whether not only ATLASGAL filaments, but also the elongated and network-like dust features overlap with a single cloud. We find 19 coherent resolved elongated structures and 24 coherent filaments (Table \ref{table: overlapping structures}). The matching of ATLASGAL structures with our molecular clouds has also led us to discover that each network of filaments overlaps with at least two clouds. This agrees with their description as a connection of several filaments \citep{ATLASGAL_filaments}, that are unlikely to be a single coherent structure.

%--------------------------------------------------------------------
\subsection{Star Formation Properties}\label{sec: star formation prop}

\subsubsection{Star Formation Efficiency and Dense Gas Fraction}\label{sec: sfe dgf prop}

Star formation plays an important role in galaxy evolution and it can be studied using dense material traced by dust in the galaxy, which can be observed at sub-mm wavelengths. Star formation in molecular clouds can be quantified to an extent using the Star Formation Efficiency (SFE) and the Dense Gas Fraction (DGF), which vary greatly from cloud to cloud \citep{eden_2012}. \citet{james_paper} have obtained the SFE and DGF for the SEDIGISM clouds with an ATLASGAL counterpart. These are calculated using the mass ($M_\mathrm{GMC}$; DC21) of SEDIGISM clouds and the mass ($M_\mathrm{clump}$) and the luminosity ($L_\mathrm{clump}$;  \citealt{Urquhart_2018}) of the clumps (belonging to the cloud) identified using the ATLASGAL survey (Eqn. \ref{eqn: dgf calc} \& \ref{ eqn: sfe calc}). %The luminosity of a clump is obtained using its distance and the observed flux. The temperatures for the clumps are derived from greybody fits to the dust emission and are further used to obtain the clump mass \citep{Urquhart_2018, urquhart_2013, hildebrand_1983}.
\add{The SFE (Eqn. \ref{ eqn: sfe calc}) is obtained under the assumption that the initial mass function is universal and completely sampled. Thus, Eqn. \ref{ eqn: sfe calc} is a proxy for the actual star formation efficiency. A detailed description of the assumption is provided in} \citet{james_paper}.

\begin{align}\label{eqn: dgf calc}
    \mathrm{DGF} &= \frac{\sum M_\mathrm{clump}}{M_\mathrm{GMC}} \; , \\ \label{ eqn: sfe calc} 
    \mathrm{SFE} &= \frac{\sum L_\mathrm{clump}}{M_\mathrm{GMC}}\;  \; \add{[\si{L_\odot/M_\odot}].}   
\end{align}

\begin{figure}
    %\hspace*{-2cm} 
    \centering
    %\begin{minipage}{1.1\textwidth}
    \includegraphics[width = .5\textwidth, keepaspectratio]{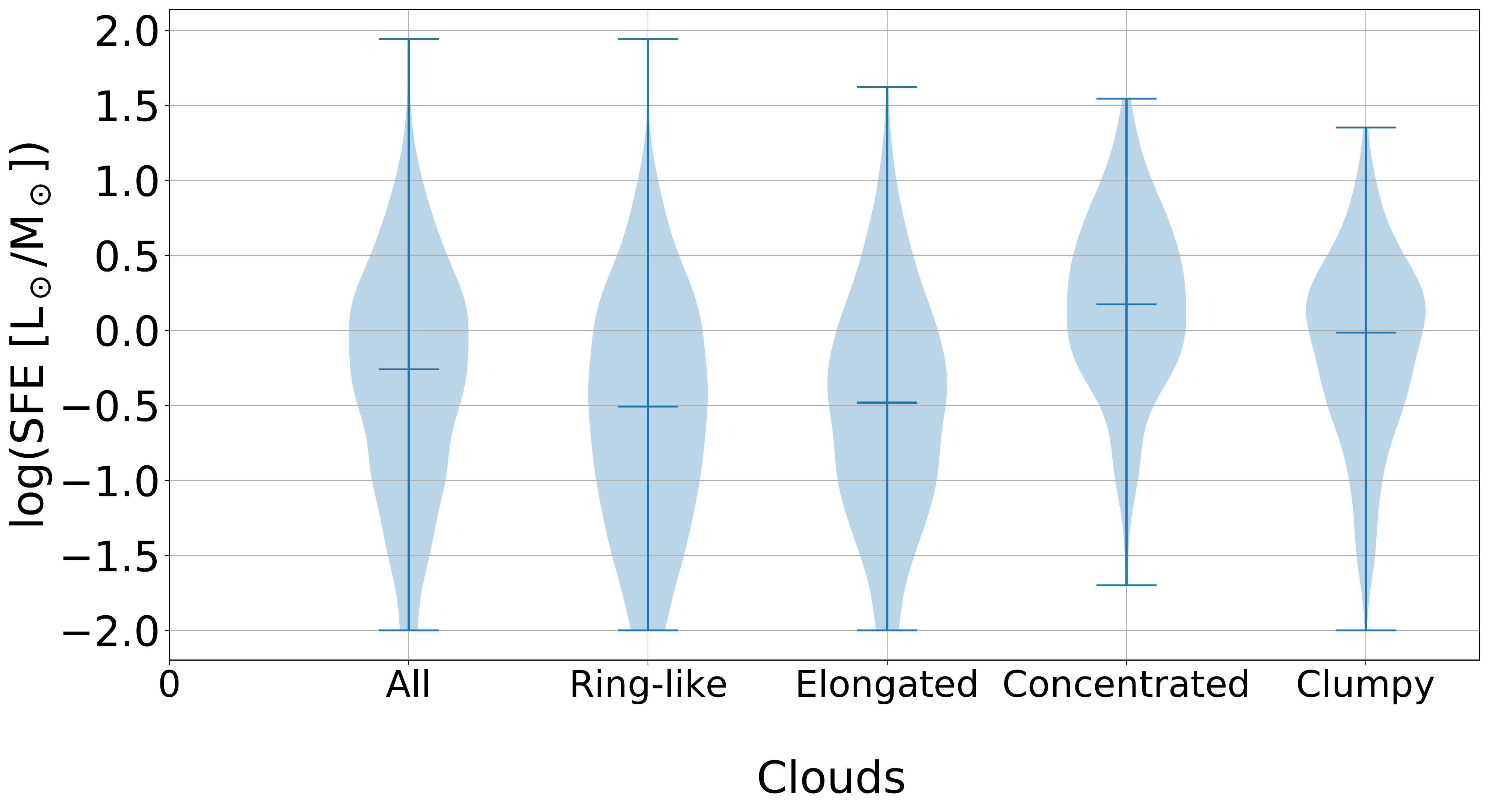}
    
    \vspace{0.5 cm}
    
    \includegraphics[width = .5\textwidth, keepaspectratio]{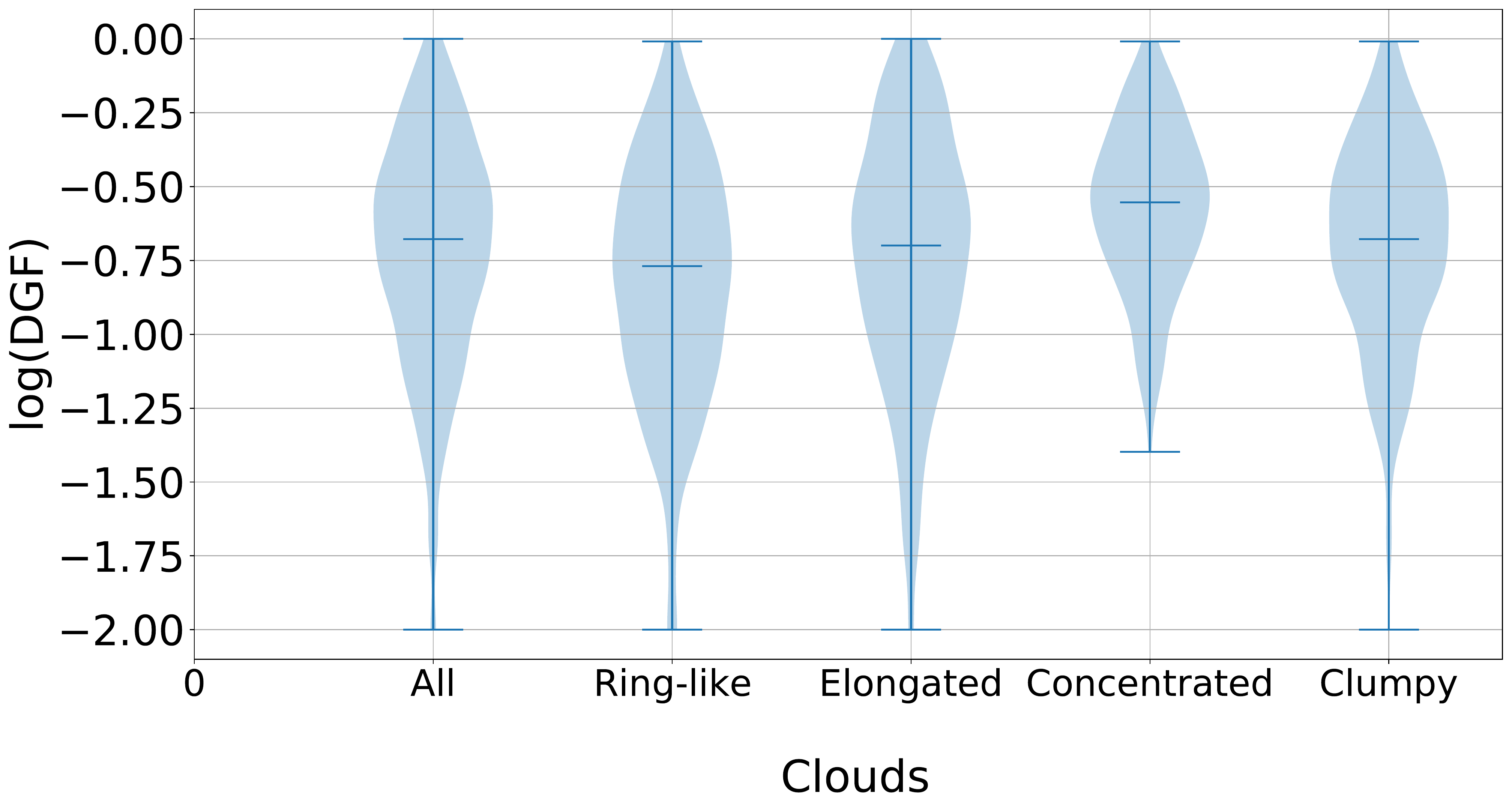}
    
    \vspace{0.5cm}
    
    \includegraphics[width = .5\textwidth, keepaspectratio]{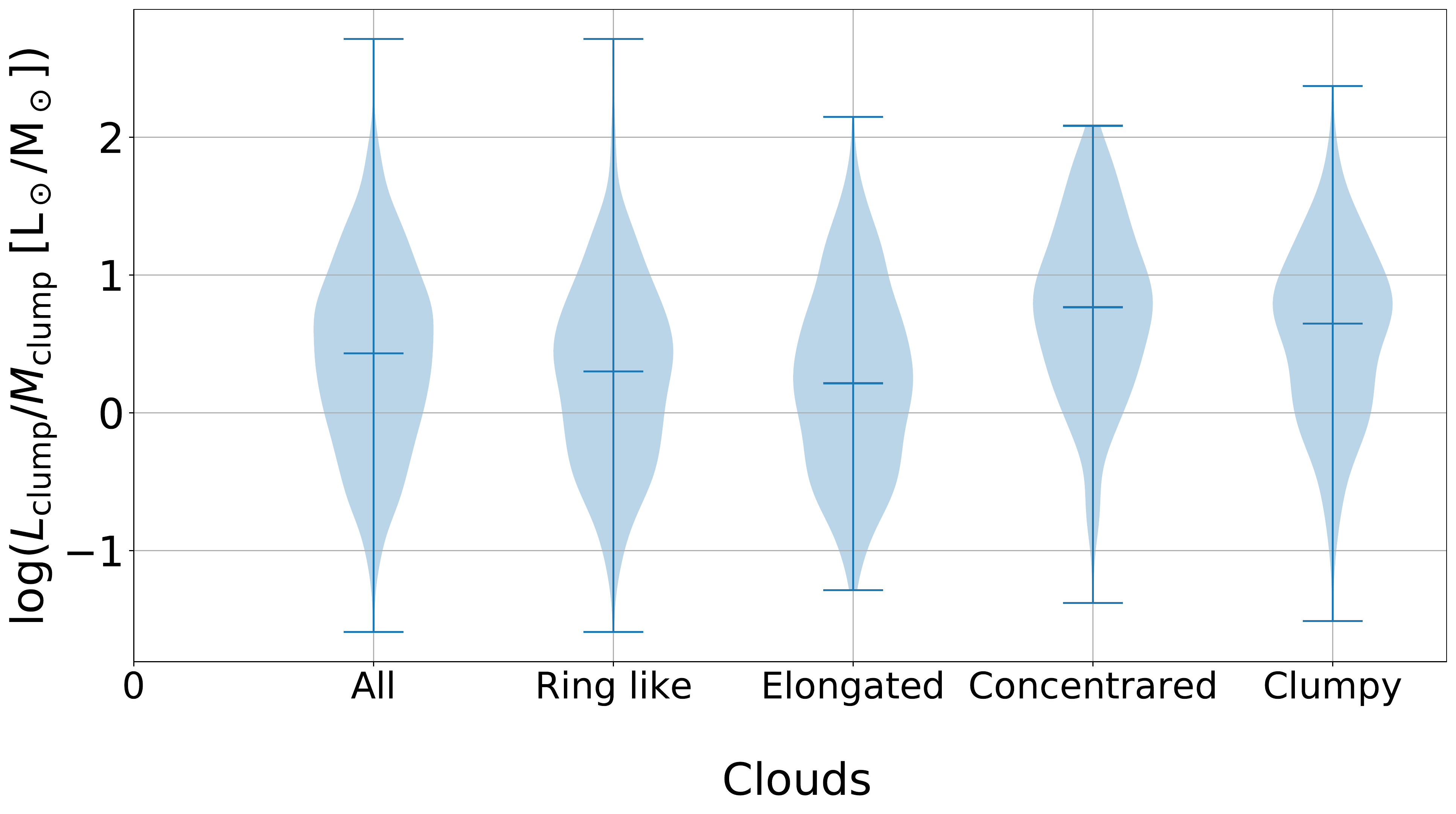}
    \caption{Top: Star formation efficiency (SFE); \add{Centre: Dense gas fraction (DGF); Bottom: Clump luminosity-to-mass ratio ($L_\mathrm{clump}/M_\mathrm{clump}$)}\remove{Bottom: Dense gas fraction (DGF)} distributions for different morphologies. The violin plots present the density of the data at different values, which is smoothed through kernel density estimator. The upper, lower and the middle horizontal lines in the plots represent the highest, lowest and the median values respectively.}
    \label{fig: SFE}
    %\end{minipage}\hfill
\end{figure}

The distributions shown in Fig. \ref{fig: SFE} for SFE and DGF as a function of cloud morphology, are obtained using the \add{1520} \remove{1588} \remove{1672} clouds with non-zero SFE and DGF values from the full sample of the SEDIGISM data. \add{We also plot the clump luminosity-to-mass ($L_\mathrm{clump}/M_\mathrm{clump}$) ratio (Fig. \ref{fig: SFE}) as an addition to the SFE and DGF distributions. It serves as measurement of cloud evolution} \citep{urquhart_2022}. In the catalogue of \citet{james_paper}, a few clouds have a SFE (or DGF) value < 0.01, which gets rounded off to zero. We exclude these clouds in our analysis, \add{which gives us 1672 clouds}. % and hence get the total clouds lower than \citet{james_paper}.
\add{The large uncertainties in the parameters involved in the calculation of cloud and clump masses lead to some clouds having extremely high DGF values (e.g. > 10). There could also be multiple clouds along a line of sight and the ATLASGAL clump could have been assigned to the wrong cloud, leading to large DGF values. We avoid such cases by excluding all the clouds with DGF > 1, and thus get the final sample of 1520 clouds. }

It is seen that the concentrated clouds have higher average SFE and DGF values compared to the other morphologies. Their compact structure causes the ATLASGAL clumps to overlap with the whole of the SEDIGISM cloud, increasing the relative clump mass and luminosity. The highest\footnote{These represents the extreme values in the distribution whereas `higher average values' refer to an entire distribution that show higher values for the given morphology.} SFE values are observed for ring-like clouds, although the number of ring-like clouds with a high SFE are low. This can be explained by comparing ring-like clouds to the infrared bubbles, where the shell of a bubble is a site for triggered star formation \citep{elmegreen_1977, zavagno_2006, jianjun_2020}. Moreover the L/M ratio acts as a proxy for the dust temperature of a clump \citep{pitts_2019} and as bubbles are formed due to stellar feedback, they are expected to have high dust temperatures. The clumpy clouds show higher average values of SFE than the elongated clouds but the distributions of the two morphologies are similar for DGF \remove{(p-value = 0.06)} \add{(Table \ref{Table: p-values for sfe} \& \ref{Table: p-values for dgf})}. The \add{average} higher SFE \add{(Table \ref{Table: p-values for sfe})} \remove{or L/M} \add{and luminosity-to-mass} ratio \add{(Fig. \ref{fig: SFE}; Table \ref{Table: p-values for l2m})}, suggest that clumpy clouds are more evolved \citep{urquhart_2022} than elongated clouds. \citet{james_paper} considers this to be a consequence of localised variations of SFE and DGF in the clouds. \remove{However, it could also be due to the ATLASGAL clumps being sparsely distributed in the elongated clouds as opposed to being densely packed in clumpy clouds, hence resulting in different SFEs.}\add{However, it could be a selection effect. High luminosity (dense) clumps could lead to a visibly denser (high-intensity) region in a cloud leading to it being classified as a clumpy cloud (Fig. \ref{fig: clumpy cloud image}). The high luminosity values could then lead to a higher average SFE for clumpy clouds as compared to elongated clouds.}
The average distributions of SFE and DGF are highest for concentrated clouds, but this might not reflect the actual star formation in them. These clouds have low mass (details in the parallel paper Neralwar et al. in prep) that is concentrated in a small region. This leads to them having high surface densities (Neralwar et al. in prep), and higher average values of SFE and DGF. Moreover, most of the concentrated clouds are associated with a single ATLASGAL clump, which combined with other properties (e.g., size) makes them comparable to the clumps.

\begin{table}[]
\caption{\add{P-values for SFE. The values are obtained by comparing SFE distributions between the different cloud morphologies. The values above the diagonal (shaded) are p-values obtained using the two sample KS test and the values below the diagonal are the p-values obtained using the MWU test. The diagonal in the table presents the mean values of the SFE distributions for the different morphologies.}}
\resizebox{\columnwidth}{!}{\begin{tabular}{lccccc}
\hline
Cloud        & All                           & Ring-like                     & Elongated                    & Concentrated                  & Clumpy                        \\ \hline
All          & \cellcolor[HTML]{C0C0C0}-0.31 & $\sim 10^{-06}$               & $\sim 10^{-06}$              & $\sim 10^{-16}$               & $\sim 10^{-05}$               \\
Ring-like    & $\sim 10^{-08}$               & \cellcolor[HTML]{C0C0C0}-0.55 & 0.69                         & $\sim 10^{-29}$               & $\sim 10^{-13}$               \\
Elongated    & $\sim 10^{-07}$               & 0.42                          & \cellcolor[HTML]{C0C0C0}-0.5 & $\sim 10^{-15}$               & $\sim 10^{-13}$               \\
Concentrated & $\sim 10^{-23}$               & $\sim 10^{-33}$               & $\sim 10^{-35}$              & \cellcolor[HTML]{C0C0C0}-0.17 & $\sim 10^{-05}$               \\
Clumpy       & $\sim 10^{-05}$               & $\sim 10^{-15}$               & $\sim 10^{-15}$              & $\sim 10^{-08}$               & \cellcolor[HTML]{C0C0C0}-0.12 \\ \hline

\end{tabular}}
\label{Table: p-values for sfe}
\end{table}

\begin{table}[]
\caption{\add{P-values for DGF distributions of the cloud morphologies. The symbols and conventions follow Table \ref{Table: p-values for sfe}.}}
\resizebox{\columnwidth}{!}{\begin{tabular}{lccccc}
\hline
Cloud        & All                                               & Ring-like                                         & Elongated                                         & Concentrated                                      & Clumpy                         \\ \hline
All          & \cellcolor[HTML]{C0C0C0}-0.72 & $\sim 10^{-05}$                                   & {0.6}                           & $\sim 10^{-08}$                                   & {0.38}       \\
Ring-like    & $\sim 10^{-06}$                                   & {\cellcolor[HTML]{C0C0C0}-0.83} & $\sim 10^{-03}$                                   & $\sim 10^{-15}$                                   & $\sim 10^{-04}$                \\
Elongated    & {0.52}                          & $\sim 10^{-03}$                                   & {\cellcolor[HTML]{C0C0C0}-0.74} & $\sim 10^{-08}$                                   & {0.58}       \\
Concentrated & $\sim 10^{-09}$                                   & $\sim 10^{-17}$                                   & $\sim 10^{-07}$                                   & {\cellcolor[HTML]{C0C0C0}-0.57} & $\sim 10^{-04}$                \\
Clumpy       & {0.37}                          & $\sim 10^{-05}$                                   & {0.21}                          & $\sim 10^{-05}$                                   & \cellcolor[HTML]{C0C0C0}-0.69 \\ \hline

\end{tabular}}
\label{Table: p-values for dgf}
\end{table}

\begin{table}[]
\caption{\add{P-values for luminosity-to-mass ratio distributions of the cloud morphologies. The symbols and conventions follow Table \ref{Table: p-values for sfe}.}}
\resizebox{\columnwidth}{!}{\begin{tabular}{lccccc}
\hline
Cloud        & All                                              & Ring-like                                        & Elongated                                        & Concentrated                                     & Clumpy                                           \\ \hline
All          & {\cellcolor[HTML]{C0C0C0}0.41} & $\sim 10^{-03}$                                  & {0.39}                         & $\sim 10^{-10}$                                  & $\sim 10^{-04}$                                  \\
Ring-like    & $\sim 10^{-04}$                                  & {\cellcolor[HTML]{C0C0C0}0.28} & {0.35}                         & $\sim 10^{-14}$                                  & $\sim 10^{-08}$                                  \\
Elongated    & $\sim 10^{-07}$                                  & {0.35}                         & {\cellcolor[HTML]{C0C0C0}0.24} & $\sim 10^{-15}$                                  & $\sim 10^{-11}$                                  \\
Concentrated & $\sim 10^{-13}$                                  & $\sim 10^{-18}$                                  & $\sim 10^{-22}$                                  & {\cellcolor[HTML]{C0C0C0}0.74} & {0.59}                         \\
Clumpy       & $\sim 10^{-05}$                                  & $\sim 10^{-10}$                                  & $\sim 10^{-12}$                                  & $\sim 10^{-03}$                                  & {\cellcolor[HTML]{C0C0C0}0.58} \\ \hline

\end{tabular}}
\label{Table: p-values for l2m}
\end{table}

\subsubsection{HMSF relation}
\begin{figure*}[ht]
    \centering
    \includegraphics[width = \columnwidth]{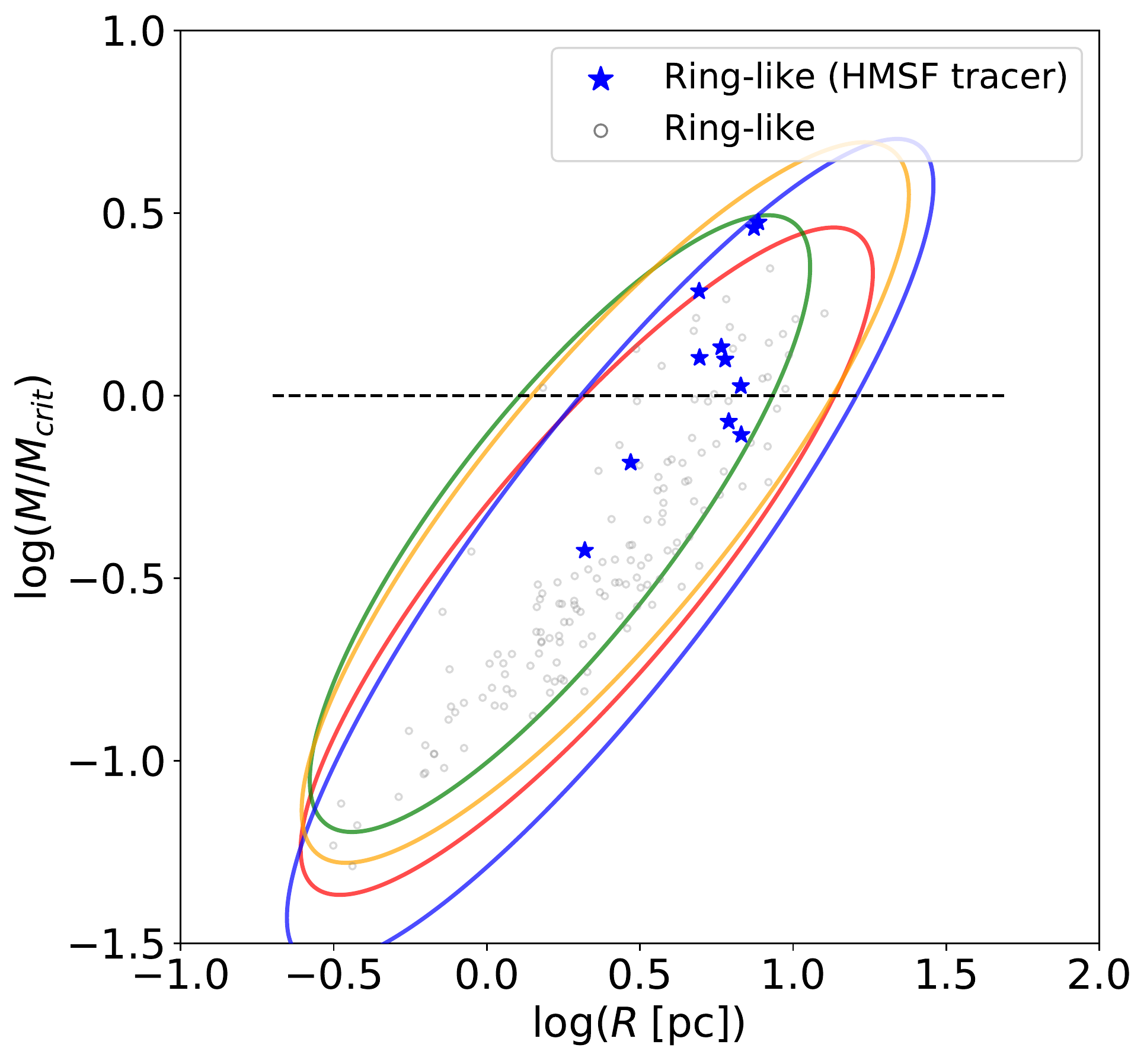}
    \includegraphics[width = \columnwidth]{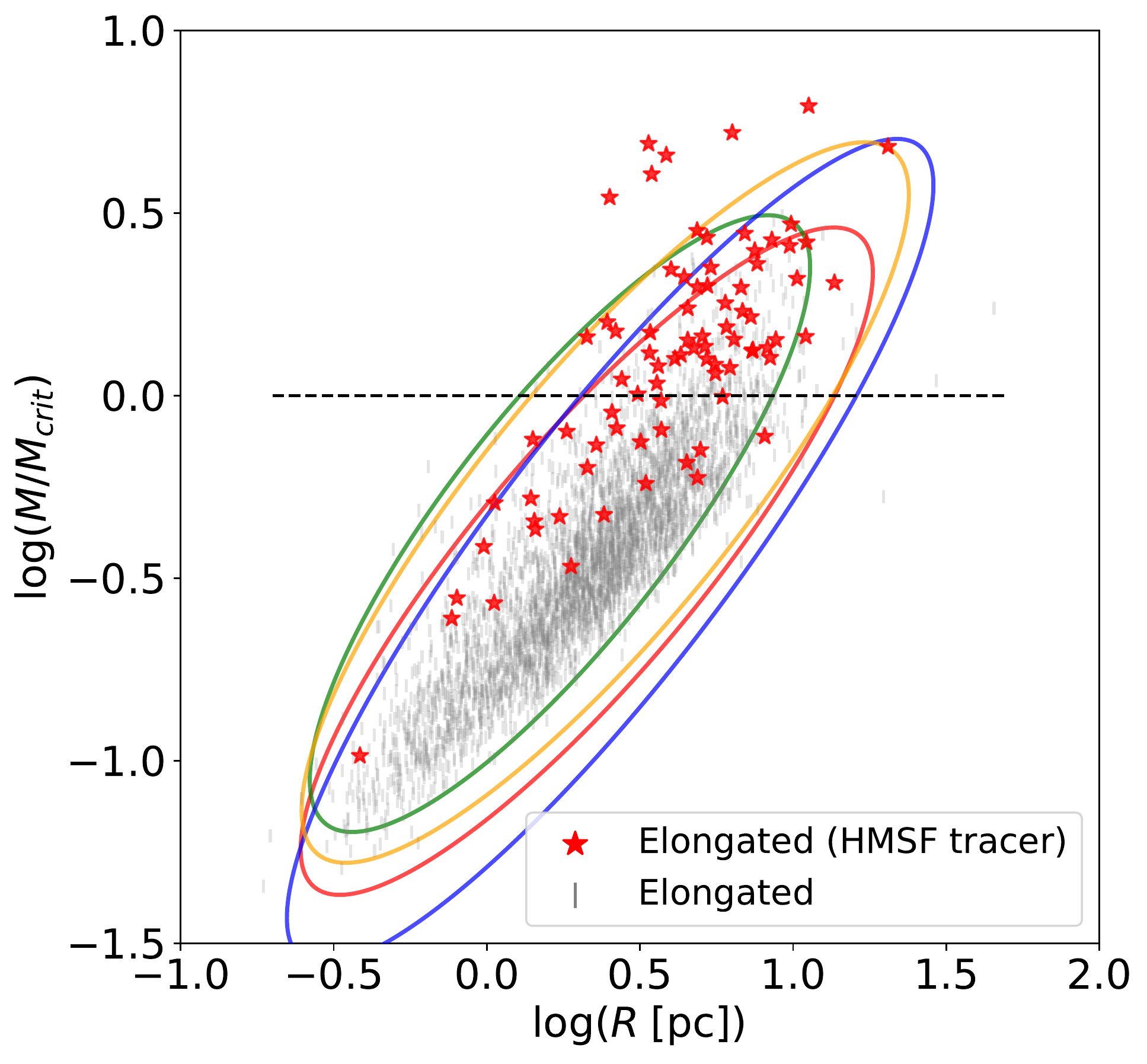}
    \includegraphics[width = \columnwidth]{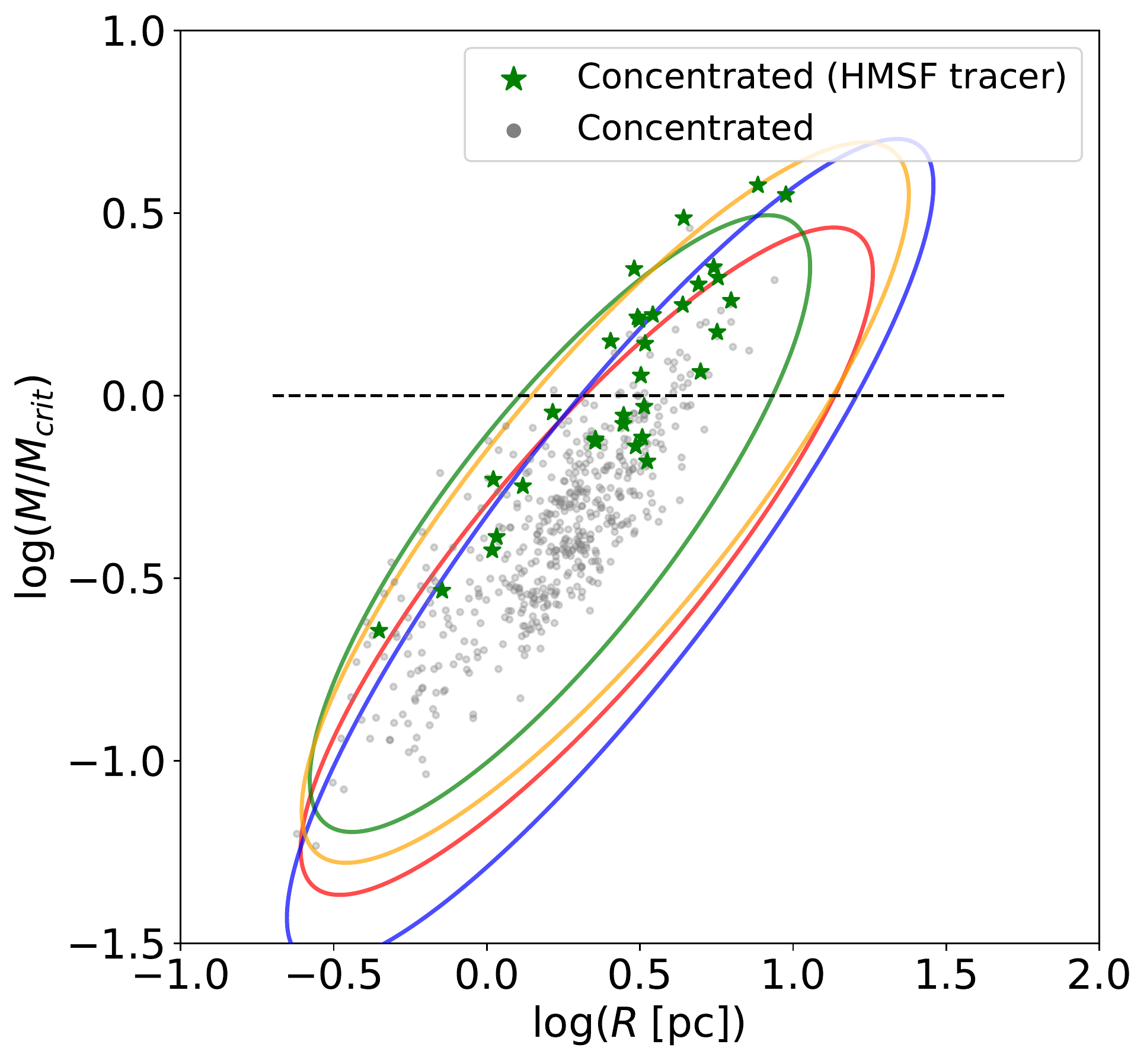}
    \includegraphics[width = \columnwidth]{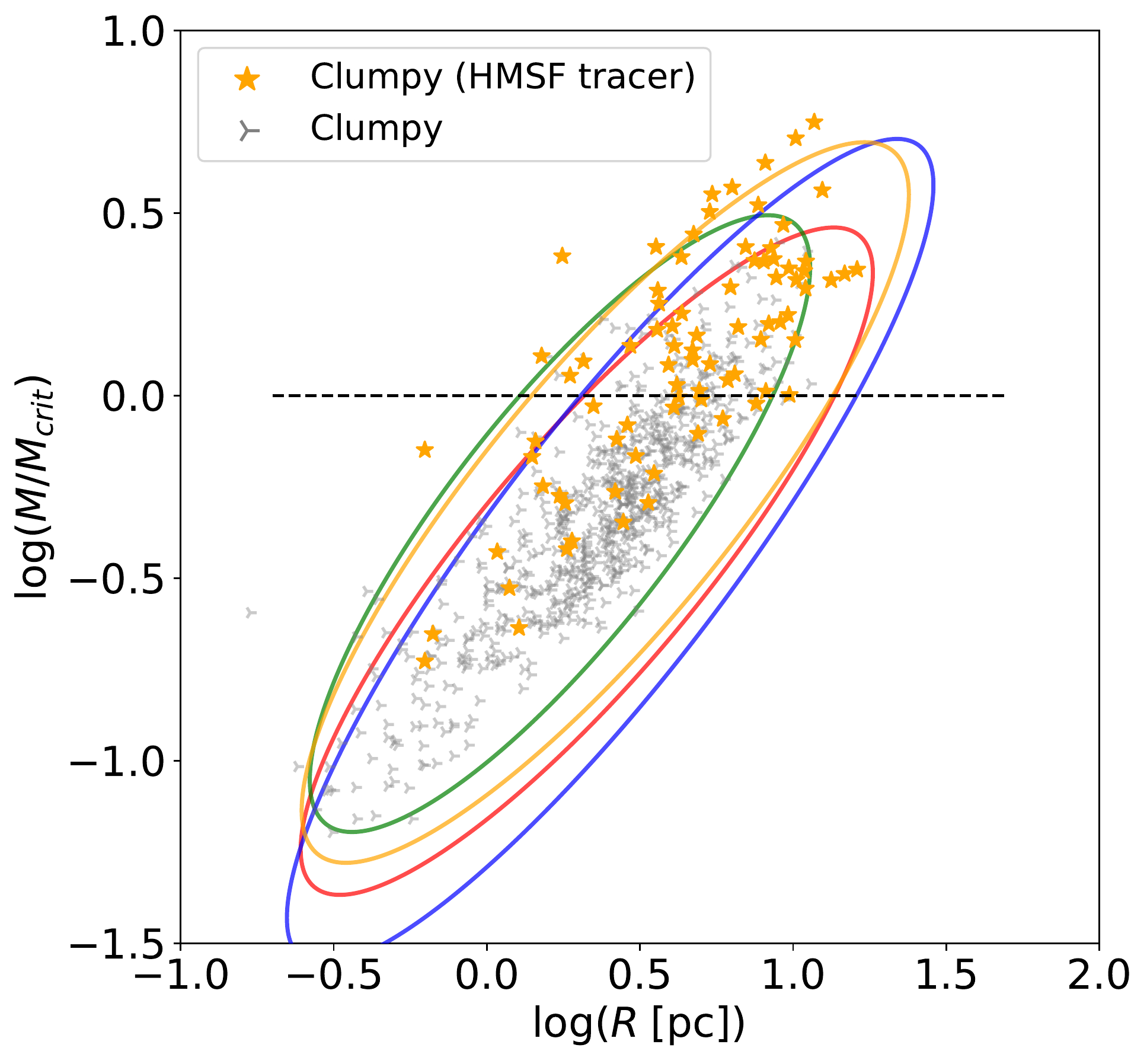}
    \caption{HMSF threshold (dashed line; $M[\mathrm{M_\odot}] = 1053 \, (R[\mathrm{pc}])^{1.33}$) for molecular clouds; using MR (science) sample. The y axis represents $M/M_{crit}$, where $M_{crit} = 1053 \, (R[\mathrm{pc}])^{1.33}$ is the HMSF threshold. The \remove{confidence} \add{coloured} ellipses \add{encompass approximately 99.7\% of the data points and} represent 3-sigma levels for the cloud morphologies\add{, which are} indicated by the legends in the four plots. The clouds with HMSF tracer (Table \ref{tab:hmsf tracer}) are denoted by a star ($\star$).}
    \label{fig: scaling relation mass radius}
\end{figure*}

The high-mass star formation in clouds cannot be understood using a single cloud property and therefore, must be studied by relating different properties. Fig. \ref{fig: scaling relation mass radius} shows the empirical relation between mass and radius for the various cloud morphologies. It highlights the HMSF threshold $M[\mathrm{M_\odot}] = 1053 \, (R[\mathrm{pc}])^{1.33}$, where $M$ is the cloud mass and $R$ is the deconvolved radius (DC21). The threshold relation was initially obtained by \citet{kauffmann_2010} for dusty clumps, and was further updated by DC21 for the molecular clouds from the SEDIGISM survey. 

The clouds above the HMSF threshold are expected to form high-mass stars. \remove{but this acts as a necessary condition rather than a sufficient one.} \add{The original HMSF threshold} \citep[for clumps;][]{kauffmann_2010} \add{is a necessary but not a sufficient condition for high-mass star formation, as it does not rule out the possibility for false positives. However, we see molecular clouds with star formation indicators (tracers) below the HMSF threshold, i.e. false negatives (or missed true positives). It suggests that the HMSF threshold might not be suitable for our sample (discussed in details in DC21).} \remove{Therefore, there could be false positives and negatives} \add{The errors in mass and radius estimations can also shift the positions of clouds with respect to the mass-radius relation (Fig. \ref{fig: scaling relation mass radius}) and lead to false positives and negatives.} Moreover, cloud radius (deconvolved radius from DC21) might not effectively represent the size of a cloud with non-spherical geometry. \add{We study the HMSF threshold as a follow up on the cloud analysis of DC21. }We have obtained the percentages of clouds above the threshold for different morphologies and they are presented in Table \ref{table: kauffmann HMSF relation}.
As we use the cloud radius which require reliable distance estimates for precise calculation, we use the clouds from the science sample (described in DC21) to plot the HMSF relation. These clouds contain reliable distance estimates, are well resolved and do not lie on the edge of the survey. %This is equivalent to the science sample of DC21.

We find that none of the morphologies show distinctly high percentage of HMSF clouds (Table \ref{table: kauffmann HMSF relation}). However, ring-like clouds and clumpy clouds show comparatively higher percentage of HMSF clouds than the other two morphologies and this is consistent with what is seen for the cloud SFEs in Sec. \ref{sec: sfe dgf prop}.
Similarly, it is also seen that the clumpy clouds have a higher percentage of high-mass star formation (HMSF) regions associated with them (using tracers mentioned in DC21) as compared to the elongated clouds (Table \ref{tab:hmsf tracer}). We only consider the clouds with an ATLASGAL counterpart while calculating the percentage of clouds with HMSF tracer as these are a sub-sample of the ATLASGAL sources. The HMSF tracers/signposts include methanol masers, HII regions and young stellar objects from various surveys and are described in detail in DC21. The clouds with that host HMSF tracers are plotted with a star ($\star$) in Fig. \ref{fig: scaling relation mass radius}. The clouds with HMSF tracers typically lie above the mass-radius HMSF threshold for the different morphology, supporting the empirical relation.

\begin{comment}

\begin{table}
\caption{Percentage of clouds above the HMSF relation shown by the dashed line in Fig. \ref{fig: scaling relation mass radius}. These clouds belong to the science sample (described in DC21).}
\label{table: kauffmann HMSF relation}
\centering
\resizebox{\columnwidth}{!}{\begin{tabular}{ccccc}
\hline
Sample  & Ring-like & Elongated & Concentrated & Clumpy \\ \hline
VC & 20\%     & 7\%      & 9\%          & 18\%      \\ 
MR      & 17\%     & 7\%      & 10\%         & 18\%   \\ \hline
\end{tabular}}

\end{table}
\end{comment}

\begin{table}
\caption{Percentage of clouds above the HMSF relation shown by the dashed line in Fig. \ref{fig: scaling relation mass radius}. These clouds belong to the science sample (described in DC21). \add{We use a binomial model to calculate the maximum likehood estimator of p (the probabilities of clouds lying above the HMSF) and the corresponding standard deviations are the errors on cloud percentages. The clouds without error values have errors < 1\%.}}
\label{table: kauffmann HMSF relation}
\centering
\resizebox{\columnwidth}{!}{\begin{tabular}{ccccc}
\hline
Sample  & Ring-like & Elongated & Concentrated & Clumpy \\ \hline
VC & 20 $\pm$ 1\%     & 7\%      & 9\%          & 18$\pm$ 1\%      \\ 
MR      & 17$\pm$ 3\%     & 7\%      & 10$\pm$ 1\%         & 18$\pm$ 1\%   \\ \hline
\end{tabular}}

\end{table}

\begin{table}[]
\caption{Percentage of clouds containing an HMSF tracer compared to the total clouds with an ATLASGAL clump (described in DC21) for MR science sample (5020 clouds).}
\label{tab:hmsf tracer}
\centering
\resizebox{\columnwidth}{!}{\begin{tabular}{ccccc}
\hline
HMSF tracer & Ring-like & Elongated & Concentrated & Clumpy \\
\hline
%Present     & 11        & 82        & 32           & 82     \\
%Absent      & 28       & 281      & 80          & 155 \\
Present & 28\%       & 23\%       & 29\%       &34\% \\ 
\hline
\end{tabular}}

\end{table}

\begin{comment}

\begin{figure}[ht]
    \centering
    \includegraphics[trim=10 30 10 40,clip,width = 0.44 \paperwidth, keepaspectratio]{figure/mass_rad.png}
    \caption{HMSF threshold (dashed line; $M[\mathrm{M_\odot}] = 1053 \, (R[\mathrm{pc}])^{1.33}$) for molecular clouds; using MR (science) sample.}
    \label{fig: scaling relation mass radius}
\end{figure}
\end{comment}

%--------------------------------------------------------------------

%\section{Discussion}\label{sec: discussions}

The HMSF threshold \citep{kauffmann_2010} provides evidence towards higher degree of star formation in ring-like clouds and clumpy clouds (Fig. \ref{fig: scaling relation mass radius}). %, that was observed with SFE and DGF (Sec. \ref{sec: sfe dgf prop}).
The number of clouds with HMSF regions belonging to these two morphologies is $> 16\%$ for both the samples, as opposed to the lower values ($< 10\%$) for elongated and concentrated clouds (Table \ref{table: kauffmann HMSF relation}). 

\section{Summary}\label{sec: summary}

In this work, we have classified molecular clouds from the SEDIGISM survey based on their morphology. This has been achieved using two classification methods. The first method -- $J$ plots -- uses the moment of inertia of structures in the integrated intensity maps to classify them into three types. The second method -- by-eye classification -- visually classifies the clouds into four groups.
The combined results from these two classifications result in the visually classified (VC) sample and the morphologically reliable (MR) sample, which are used to affirm the reliability of the $J$ plots classification. The VC sample (10365 clouds) is a subset of full sample in which we exclude the unclassified clouds. The MR sample (8086 clouds) is a subset of the VC sample that contains only the clouds for which the morphologies are consistent for the two methods.
We thus present the MR sample as our most reliable and robust sample whereas the VC sample should be used when a larger sample size needs to be prioritised over the robustness of data.

We compare the positions of SEDIGISM molecular clouds with their submillimeter/infrared counterparts from two continuum surveys -- ATLASGAL and MWP. We use the ATLASGAL survey to see how the lower density SEDIGISM elongated clouds compare to the dense dust continuum structures. The Milky Way Project (MWP) is used to see how our ring-like clouds fare in comparison to the dust bubbles and stellar feedback regions. Almost all the ATLASGAL elongated structures (AG-El) and more than 90\% MWP bubbles in the SEDIGISM coverage have a molecular cloud counterpart. However, $\approx$ 64\% SEDIGISM clouds (full sample) neither have an ATLASGAL or a MWP counterpart. We find that 80\% elongated clouds and 71\% clumpy clouds in the MR sample %(with edge = 0)
lack an ATLASGAL elongated counterpart. The high percentage is in agreement with other findings in the literature since ATLASGAL traces high density gas which is accompanied by low density gas traced by SEDIGISM clouds but not vice-versa. There might also be subthermally-excited gas traced by SEDIGISM, which has a high column density \add{but not sufficiently high volume density to be detected by ATLASGAL in continuum.} \remove{and thus might not be detected by ATLASGAL.}
We also find that 66\% ring-like clouds lack a MWP bubble counterpart. MWP probes the HII and PAH regions while SEDIGISM traces the molecular gas in the ISM and thus a comparison between the column densities traced by the two surveys is complex. \add{Large scale shocks (HI flows) and supersonic turbulence could also lead to creation of ring-like structures/clouds which are not detected at mid-infrared wavelengths.}
 
We also study the star formation activity for clouds in different morphology classes using two methods. The first method uses SFE and DGF for SEDIGISM clouds obtained by \citet{james_paper} and the second method uses HMSF threshold \citep[][DC21]{kauffmann_2010} for molecular clouds. These methods show that although none of the morphologies show very high star formation, ring-like and clumpy clouds show higher star formation when compared to elongated and concentrated clouds.

In conclusion, we classify molecular clouds based on their morphology and these morphologies show variations in star formation properties compared to the global cloud distribution. The various cloud morphologies closely resemble similar structures from other continuum surveys. Furthermore, ring-like clouds and especially clumpy clouds show evidence towards higher star formation activity as compared to the other morphologies. A major observation across all the samples is that most of the molecular clouds are elongated. Finally, we also conclude that the automated cloud morphology classification based on $J$ plots alone is not completely reliable.

%The sample of molecular clouds in this work covers less than a quarter of the inner Galaxy. An extension to this is possible by using data from upcoming Galaxy surveys; e.g. Outer Galaxy High Resolution Survey (OGHReS), LASMAGAL project, Global View on Star formation in the Milky Way (GLOSTAR) project. This would allow us to verify the trends in properties of clouds in different Galactic environments (e.g. the outer Galaxy). 
%In addition, most of the Milky Way is fully mapped by high resolution CO observations \citep[][and references therein]{scimes_2}. %give reference for these surveys

\begin{acknowledgement}
\add{The authors thank the referee for his/her valuable comments on the draft which helped improve the discussion of the paper and strengthened the results.}
This publication is based on data acquired with the Atacama Pathfinder Experiment (APEX) under programmes 092.F-9315 and 193.C-0584. APEX is a collaboration among the Max-Planck-Institut f\"ur Radioastronomie, the European Southern Observatory, and the Onsala Space Observatory. 
This publication uses data generated via the Zooniverse.org platform, development of which is funded by generous support, including a Global Impact Award from Google, and by a grant from the Alfred P. Sloan Foundation. A part of this work is based on observations made with the Spitzer Space Telescope, which is operated by the Jet Propulsion Laboratory, California Institute of Technology under a contract with NASA.
DC acknowledges support by the German \emph{Deut\-sche For\-schungs\-ge\-mein\-schaft, DFG\/} project number SFB956A.
ADC acknowledges the support from the Royal Society University Research Fellowship (URF/R1/191609). HB acknowledges support from the European Research Council under the Horizon 2020 Framework Programme via the ERC Consol-idator Grant CSF-648505. HB also acknowledges support from the Deutsche Forschungsgemeinschaft in the Collaborative Research Center (SFB 881) “The Milky Way System” (subproject B1). CLD acknowledges funding from the European Research Council for the Horizon 2020 ERC consolidator grant project ICYBOB, grant number 818940.
\end{acknowledgement}

\footnotesize{
\bibliographystyle{aa}
\bibliography{reference}

\begin{thebibliography}{104}
\expandafter\ifx\csname natexlab\endcsname\relax\def\natexlab#1{#1}\fi

\bibitem[{{Abe} {et~al.}(2021){Abe}, {Inoue}, {Inutsuka}, \&
  {Matsumoto}}]{abe_2021}
{Abe}, D., {Inoue}, T., {Inutsuka}, S.-i., \& {Matsumoto}, T. 2021, \apj, 916,
  83

\bibitem[{{Ali} {et~al.}(2022){Ali}, {Bending}, \& {Dobbs}}]{ali_2022}
{Ali}, A.~A., {Bending}, T. J.~R., \& {Dobbs}, C.~L. 2022, \mnras, 510, 5592

\bibitem[{{Ali} \& {Harries}(2019)}]{ali_2019}
{Ali}, A.~A. \& {Harries}, T.~J. 2019, \mnras, 487, 4890

\bibitem[{{Andr{\'e}} {et~al.}(2014){Andr{\'e}}, {Di Francesco},
  {Ward-Thompson}, {Inutsuka}, {Pudritz}, \& {Pineda}}]{andre_2014}
{Andr{\'e}}, P., {Di Francesco}, J., {Ward-Thompson}, D., {et~al.} 2014, in
  Protostars and Planets VI, ed. H.~{Beuther}, R.~S. {Klessen}, C.~P.
  {Dullemond}, \& T.~{Henning}, 27

\bibitem[{{Andr{\'e}} {et~al.}(2010){Andr{\'e}}, {Men'shchikov}, {Bontemps},
  {K{\"o}nyves}, {Motte}, {Schneider}, {Didelon}, {Minier}, {Saraceno},
  {Ward-Thompson}, {di Francesco}, {White}, {Molinari}, {Testi}, {Abergel},
  {Griffin}, {Henning}, {Royer}, {Mer{\'\i}n}, {Vavrek}, {Attard},
  {Arzoumanian}, {Wilson}, {Ade}, {Aussel}, {Baluteau}, {Benedettini},
  {Bernard}, {Blommaert}, {Cambr{\'e}sy}, {Cox}, {di Giorgio}, {Hargrave},
  {Hennemann}, {Huang}, {Kirk}, {Krause}, {Launhardt}, {Leeks}, {Le Pennec},
  {Li}, {Martin}, {Maury}, {Olofsson}, {Omont}, {Peretto}, {Pezzuto}, {Prusti},
  {Roussel}, {Russeil}, {Sauvage}, {Sibthorpe}, {Sicilia-Aguilar}, {Spinoglio},
  {Waelkens}, {Woodcraft}, \& {Zavagno}}]{andre_2010}
{Andr{\'e}}, P., {Men'shchikov}, A., {Bontemps}, S., {et~al.} 2010, \aap, 518,
  L102

\bibitem[{{Arzoumanian}(2017)}]{arzoumanian_2017_b}
{Arzoumanian}, D. 2017, arXiv e-prints, arXiv:1712.00604

\bibitem[{{Arzoumanian} {et~al.}(2019){Arzoumanian}, {Andr{\'e}},
  {K{\"o}nyves}, {Palmeirim}, {Roy}, {Schneider}, {Benedettini}, {Didelon}, {Di
  Francesco}, {Kirk}, \& {Ladjelate}}]{arzoumanian_2019}
{Arzoumanian}, D., {Andr{\'e}}, P., {K{\"o}nyves}, V., {et~al.} 2019, \aap,
  621, A42

\bibitem[{{Arzoumanian} {et~al.}(2018){Arzoumanian}, {Shimajiri}, {Inutsuka},
  {Inoue}, \& {Tachihara}}]{arzoumanian_2018}
{Arzoumanian}, D., {Shimajiri}, Y., {Inutsuka}, S.-i., {Inoue}, T., \&
  {Tachihara}, K. 2018, \pasj, 70, 96

\bibitem[{{Arzoumanian} {et~al.}(2017){Arzoumanian}, {Shimajiri}, {Roy},
  {Andr{\'e}}, {K{\"o}nyves}, \& {Bracco}}]{arzoumanian_2017}
{Arzoumanian}, D., {Shimajiri}, Y., {Roy}, A., {et~al.} 2017, \memsai, 88, 720

\bibitem[{{Balfour} {et~al.}(2015){Balfour}, {Whitworth}, {Hubber}, \&
  {Jaffa}}]{balfour_2015}
{Balfour}, S.~K., {Whitworth}, A.~P., {Hubber}, D.~A., \& {Jaffa}, S.~E. 2015,
  \mnras, 453, 2471

\bibitem[{{Beaumont} \& {Williams}(2010)}]{Beaumont_2010}
{Beaumont}, C.~N. \& {Williams}, J.~P. 2010, \apj, 709, 791

\bibitem[{{Bending} {et~al.}(2020){Bending}, {Dobbs}, \& {Bate}}]{bending_2020}
{Bending}, T. J.~R., {Dobbs}, C.~L., \& {Bate}, M.~R. 2020, \mnras, 495, 1672

\bibitem[{{Benedettini} {et~al.}(2021){Benedettini}, {Traficante}, {Olmi},
  {Pezzuto}, {Baldeschi}, {Molinari}, {Elia}, {Schisano}, {Merello}, {Fontani},
  {Rygl}, {Brand}, {Beltran}, {Cesaroni}, {Liu}, \& {Testi}}]{benedettini_2021}
{Benedettini}, M., {Traficante}, A., {Olmi}, L., {et~al.} 2021, arXiv e-prints,
  arXiv:2109.00950

\bibitem[{{Benjamin} {et~al.}(2003){Benjamin}, {Churchwell}, {Babler}, {Bania},
  {Clemens}, {Cohen}, {Dickey}, {Indebetouw}, {Jackson}, {Kobulnicky},
  {Lazarian}, {Marston}, {Mathis}, {Meade}, {Seager}, {Stolovy}, {Watson},
  {Whitney}, {Wolff}, \& {Wolfire}}]{benjamin_2003}
{Benjamin}, R.~A., {Churchwell}, E., {Babler}, B.~L., {et~al.} 2003, \pasp,
  115, 953

\bibitem[{{Bialy} {et~al.}(2021){Bialy}, {Zucker}, {Goodman}, {Foley}, {Alves},
  {Semenov}, {Benjamin}, {Leike}, \& {En{\ss}lin}}]{bialy_2021}
{Bialy}, S., {Zucker}, C., {Goodman}, A., {et~al.} 2021, \apjl, 919, L5

\bibitem[{{Bonne} {et~al.}(2020{\natexlab{a}}){Bonne}, {Bontemps}, {Schneider},
  {Clarke}, {Arzoumanian}, {Fukui}, {Tachihara}, {Csengeri}, {Guesten},
  {Ohama}, {Okamoto}, {Simon}, {Yahia}, \& {Yamamoto}}]{bonne_2020_b}
{Bonne}, L., {Bontemps}, S., {Schneider}, N., {et~al.} 2020{\natexlab{a}},
  \aap, 644, A27

\bibitem[{{Bonne} {et~al.}(2020{\natexlab{b}}){Bonne}, {Schneider}, {Bontemps},
  {Clarke}, {Gusdorf}, {Lehmann}, {Steinke}, {Csengeri}, {Kabanovic}, {Simon},
  {Buchbender}, \& {G{\"u}sten}}]{bonne_2020}
{Bonne}, L., {Schneider}, N., {Bontemps}, S., {et~al.} 2020{\natexlab{b}},
  \aap, 641, A17

\bibitem[{{Carey}(2016)}]{Hi-Gal_bubbles}
{Carey}, D. 2016, {Via Lactea Galactic Plane bubble Catalogue}

\bibitem[{{Carey} {et~al.}(2008){Carey}, {Ali}, {Berriman}, {Boulanger},
  {Brunt}, {Cutri}, {Flagey}, {Gibson}, {Heyer}, {Hora}, {Indebetouw},
  {Kraemer}, {Kuchar}, {Latter}, {Marleau}, {Miville-Deschenes}, {Mizuno},
  {Molinari}, {Noriega-Crespo}, {Padgett}, {Paladini}, {Price}, {Rebull},
  {Rottler}, {Shenoy}, {Shipman}, \& {Testi}}]{carey_2008}
{Carey}, S., {Ali}, B., {Berriman}, B., {et~al.} 2008, {Spitzer Mapping of the
  Outer Galaxy (SMOG)}, Spitzer Proposal

\bibitem[{{Carey} {et~al.}(2009){Carey}, {Noriega-Crespo}, {Mizuno}, {Shenoy},
  {Paladini}, {Kraemer}, {Price}, {Flagey}, {Ryan}, {Ingalls}, {Kuchar},
  {Pinheiro Gon{\c{c}}alves}, {Indebetouw}, {Billot}, {Marleau}, {Padgett},
  {Rebull}, {Bressert}, {Ali}, {Molinari}, {Martin}, {Berriman}, {Boulanger},
  {Latter}, {Miville-Deschenes}, {Shipman}, \& {Testi}}]{carey_2009}
{Carey}, S.~J., {Noriega-Crespo}, A., {Mizuno}, D.~R., {et~al.} 2009, \pasp,
  121, 76

\bibitem[{{Churchwell} {et~al.}(2009){Churchwell}, {Babler}, {Meade},
  {Whitney}, {Benjamin}, {Indebetouw}, {Cyganowski}, {Robitaille}, {Povich},
  {Watson}, \& {Bracker}}]{churchwell_2009}
{Churchwell}, E., {Babler}, B.~L., {Meade}, M.~R., {et~al.} 2009, \pasp, 121,
  213

\bibitem[{{Churchwell} {et~al.}(2006){Churchwell}, {Povich}, {Allen}, {Taylor},
  {Meade}, {Babler}, {Indebetouw}, {Watson}, {Whitney}, {Wolfire}, {Bania},
  {Benjamin}, {Clemens}, {Cohen}, {Cyganowski}, {Jackson}, {Kobulnicky},
  {Mathis}, {Mercer}, {Stolovy}, {Uzpen}, {Watson}, \&
  {Wolff}}]{churchwell_2006}
{Churchwell}, E., {Povich}, M.~S., {Allen}, D., {et~al.} 2006, \apj, 649, 759

\bibitem[{{Churchwell} {et~al.}(2007){Churchwell}, {Watson}, {Povich},
  {Taylor}, {Babler}, {Meade}, {Benjamin}, {Indebetouw}, \&
  {Whitney}}]{churchwell_2007}
{Churchwell}, E., {Watson}, D.~F., {Povich}, M.~S., {et~al.} 2007, \apj, 670,
  428

\bibitem[{{Clarke} {et~al.}(2017){Clarke}, {Whitworth}, {Duarte-Cabral}, \&
  {Hubber}}]{Clarke_2017}
{Clarke}, S.~D., {Whitworth}, A.~P., {Duarte-Cabral}, A., \& {Hubber}, D.~A.
  2017, \mnras, 468, 2489

\bibitem[{{Clarke} {et~al.}(2020){Clarke}, {Williams}, \&
  {Walch}}]{clarke_2020}
{Clarke}, S.~D., {Williams}, G.~M., \& {Walch}, S. 2020, \mnras, 497, 4390

\bibitem[{{Colombo} {et~al.}(2021){Colombo}, {K{\"o}nig}, {Urquhart},
  {Wyrowski}, {Mattern}, {Menten}, {Lee}, {Brand}, {Wienen}, {Mazumdar},
  {Schuller}, \& {Leurini}}]{colombo2021}
{Colombo}, D., {K{\"o}nig}, C., {Urquhart}, J.~S., {et~al.} 2021, \aap, 655, L2

\bibitem[{{Colombo} {et~al.}(2019){Colombo}, {Rosolowsky}, {Duarte-Cabral},
  {Ginsburg}, {Glenn}, {Zetterlund}, {Hernand ez}, {Dempsey}, \&
  {Currie}}]{scimes_2}
{Colombo}, D., {Rosolowsky}, E., {Duarte-Cabral}, A., {et~al.} 2019, \mnras,
  483, 4291

\bibitem[{{Colombo} {et~al.}(2015){Colombo}, {Rosolowsky}, {Ginsburg},
  {Duarte-Cabral}, \& {Hughes}}]{SCIMES}
{Colombo}, D., {Rosolowsky}, E., {Ginsburg}, A., {Duarte-Cabral}, A., \&
  {Hughes}, A. 2015, \mnras, 454, 2067

\bibitem[{{Contreras} {et~al.}(2013){Contreras}, {Schuller}, {Urquhart},
  {Csengeri}, {Wyrowski}, {Beuther}, {Bontemps}, {Bronfman}, {Henning},
  {Menten}, {Schilke}, {Walmsley}, {Wienen}, {Tackenberg}, \&
  {Linz}}]{contrearas_2013}
{Contreras}, Y., {Schuller}, F., {Urquhart}, J.~S., {et~al.} 2013, \aap, 549,
  A45

\bibitem[{{Csengeri} {et~al.}(2014){Csengeri}, {Urquhart}, {Schuller}, {Motte},
  {Bontemps}, {Wyrowski}, {Menten}, {Bronfman}, {Beuther}, {Henning}, {Testi},
  {Zavagno}, \& {Walmsley}}]{Csengeri_2014}
{Csengeri}, T., {Urquhart}, J.~S., {Schuller}, F., {et~al.} 2014, \aap, 565,
  A75

\bibitem[{{Dale} {et~al.}(2005){Dale}, {Bonnell}, {Clarke}, \&
  {Bate}}]{dale_2005}
{Dale}, J.~E., {Bonnell}, I.~A., {Clarke}, C.~J., \& {Bate}, M.~R. 2005,
  \mnras, 358, 291

\bibitem[{{Dale} {et~al.}(2014){Dale}, {Ngoumou}, {Ercolano}, \&
  {Bonnell}}]{dale_2014}
{Dale}, J.~E., {Ngoumou}, J., {Ercolano}, B., \& {Bonnell}, I.~A. 2014, \mnras,
  442, 694

\bibitem[{{Deharveng} {et~al.}(2010){Deharveng}, {Schuller}, {Anderson},
  {Zavagno}, {Wyrowski}, {Menten}, {Bronfman}, {Testi}, {Walmsley}, \&
  {Wienen}}]{deharveng_2010}
{Deharveng}, L., {Schuller}, F., {Anderson}, L.~D., {et~al.} 2010, \aap, 523,
  A6

\bibitem[{{Deharveng} {et~al.}(2009){Deharveng}, {Zavagno}, {Schuller},
  {Caplan}, {Pomar{\`e}s}, \& {De Breuck}}]{deharveng_2009}
{Deharveng}, L., {Zavagno}, A., {Schuller}, F., {et~al.} 2009, \aap, 496, 177

\bibitem[{{Dobbs} \& {Wurster}(2021)}]{dobbs_2021}
{Dobbs}, C.~L. \& {Wurster}, J. 2021, \mnras, 502, 2285

\bibitem[{{Duarte-Cabral} {et~al.}(2021){Duarte-Cabral}, {Colombo}, {Urquhart},
  {Ginsburg}, {Russeil}, {Schuller}, {Anderson}, {Barnes}, {Beltr{\'a}n},
  {Beuther}, {Bontemps}, {Bronfman}, {Csengeri}, {Dobbs}, {Eden}, {Giannetti},
  {Kauffmann}, {Mattern}, {Medina}, {Menten}, {Lee}, {Pettitt}, {Riener},
  {Rigby}, {Traficante}, {Veena}, {Wienen}, {Wyrowski}, {Agurto}, {Azagra},
  {Cesaroni}, {Finger}, {Gonzalez}, {Henning}, {Hernandez}, {Kainulainen},
  {Leurini}, {Lopez}, {Mac-Auliffe}, {Mazumdar}, {Molinari}, {Motte}, {Muller},
  {Nguyen-Luong}, {Parra}, {Perez-Beaupuits}, {Montenegro-Montes}, {Moore},
  {Ragan}, {S{\'a}nchez-Monge}, {Sanna}, {Schilke}, {Schisano}, {Schneider},
  {Suri}, {Testi}, {Torstensson}, {Venegas}, {Wang}, \& {Zavagno}}]{ana_paper}
{Duarte-Cabral}, A., {Colombo}, D., {Urquhart}, J.~S., {et~al.} 2021, \mnras,
  500, 3027

\bibitem[{{Duarte-Cabral} \& {Dobbs}(2017)}]{ana_2017}
{Duarte-Cabral}, A. \& {Dobbs}, C.~L. 2017, \mnras, 470, 4261

\bibitem[{{Eden} {et~al.}(2012){Eden}, {Moore}, {Plume}, \&
  {Morgan}}]{eden_2012}
{Eden}, D.~J., {Moore}, T.~J.~T., {Plume}, R., \& {Morgan}, L.~K. 2012, \mnras,
  422, 3178

\bibitem[{{Elmegreen} \& {Lada}(1977)}]{elmegreen_1977}
{Elmegreen}, B.~G. \& {Lada}, C.~J. 1977, \apj, 214, 725

\bibitem[{Fay \& Proschan(2010)}]{fay_2010}
Fay, M. \& Proschan, M. 2010, Statistics surveys, 4, 1

\bibitem[{{Francis} {et~al.}(1998){Francis}, {Boffin}, {Watkins}, \&
  {Whitworth}}]{francis_1998}
{Francis}, N., {Boffin}, H.~M.~J., {Watkins}, S.~J., \& {Whitworth}, A.~P.
  1998, in Astronomical Society of the Pacific Conference Series, Vol. 132,
  Star Formation with the Infrared Space Observatory, ed. J.~{Yun} \&
  L.~{Liseau}, 346

\bibitem[{{Fukui} {et~al.}(2021){Fukui}, {Inoue}, {Hayakawa}, \&
  {Torii}}]{fukui_2021}
{Fukui}, Y., {Inoue}, T., {Hayakawa}, T., \& {Torii}, K. 2021, \pasj, 73, S405

\bibitem[{{Fukushima} \& {Yajima}(2021)}]{fukushima_2021}
{Fukushima}, H. \& {Yajima}, H. 2021, \mnras, 506, 5512

\bibitem[{{Geen} {et~al.}(2021){Geen}, {Bieri}, {Rosdahl}, \& {de
  Koter}}]{geen_2021}
{Geen}, S., {Bieri}, R., {Rosdahl}, J., \& {de Koter}, A. 2021, \mnras, 501,
  1352

\bibitem[{{Geen} {et~al.}(2015){Geen}, {Hennebelle}, {Tremblin}, \&
  {Rosdahl}}]{geen_2015}
{Geen}, S., {Hennebelle}, P., {Tremblin}, P., \& {Rosdahl}, J. 2015, \mnras,
  454, 4484

\bibitem[{{Grudi{\'c}} {et~al.}(2021){Grudi{\'c}}, {Kruijssen},
  {Faucher-Gigu{\`e}re}, {Hopkins}, {Ma}, {Quataert}, \&
  {Boylan-Kolchin}}]{grudic_2021}
{Grudi{\'c}}, M.~Y., {Kruijssen}, J.~M.~D., {Faucher-Gigu{\`e}re}, C.-A.,
  {et~al.} 2021, \mnras, 506, 3239

\bibitem[{{G{\"u}sten} {et~al.}(2006){G{\"u}sten}, {Nyman}, {Schilke},
  {Menten}, {Cesarsky}, \& {Booth}}]{gusten_2006}
{G{\"u}sten}, R., {Nyman}, L.~{\r{A}}., {Schilke}, P., {et~al.} 2006, \aap,
  454, L13

\bibitem[{{Habing} {et~al.}(1972){Habing}, {Israel}, \& {de
  Jong}}]{habing_1972}
{Habing}, H.~J., {Israel}, F.~P., \& {de Jong}, T. 1972, \aap, 17, 329

\bibitem[{{Haid} {et~al.}(2018){Haid}, {Walch}, {Seifried}, {W{\"u}nsch},
  {Dinnbier}, \& {Naab}}]{haid_2018}
{Haid}, S., {Walch}, S., {Seifried}, D., {et~al.} 2018, \mnras, 478, 4799

\bibitem[{{Haid} {et~al.}(2019){Haid}, {Walch}, {Seifried}, {W{\"u}nsch},
  {Dinnbier}, \& {Naab}}]{haid_2019}
{Haid}, S., {Walch}, S., {Seifried}, D., {et~al.} 2019, \mnras, 482, 4062

\bibitem[{{Heitsch} {et~al.}(2008){Heitsch}, {Hartmann}, {Slyz}, {Devriendt},
  \& {Burkert}}]{fil_sim_6}
{Heitsch}, F., {Hartmann}, L.~W., {Slyz}, A.~D., {Devriendt}, J. E.~G., \&
  {Burkert}, A. 2008, \apj, 674, 316

\bibitem[{{Hennebelle} {et~al.}(2008){Hennebelle}, {Banerjee},
  {V{\'a}zquez-Semadeni}, {Klessen}, \& {Audit}}]{fil_sim_4}
{Hennebelle}, P., {Banerjee}, R., {V{\'a}zquez-Semadeni}, E., {Klessen}, R.~S.,
  \& {Audit}, E. 2008, \aap, 486, L43

\bibitem[{{Hora} {et~al.}(2009){Hora}, {Bontemps}, {Megeath}, {Schneider},
  {Motte}, {Carey}, {Simon}, {Keto}, {Smith}, {Allen}, {Gutermuth}, {Fazio},
  {Adams}, \& {Cygnus-X 24 Micron Data Processing Team}}]{hora_2009}
{Hora}, J.~L., {Bontemps}, S., {Megeath}, S.~T., {et~al.} 2009, in American
  Astronomical Society Meeting Abstracts, Vol. 213, American Astronomical
  Society Meeting Abstracts \#213, 356.01

\bibitem[{{Jackson} {et~al.}(2010){Jackson}, {Finn}, {Chambers}, {Rathborne},
  \& {Simon}}]{nessie_1}
{Jackson}, J.~M., {Finn}, S.~C., {Chambers}, E.~T., {Rathborne}, J.~M., \&
  {Simon}, R. 2010, \apjl, 719, L185

\bibitem[{{Jaffa} {et~al.}(2018){Jaffa}, {Whitworth}, {Clarke}, \&
  {Howard}}]{jaffa_2018}
{Jaffa}, S.~E., {Whitworth}, A.~P., {Clarke}, S.~D., \& {Howard}, A.~D.~P.
  2018, \mnras, 477, 1940

\bibitem[{{Jayasinghe} {et~al.}(2019){Jayasinghe}, {Dixon}, {Povich}, {Binder},
  {Velasco}, {Lepore}, {Xu}, {Offner}, {Kobulnicky}, {Anderson}, {Kendrew}, \&
  {Simpson}}]{jayasingghe_2019}
{Jayasinghe}, T., {Dixon}, D., {Povich}, M.~S., {et~al.} 2019, \mnras, 488,
  1141

\bibitem[{{Kauffmann} \& {Pillai}(2010)}]{kauffmann_2010}
{Kauffmann}, J. \& {Pillai}, T. 2010, \apjl, 723, L7

\bibitem[{{Klessen} \& {Burkert}(2000)}]{fil_sim_5}
{Klessen}, R.~S. \& {Burkert}, A. 2000, \apjs, 128, 287

\bibitem[{{K{\"o}nyves} {et~al.}(2015){K{\"o}nyves}, {Andr{\'e}}, \&
  {Maury}}]{konyves_2015}
{K{\"o}nyves}, V., {Andr{\'e}}, P., \& {Maury}, A. 2015, in IAU General
  Assembly, Vol.~29, 2257481

\bibitem[{{Krumholz}(2006)}]{krumholz_2006}
{Krumholz}, M.~R. 2006, arXiv e-prints, astro

\bibitem[{{Li} {et~al.}(2016){Li}, {Urquhart}, {Leurini}, {Csengeri},
  {Wyrowski}, {Menten}, \& {Schuller}}]{ATLASGAL_filaments}
{Li}, G.-X., {Urquhart}, J.~S., {Leurini}, S., {et~al.} 2016, \aap, 591, A5

\bibitem[{{Li} {et~al.}(2019){Li}, {Vogelsberger}, {Marinacci}, \&
  {Gnedin}}]{li_2019}
{Li}, H., {Vogelsberger}, M., {Marinacci}, F., \& {Gnedin}, O.~Y. 2019, \mnras,
  487, 364

\bibitem[{{Lin} {et~al.}(2020){Lin}, {Wang}, {Su}, {Li}, \& {Yang}}]{lin_2020}
{Lin}, L.-H., {Wang}, H.-C., {Su}, Y., {Li}, C., \& {Yang}, J. 2020, Research
  in Astronomy and Astrophysics, 20, 143

\bibitem[{{Liow} \& {Dobbs}(2020)}]{liow_2020}
{Liow}, K.~Y. \& {Dobbs}, C.~L. 2020, \mnras, 499, 1099

\bibitem[{{Marsh} {et~al.}(2016){Marsh}, {Kirk}, {Andr{\'e}}, {Griffin},
  {K{\"o}nyves}, {Palmeirim}, {Men'shchikov}, {Ward-Thompson}, {Benedettini},
  {Bresnahan}, {di Francesco}, {Elia}, {Motte}, {Peretto}, {Pezzuto}, {Roy},
  {Sadavoy}, {Schneider}, {Spinoglio}, \& {White}}]{marsh_2016}
{Marsh}, K.~A., {Kirk}, J.~M., {Andr{\'e}}, P., {et~al.} 2016, \mnras, 459, 342

\bibitem[{{Mattern} {et~al.}(2018{\natexlab{a}}){Mattern}, {Kainulainen},
  {Zhang}, \& {Beuther}}]{filament_intro_2_nessie_strucuture}
{Mattern}, M., {Kainulainen}, J., {Zhang}, M., \& {Beuther}, H.
  2018{\natexlab{a}}, \aap, 616, A78

\bibitem[{{Mattern} {et~al.}(2018{\natexlab{b}}){Mattern}, {Kauffmann},
  {Csengeri}, {Urquhart}, {Leurini}, {Wyrowski}, {Giannetti}, {Barnes},
  {Beuther}, {Bronfman}, {Duarte-Cabral}, {Henning}, {Kainulainen}, {Menten},
  {Schisano}, \& {Schuller}}]{mass_vel_scaling_relation}
{Mattern}, M., {Kauffmann}, J., {Csengeri}, T., {et~al.} 2018{\natexlab{b}},
  \aap, 619, A166

\bibitem[{{Mazumdar} {et~al.}(2021){Mazumdar}, {Wyrowski}, {Colombo},
  {Urquhart}, {Thompson}, \& {Menten}}]{mazumdar_2021}
{Mazumdar}, P., {Wyrowski}, F., {Colombo}, D., {et~al.} 2021, \aap, 650, A164

\bibitem[{{Molinari} {et~al.}(2010){Molinari}, {Swinyard}, {Bally}, {Barlow},
  {Bernard}, {Martin}, {Moore}, {Noriega-Crespo}, {Plume}, {Testi}, {Zavagno},
  {Abergel}, {Ali}, {Anderson}, {Andr{\'e}}, {Baluteau}, {Battersby},
  {Beltr{\'a}n}, {Benedettini}, {Billot}, {Blommaert}, {Bontemps}, {Boulanger},
  {Brand}, {Brunt}, {Burton}, {Calzoletti}, {Carey}, {Caselli}, {Cesaroni},
  {Cernicharo}, {Chakrabarti}, {Chrysostomou}, {Cohen}, {Compiegne}, {de
  Bernardis}, {de Gasperis}, {di Giorgio}, {Elia}, {Faustini}, {Flagey},
  {Fukui}, {Fuller}, {Ganga}, {Garcia-Lario}, {Glenn}, {Goldsmith}, {Griffin},
  {Hoare}, {Huang}, {Ikhenaode}, {Joblin}, {Joncas}, {Juvela}, {Kirk},
  {Lagache}, {Li}, {Lim}, {Lord}, {Marengo}, {Marshall}, {Masi}, {Massi},
  {Matsuura}, {Minier}, {Miville-Desch{\^e}nes}, {Montier}, {Morgan}, {Motte},
  {Mottram}, {M{\"u}ller}, {Natoli}, {Neves}, {Olmi}, {Paladini}, {Paradis},
  {Parsons}, {Peretto}, {Pestalozzi}, {Pezzuto}, {Piacentini}, {Piazzo},
  {Polychroni}, {Pomar{\`e}s}, {Popescu}, {Reach}, {Ristorcelli}, {Robitaille},
  {Robitaille}, {Rod{\'o}n}, {Roy}, {Royer}, {Russeil}, {Saraceno}, {Sauvage},
  {Schilke}, {Schisano}, {Schneider}, {Schuller}, {Schulz}, {Sibthorpe},
  {Smith}, {Smith}, {Spinoglio}, {Stamatellos}, {Strafella}, {Stringfellow},
  {Sturm}, {Taylor}, {Thompson}, {Traficante}, {Tuffs}, {Umana}, {Valenziano},
  {Vavrek}, {Veneziani}, {Viti}, {Waelkens}, {Ward-Thompson}, {White},
  {Wilcock}, {Wyrowski}, {Yorke}, \& {Zhang}}]{higal_MW}
{Molinari}, S., {Swinyard}, B., {Bally}, J., {et~al.} 2010, \aap, 518, L100

\bibitem[{{Nakamura} \& {Li}(2008)}]{fil_sim_3}
{Nakamura}, F. \& {Li}, Z.-Y. 2008, \apj, 687, 354

\bibitem[{{Pabst} {et~al.}(2020){Pabst}, {Goicoechea}, {Teyssier}, {Bern{\'e}},
  {Higgins}, {Chambers}, {Kabanovic}, {G{\"u}sten}, {Stutzki}, \&
  {Tielens}}]{pabst_2020}
{Pabst}, C.~H.~M., {Goicoechea}, J.~R., {Teyssier}, D., {et~al.} 2020, \aap,
  639, A2

\bibitem[{{Padoan} {et~al.}(2001){Padoan}, {Juvela}, {Goodman}, \&
  {Nordlund}}]{fil_sim_1}
{Padoan}, P., {Juvela}, M., {Goodman}, A.~A., \& {Nordlund}, {\r{A}}. 2001,
  \apj, 553, 227

\bibitem[{{Peretto} \& {Fuller}(2010)}]{peretto_2010}
{Peretto}, N. \& {Fuller}, G.~A. 2010, \apj, 723, 555

\bibitem[{{Pitts} {et~al.}(2019){Pitts}, {Barnes}, \& {Varosi}}]{pitts_2019}
{Pitts}, R.~L., {Barnes}, P.~J., \& {Varosi}, F. 2019, \mnras, 484, 305

\bibitem[{{Priestley} \& {Whitworth}(2021)}]{priestley_2021}
{Priestley}, F.~D. \& {Whitworth}, A.~P. 2021, arXiv e-prints, arXiv:2109.13277

\bibitem[{{Ragan} {et~al.}(2014){Ragan}, {Henning}, {Tackenberg}, {Beuther},
  {Johnston}, {Kainulainen}, \& {Linz}}]{GMF_fil_1}
{Ragan}, S.~E., {Henning}, T., {Tackenberg}, J., {et~al.} 2014, \aap, 568, A73

\bibitem[{{Rosen} {et~al.}(2020){Rosen}, {Offner}, {Sadavoy}, {Bhandare},
  {V{\'a}zquez-Semadeni}, \& {Ginsburg}}]{rosen_2020}
{Rosen}, A.~L., {Offner}, S. S.~R., {Sadavoy}, S.~I., {et~al.} 2020, \ssr, 216,
  62

\bibitem[{{Rosolowsky} {et~al.}(2021){Rosolowsky}, {Hughes}, {Leroy}, {Sun},
  {Querejeta}, {Schruba}, {Usero}, {Herrera}, {Liu}, {Pety}, {Saito},
  {Be{\v{s}}li{\'c}}, {Bigiel}, {Blanc}, {Chevance}, {Dale}, {Deger}, {Faesi},
  {Glover}, {Henshaw}, {Klessen}, {Kruijssen}, {Larson}, {Lee}, {Meidt}, {Mok},
  {Schinnerer}, {Thilker}, \& {Williams}}]{rosolowsky_2021}
{Rosolowsky}, E., {Hughes}, A., {Leroy}, A.~K., {et~al.} 2021, \mnras, 502,
  1218

\bibitem[{{Rosolowsky} \& {Leroy}(2006)}]{crops}
{Rosolowsky}, E. \& {Leroy}, A. 2006, \pasp, 118, 590

\bibitem[{{Rosolowsky} {et~al.}(2008){Rosolowsky}, {Pineda}, {Kauffmann}, \&
  {Goodman}}]{dendrograms}
{Rosolowsky}, E.~W., {Pineda}, J.~E., {Kauffmann}, J., \& {Goodman}, A.~A.
  2008, \apj, 679, 1338

\bibitem[{{Schisano} {et~al.}(2020){Schisano}, {Molinari}, {Elia},
  {Benedettini}, {Olmi}, {Pezzuto}, {Traficante}, {Brescia}, {Cavuoti}, {di
  Giorgio}, {Liu}, {Moore}, {Noriega-Crespo}, {Riccio}, {Baldeschi},
  {Becciani}, {Peretto}, {Merello}, {Vitello}, {Zavagno}, {Beltr{\'a}n},
  {Cambr{\'e}sy}, {Eden}, {Li Causi}, {Molinaro}, {Palmeirim}, {Sciacca},
  {Testi}, {Umana}, \& {Whitworth}}]{schisano_2020}
{Schisano}, E., {Molinari}, S., {Elia}, D., {et~al.} 2020, \mnras, 492, 5420

\bibitem[{{Schisano} {et~al.}(2014){Schisano}, {Rygl}, {Molinari}, {Busquet},
  {Elia}, {Pestalozzi}, {Polychroni}, {Billot}, {Carey}, {Paladini},
  {Noriega-Crespo}, {Moore}, {Plume}, {Glover}, \&
  {V{\'a}zquez-Semadeni}}]{schisano_2014}
{Schisano}, E., {Rygl}, K.~L.~J., {Molinari}, S., {et~al.} 2014, \apj, 791, 27

\bibitem[{{Schneider} {et~al.}(2020){Schneider}, {Simon}, {Guevara},
  {Buchbender}, {Higgins}, {Okada}, {Stutzki}, {G{\"u}sten}, {Anderson},
  {Bally}, {Beuther}, {Bonne}, {Bontemps}, {Chambers}, {Csengeri}, {Graf},
  {Gusdorf}, {Jacobs}, {Justen}, {Kabanovic}, {Karim}, {Luisi}, {Menten},
  {Mertens}, {Mookerjea}, {Ossenkopf-Okada}, {Pabst}, {Pound}, {Richter},
  {Reyes}, {Ricken}, {R{\"o}llig}, {Russeil}, {S{\'a}nchez-Monge}, {Sandell},
  {Tiwari}, {Wiesemeyer}, {Wolfire}, {Wyrowski}, {Zavagno}, \&
  {Tielens}}]{schneider_2020}
{Schneider}, N., {Simon}, R., {Guevara}, C., {et~al.} 2020, \pasp, 132, 104301

\bibitem[{{Schneider} \& {Elmegreen}(1979)}]{globular_filament_ref_nessie_Stru}
{Schneider}, S. \& {Elmegreen}, B.~G. 1979, \apjs, 41, 87

\bibitem[{{Schuller} {et~al.}(2017){Schuller}, {Csengeri}, {Urquhart},
  {Duarte-Cabral}, {Barnes}, {Giannetti}, {Hernandez}, {Leurini}, {Mattern},
  {Medina}, {Agurto}, {Azagra}, {Anderson}, {Beltr{\'a}n}, {Beuther},
  {Bontemps}, {Bronfman}, {Dobbs}, {Dumke}, {Finger}, {Ginsburg}, {Gonzalez},
  {Henning}, {Kauffmann}, {Mac-Auliffe}, {Menten}, {Montenegro-Montes},
  {Moore}, {Muller}, {Parra}, {Perez-Beaupuits}, {Pettitt}, {Russeil},
  {S{\'a}nchez-Monge}, {Schilke}, {Schisano}, {Suri}, {Testi}, {Torstensson},
  {Venegas}, {Wang}, {Wienen}, {Wyrowski}, \& {Zavagno}}]{schuller_2017}
{Schuller}, F., {Csengeri}, T., {Urquhart}, J.~S., {et~al.} 2017, \aap, 601,
  A124

\bibitem[{{Schuller} {et~al.}(2009){Schuller}, {Menten}, {Contreras},
  {Wyrowski}, {Schilke}, {Bronfman}, {Henning}, {Walmsley}, {Beuther},
  {Bontemps}, {Cesaroni}, {Deharveng}, {Garay}, {Herpin}, {Lefloch}, {Linz},
  {Mardones}, {Minier}, {Molinari}, {Motte}, {Nyman}, {Reveret}, {Risacher},
  {Russeil}, {Schneider}, {Testi}, {Troost}, {Vasyunina}, {Wienen}, {Zavagno},
  {Kovacs}, {Kreysa}, {Siringo}, \& {Wei{\ss}}}]{ATLASGAL_survey}
{Schuller}, F., {Menten}, K.~M., {Contreras}, Y., {et~al.} 2009, \aap, 504, 415

\bibitem[{{Schuller} {et~al.}(2021){Schuller}, {Urquhart}, {Csengeri},
  {Colombo}, {Duarte-Cabral}, {Mattern}, {Ginsburg}, {Pettitt}, {Wyrowski},
  {Anderson}, {Azagra}, {Barnes}, {Beltran}, {Beuther}, {Billington},
  {Bronfman}, {Cesaroni}, {Dobbs}, {Eden}, {Lee}, {Medina}, {Menten}, {Moore},
  {Montenegro-Montes}, {Ragan}, {Rigby}, {Riener}, {Russeil}, {Schisano},
  {Sanchez-Monge}, {Traficante}, {Zavagno}, {Agurto}, {Bontemps}, {Finger},
  {Giannetti}, {Gonzalez}, {Hernandez}, {Henning}, {Kainulainen}, {Kauffmann},
  {Leurini}, {Lopez}, {Mac-Auliffe}, {Mazumdar}, {Molinari}, {Motte}, {Muller},
  {Nguyen-Luong}, {Parra}, {Perez-Beaupuits}, {Schilke}, {Schneider}, {Suri},
  {Testi}, {Torstensson}, {Veena}, {Venegas}, {Wang}, \& {Wienen}}]{SEDIGISM_1}
{Schuller}, F., {Urquhart}, J.~S., {Csengeri}, T., {et~al.} 2021, \mnras, 500,
  3064

\bibitem[{{Simpson} {et~al.}(2012){Simpson}, {Povich}, {Kendrew}, {Lintott},
  {Bressert}, {Arvidsson}, {Cyganowski}, {Maddison}, {Schawinski}, {Sherman},
  {Smith}, \& {Wolf-Chase}}]{simpson_2012}
{Simpson}, R.~J., {Povich}, M.~S., {Kendrew}, S., {et~al.} 2012, \mnras, 424,
  2442

\bibitem[{{Siringo} {et~al.}(2009){Siringo}, {Kreysa}, {Kov{\'a}cs},
  {Schuller}, {Wei{\ss}}, {Esch}, {Gem{\"u}nd}, {Jethava}, {Lundershausen},
  {Colin}, {G{\"u}sten}, {Menten}, {Beelen}, {Bertoldi}, {Beeman}, \&
  {Haller}}]{siringo_2009}
{Siringo}, G., {Kreysa}, E., {Kov{\'a}cs}, A., {et~al.} 2009, \aap, 497, 945

\bibitem[{{Suri} {et~al.}(2019){Suri}, {S{\'a}nchez-Monge}, {Schilke},
  {Clarke}, {Smith}, {Ossenkopf-Okada}, {Klessen}, {Padoan}, {Goldsmith},
  {Arce}, {Bally}, {Carpenter}, {Ginsburg}, {Johnstone}, {Kauffmann}, {Kong},
  {Lis}, {Mairs}, {Pillai}, {Pineda}, \& {Duarte-Cabral}}]{suri_2019}
{Suri}, S., {S{\'a}nchez-Monge}, {\'A}., {Schilke}, P., {et~al.} 2019, \aap,
  623, A142

\bibitem[{{Tiwari} {et~al.}(2021){Tiwari}, {Karim}, {Pound}, {Wolfire},
  {Jacob}, {Buchbender}, {G{\"u}sten}, {Guevara}, {Higgins}, {Kabanovic},
  {Pabst}, {Ricken}, {Schneider}, {Simon}, {Stutzki}, \&
  {Tielens}}]{tiwari_2021}
{Tiwari}, M., {Karim}, R., {Pound}, M.~W., {et~al.} 2021, \apj, 914, 117

\bibitem[{{Urquhart} {et~al.}(2014){Urquhart}, {Csengeri}, {Wyrowski},
  {Schuller}, {Bontemps}, {Bronfman}, {Menten}, {Walmsley}, {Contreras},
  {Beuther}, {Wienen}, \& {Linz}}]{urquhart_2014}
{Urquhart}, J.~S., {Csengeri}, T., {Wyrowski}, F., {et~al.} 2014, \aap, 568,
  A41

\bibitem[{{Urquhart} {et~al.}(2021){Urquhart}, {Figura}, {Cross}, {Wells},
  {Moore}, {Eden}, {Ragan}, {Pettitt}, {Duarte-Cabral}, {Colombo}, {Schuller},
  {Csengeri}, {Mattern}, {Beuther}, {Menten}, {Wyrowski}, {Anderson}, {Barnes},
  {Beltr{\'a}n}, {Billington}, {Bronfman}, {Giannetti}, {Kainulainen},
  {Kauffmann}, {Lee}, {Leurini}, {Medina}, {Montenegro-Montes}, {Riener},
  {Rigby}, {S{\'a}nchez-Monge}, {Schilke}, {Schisano}, {Traficante}, \&
  {Wienen}}]{james_paper}
{Urquhart}, J.~S., {Figura}, C., {Cross}, J.~R., {et~al.} 2021, \mnras, 500,
  3050

\bibitem[{{Urquhart} {et~al.}(2018){Urquhart}, {K{\"o}nig}, {Giannetti},
  {Leurini}, {Moore}, {Eden}, {Pillai}, {Thompson}, {Braiding}, {Burton},
  {Csengeri}, {Dempsey}, {Figura}, {Froebrich}, {Menten}, {Schuller}, {Smith},
  \& {Wyrowski}}]{Urquhart_2018}
{Urquhart}, J.~S., {K{\"o}nig}, C., {Giannetti}, A., {et~al.} 2018, \mnras,
  473, 1059

\bibitem[{{Urquhart} {et~al.}(2022){Urquhart}, {Wells}, {Pillai}, {Leurini},
  {Giannetti}, {Moore}, {Thompson}, {Figura}, {Colombo}, {Yang}, {K{\"o}nig},
  {Wyrowski}, {Menten}, {Rigby}, {Eden}, \& {Ragan}}]{urquhart_2022}
{Urquhart}, J.~S., {Wells}, M.~R.~A., {Pillai}, T., {et~al.} 2022, \mnras, 510,
  3389

\bibitem[{{V{\'a}zquez-Semadeni} {et~al.}(2011){V{\'a}zquez-Semadeni},
  {Banerjee}, {G{\'o}mez}, {Hennebelle}, {Duffin}, \& {Klessen}}]{fil_sim_2}
{V{\'a}zquez-Semadeni}, E., {Banerjee}, R., {G{\'o}mez}, G.~C., {et~al.} 2011,
  \mnras, 414, 2511

\bibitem[{{Wang} {et~al.}(2015){Wang}, {Testi}, {Ginsburg}, {Walmsley},
  {Molinari}, \& {Schisano}}]{Herschel_fil_1}
{Wang}, K., {Testi}, L., {Ginsburg}, A., {et~al.} 2015, \mnras, 450, 4043

\bibitem[{{Wang} {et~al.}(2020){Wang}, {Bihr}, {Beuther}, {Rugel}, {Soler},
  {Ott}, {Kainulainen}, {Schneider}, {Klessen}, {Glover}, {McClure-Griffiths},
  {Goldsmith}, {Johnston}, {Menten}, {Ragan}, {Anderson}, {Urquhart}, {Linz},
  {Roy}, {Smith}, {Bigiel}, {Henning}, \& {Longmore}}]{wang_2020}
{Wang}, Y., {Bihr}, S., {Beuther}, H., {et~al.} 2020, \aap, 634, A139

\bibitem[{{Yuan} {et~al.}(2021){Yuan}, {Yang}, {Du}, {Liu}, {Zhang}, {Lin},
  {Sun}, {Yan}, {Ma}, {Su}, {Sun}, \& {Zhou}}]{Yuan_2021}
{Yuan}, L., {Yang}, J., {Du}, F., {et~al.} 2021, arXiv e-prints,
  arXiv:2108.12110

\bibitem[{{Zavagno} {et~al.}(2010){Zavagno}, {Anderson}, {Russeil}, {Morgan},
  {Stringfellow}, {Deharveng}, {Rod{\'o}n}, {Robitaille}, {Mottram},
  {Schuller}, {Testi}, {Billot}, {Molinari}, {di Gorgio}, {Kirk}, {Brunt},
  {Ward-Thompson}, {Traficante}, {Veneziani}, {Faustini}, \&
  {Calzoletti}}]{zavagno_2010}
{Zavagno}, A., {Anderson}, L.~D., {Russeil}, D., {et~al.} 2010, \aap, 518, L101

\bibitem[{{Zavagno} {et~al.}(2006){Zavagno}, {Deharveng}, {Comer{\'o}n},
  {Brand}, {Massi}, {Caplan}, \& {Russeil}}]{zavagno_2006}
{Zavagno}, A., {Deharveng}, L., {Comer{\'o}n}, F., {et~al.} 2006, \aap, 446,
  171

\bibitem[{{Zhang} {et~al.}(2019){Zhang}, {Kainulainen}, {Mattern}, {Fang}, \&
  {Henning}}]{zhang_2019}
{Zhang}, M., {Kainulainen}, J., {Mattern}, M., {Fang}, M., \& {Henning}, T.
  2019, \aap, 622, A52

\bibitem[{{Zhou} {et~al.}(2020){Zhou}, {Zhou}, {Esimbek}, {Baan}, {Wu}, {Ji},
  {He}, {Li}, {Sailanbek}, {Komesh}, \& {Tang}}]{jianjun_2020}
{Zhou}, J., {Zhou}, D., {Esimbek}, J., {et~al.} 2020, \apj, 897, 74

\bibitem[{{Zucker} {et~al.}(2018){Zucker}, {Battersby}, \&
  {Goodman}}]{filaments_catherine_zucker}
{Zucker}, C., {Battersby}, C., \& {Goodman}, A. 2018, \apj, 864, 153

\end{thebibliography}
}

\begin{appendix}

\section{\add{Data products (catalogues)} \remove{Cloud morphology catalogue}}\label{app: cloud catalogue}

We have obtained the morphologies for the 10663 SEDIGISM molecular clouds and compiled them as a catalogue. The description for the columns in the catalogue is provided in Table \ref{table: cloud catalog}. It contains the structure of the clouds determined by the $J$ plots (along with $J$ moments) and by using by-eye classification. The table also mentions if a cloud is a part of the science, VC and MR samples. We also provide the number of overlapping structures from the ATLASGAL and MWP surveys (Sec. \ref{sec: counterparts}) for each SEDIGISM molecular cloud. These structures strictly fall in the SEDIGISM coverage i.e. they are excluded if they overlap with the survey edge (e.g. $l$ = 18$\degree$). The complete catalogue of SEDIGISM molecular clouds (by DC21) is provided at \url{https://sedigism.mpifr-bonn.mpg.de/index.html}. 

\add{We also provide catalogues of the structures from ATLASGAL and MWP surveys that overlap with the SEDIGISM clouds. The ID numbers for AG-fil, AG-El} \citep{ATLASGAL_filaments} \add{and MWP bubbles} \citep{jayasingghe_2019} \add{overlapping with the SEDIGISM clouds are provided in Table \ref{table: cloud filament overlap app}, \ref{table: cloud agal overlap app} \& \ref{table: cloud mwp overlap app} respectively.}

\begin{table*}[!h]
\caption{Description of the generated catalogue, which represents the morphology of molecular clouds and their overlapping counterparts.}
\label{table: cloud catalog}
\centering
\begin{tabular}{p{4cm}p{12cm}}
\hline
Catalogue column           & Description                                                          \\ \hline
cloud\_id                & Unique cloud ID number described by DC20                               \\
cloud\_name              & Cloud name, SDG followed by Galactic coordinates as described by DC20  \\
j1                       & $J$ moment of the cloud corresponding to principal inertial axis $I_1$ \\
j2                       & $J$ moment of the cloud corresponding to principal inertial axis $I_2$ \\
j\_structure             & Structure of the cloud determined by $J$ plot analysis                 \\
\add{by\_eye\_subclass}     & Original sub-class of cloud estimated using by-eye analysis (App. \ref{App: ks test}) \\
by\_eye\_structure       & Structure of cloud determined using by-eye analysis                    \\
science\_samp            & tag identifying if the cloud belongs to science sample (described in DC20) (yes = 1, no = 0)\\
vc\_samp                 & tag identifying if the cloud belongs to visually classified (VC) sample (yes = 1, no = 0)\\
mr\_samp                 & tag identifying if the cloud belongs to morphologically reliable (MR) sample (yes = 1, no = 0)\\
agal\_elon\_count   & Number of AG-El overlapping with the cloud \\
agal\_fil\_count         & Number of AG-Fil overlapping with the cloud           \\
mwp\_bub\_count          & Number of MWP bubbles overlapping with the cloud     \\      
 \hline
\end{tabular}

\end{table*}

\begin{table*}[!h]
\caption{\add{Description of the ATLASGAL filaments (AG-Fil) overlapping with the SEDIGISM clouds.}}
\label{table: cloud filament overlap app}
\centering
\begin{tabular}{p{4cm}p{12cm}}
\hline
Catalogue column           & Description                                                          \\ \hline
cloud\_id        &   ID number of SEDIGISM cloud\\
fil\_name     & Name of the ATLASGAL filament \citep{ATLASGAL_filaments} overlapping with the SEDIGISM cloud   \\
 \hline
\end{tabular}

\end{table*}

\begin{table*}[!h]
\caption{\add{Description of the ATLASGAL elongated structures (AG-El) overlapping with the SEDIGISM clouds.}}
\label{table: cloud agal overlap app}
\centering
\begin{tabular}{p{4cm}p{12cm}}
\hline
Catalogue column           & Description                                                          \\ \hline
cloud\_id        &   ID number of SEDIGISM cloud\\
agal\_name     & Name of the ATLASGAL structure \citep{ATLASGAL_filaments}  overlapping with the SEDIGISM cloud  \\
 \hline
\end{tabular}

\end{table*}

\begin{table*}[!h]
\caption{\add{Description of the MWP bubbles overlapping with the SEDIGISM clouds.}}
\label{table: cloud mwp overlap app}
\centering
\begin{tabular}{p{4cm}p{12cm}}
\hline
Catalogue column           & Description                                                          \\ \hline
cloud\_id        & ID number of SEDIGISM cloud         \\
mwp\_name     & Name of the MWP bubble \citep{jayasingghe_2019} overlapping with the SEDIGISM cloud \\
 \hline
\end{tabular}
\end{table*}

\section{Overlap of SEDIGISM clouds with ATLASGAL elongated structures and MWP bubbles} \label{app: cloud overlap}

\begin{figure*}[]
    \centering
    \begin{minipage}{\textwidth}
    \includegraphics[width = \textwidth, keepaspectratio]{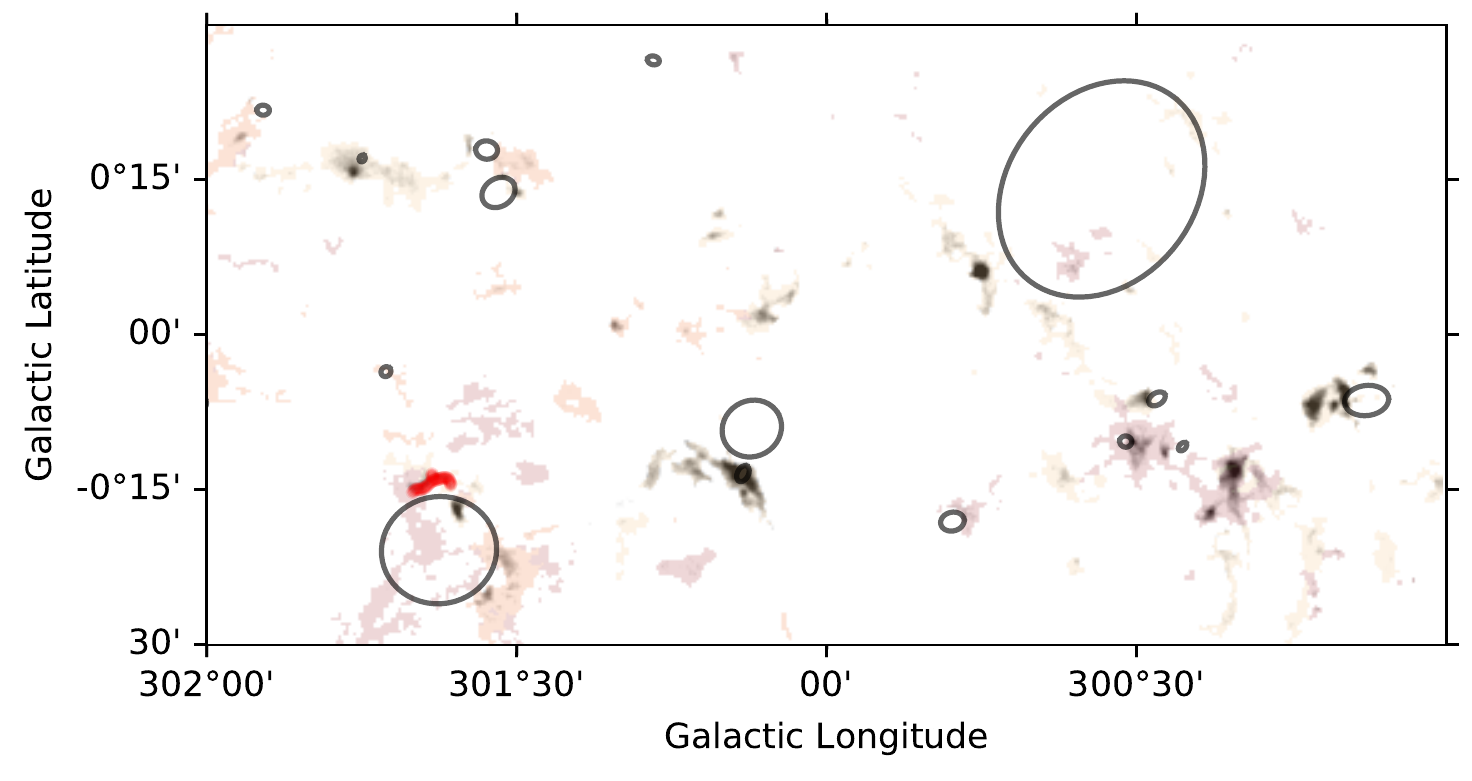}
    \caption{Elongated structures from ATLASGAL survey (filaments in red, networks of filaments in blue and resolved elongated structures in green) and bubbles (black ellipses) from MWP survey overlaid on SEDIGISM clouds for $300 \degree \leq l \leq 302 \degree$.}
    \label{fig: SED overlap 300}
    \end{minipage}\hfill
\end{figure*}

\begin{figure*}[]
    \centering
    \begin{minipage}{\textwidth}
    \includegraphics[width = \textwidth, keepaspectratio]{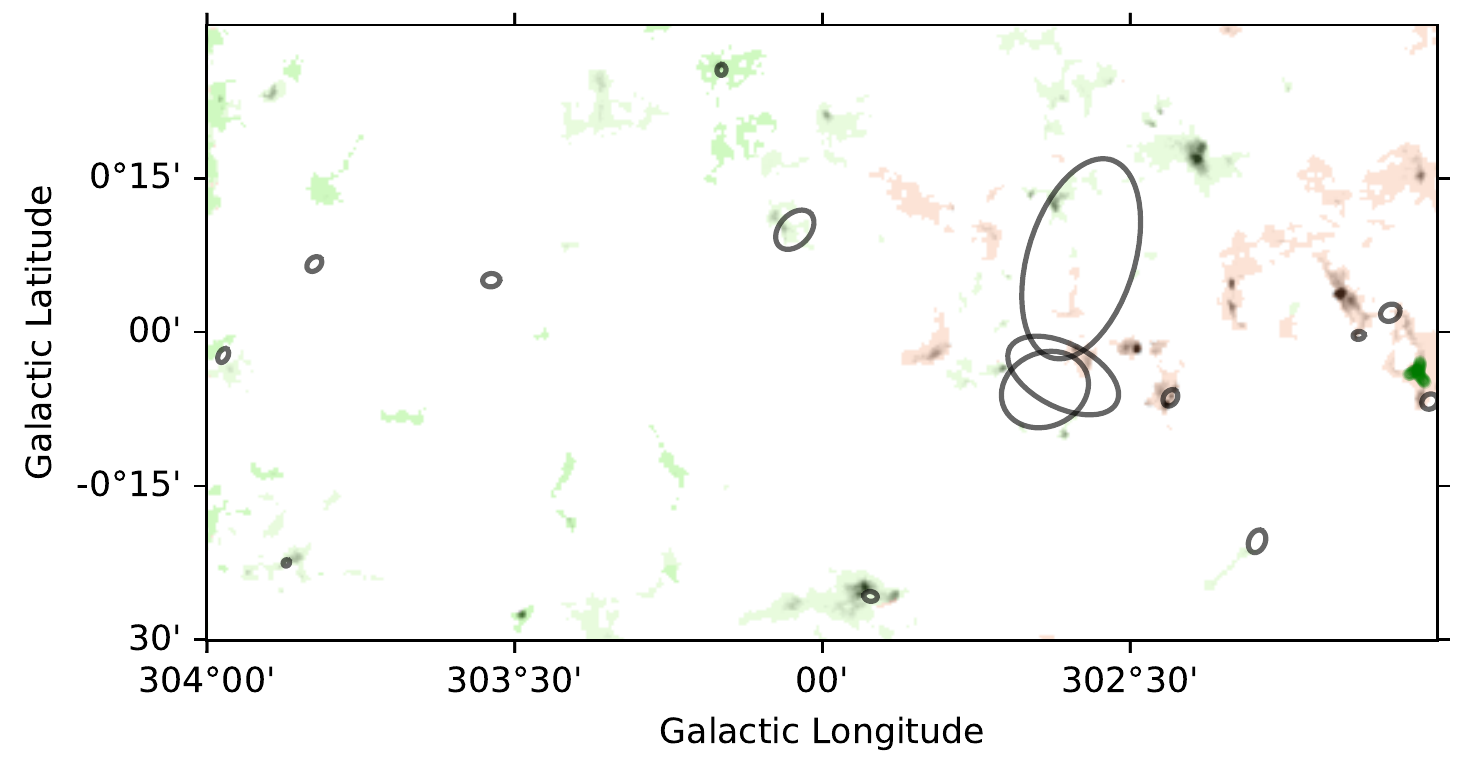}
    \caption{Elongated structures from ATLASGAL survey and bubbles from MWP survey overlaid on SEDIGISM clouds for $302 \degree \leq l \leq 304 \degree$.}
    \label{fig: SED overlap 302}
    \end{minipage}\hfill
\end{figure*}

\begin{figure*}[]
    \centering
    \begin{minipage}{\textwidth}
    \includegraphics[width = \textwidth, keepaspectratio]{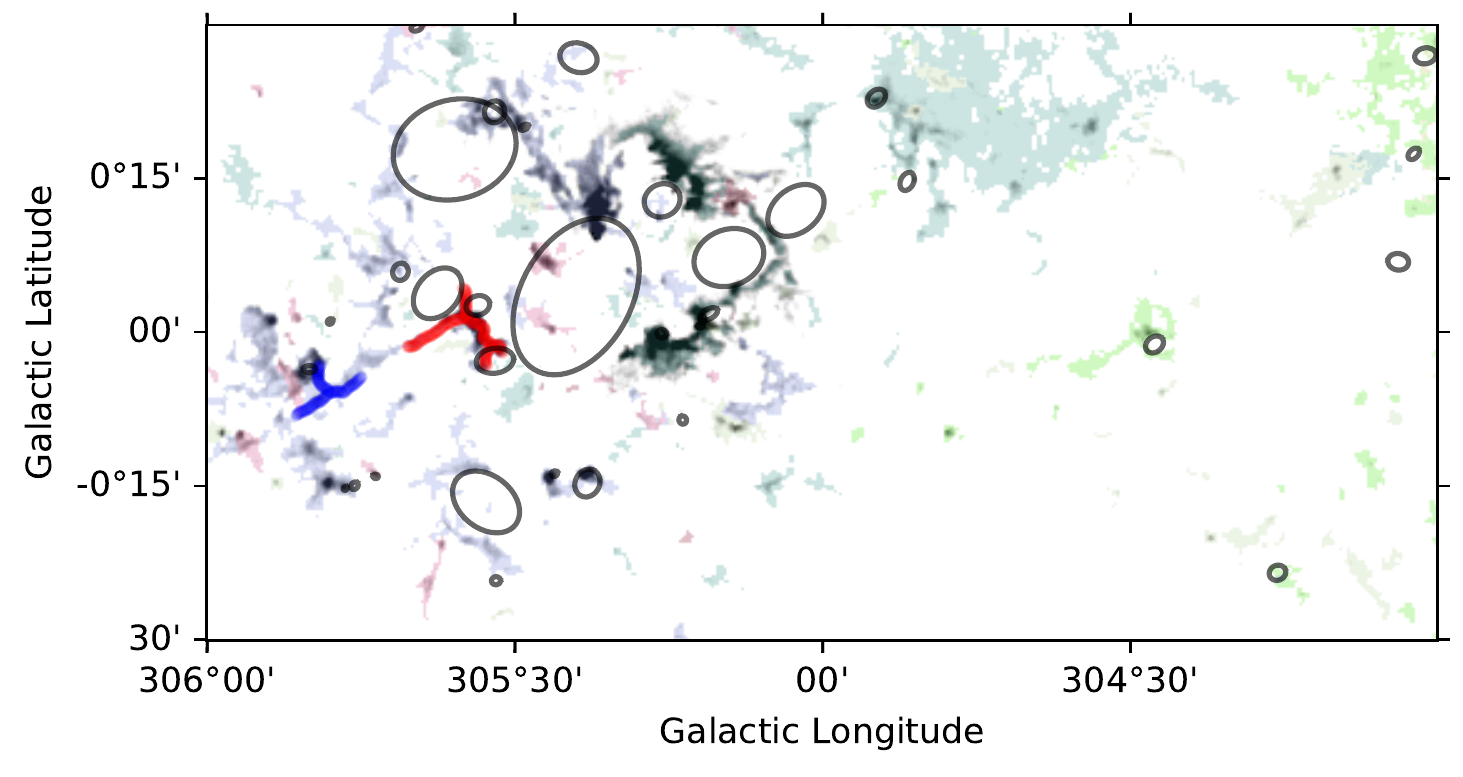}
    \caption{Elongated structures from ATLASGAL survey and bubbles from MWP survey overlaid on SEDIGISM clouds for $304 \degree \leq l \leq 306 \degree$.}
    \label{fig: SED overlap 304}
    \end{minipage}\hfill
\end{figure*}

\begin{figure*}[]
    \centering
    \begin{minipage}{\textwidth}
    \includegraphics[width = \textwidth, keepaspectratio]{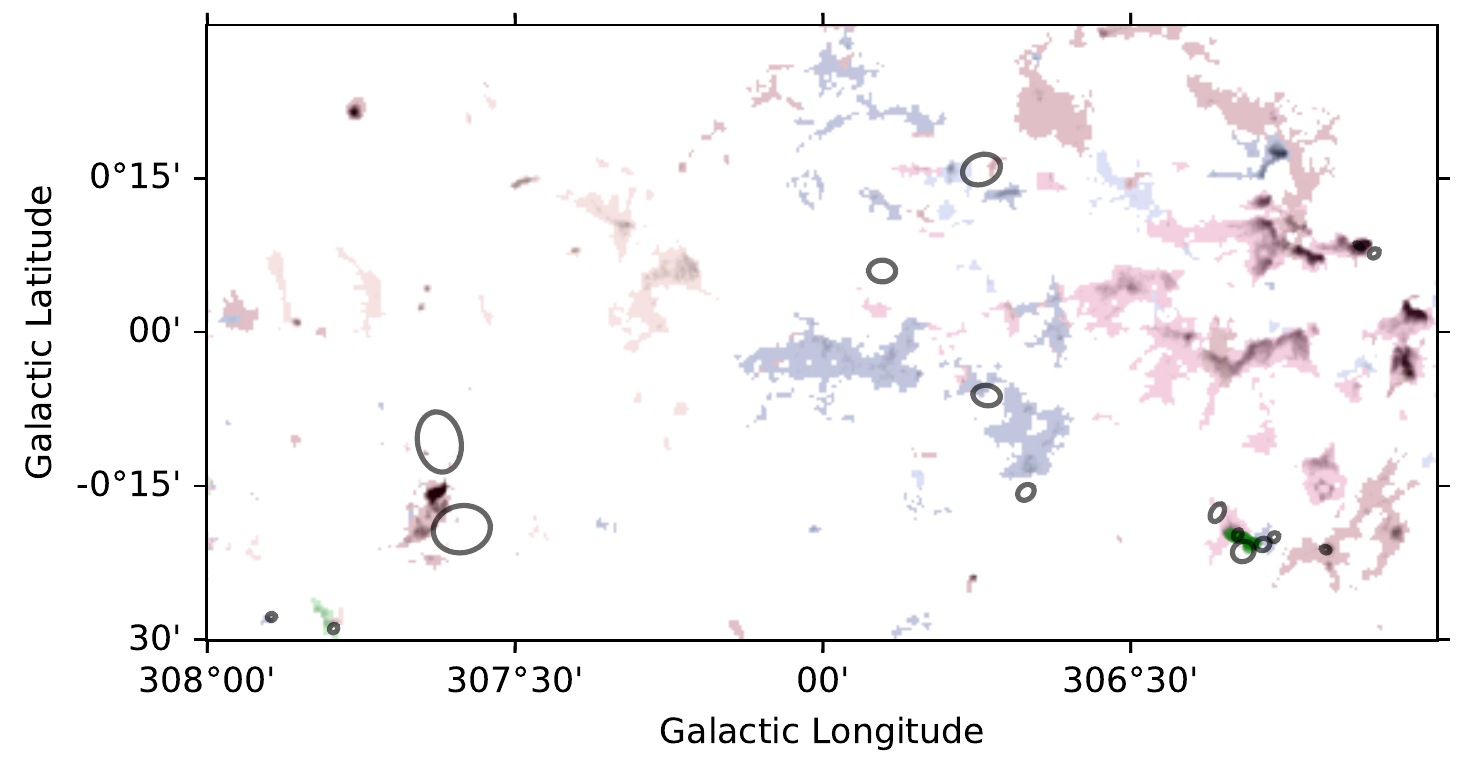}
    \caption{Elongated structures from ATLASGAL survey and bubbles from MWP survey overlaid on SEDIGISM clouds for $306 \degree \leq l \leq 308 \degree$.}
    \label{fig: SED overlap 306}
    \end{minipage}\hfill
\end{figure*}

\begin{figure*}[]
    \centering
    \begin{minipage}{\textwidth}
    \includegraphics[width = \textwidth, keepaspectratio]{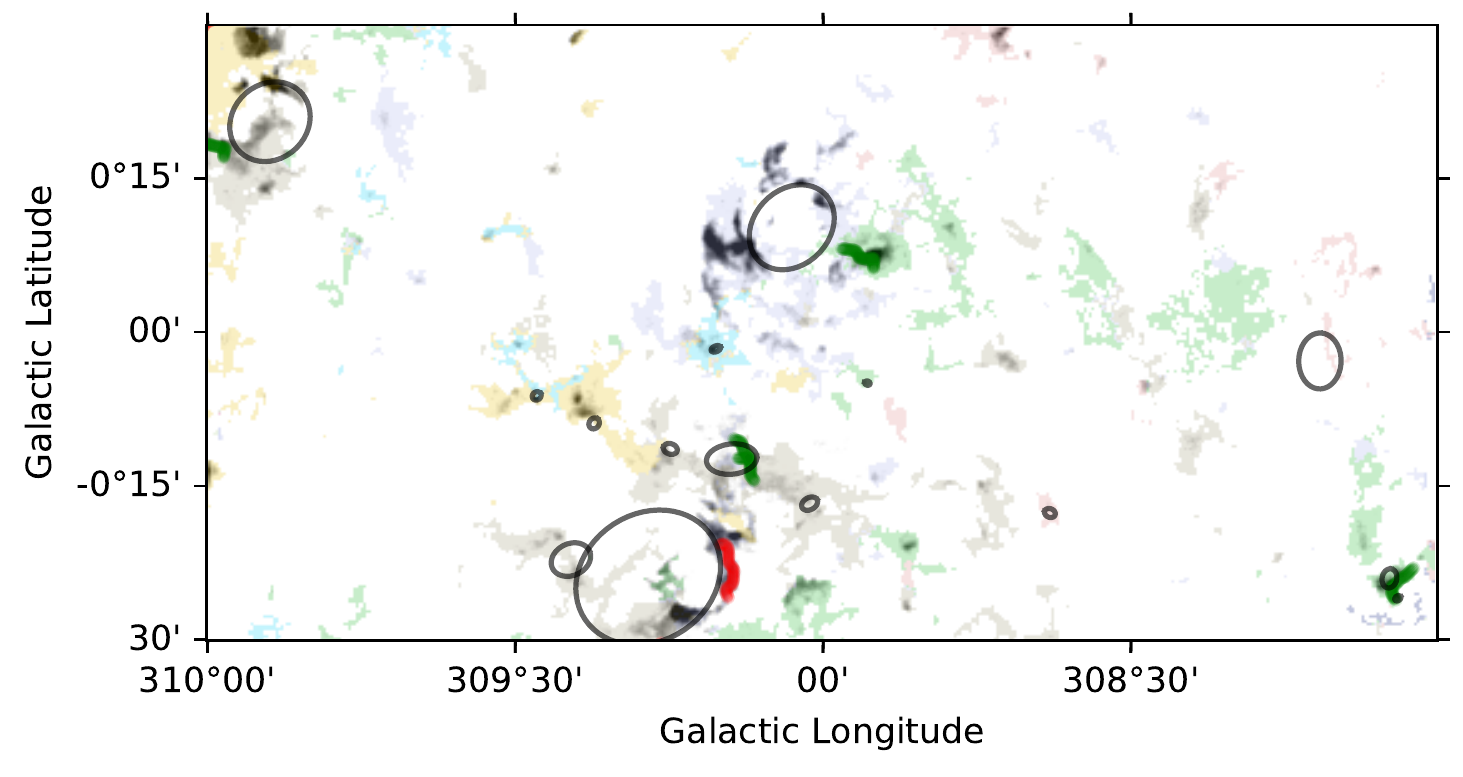}
    \caption{Elongated structures from ATLASGAL survey and bubbles from MWP survey overlaid on SEDIGISM clouds for $308 \degree \leq l \leq 310 \degree$.}
    \label{fig: SED overlap 308}
    \end{minipage}\hfill
\end{figure*}

\begin{figure*}[]
    \centering
    \begin{minipage}{\textwidth}
    \includegraphics[width = \textwidth, keepaspectratio]{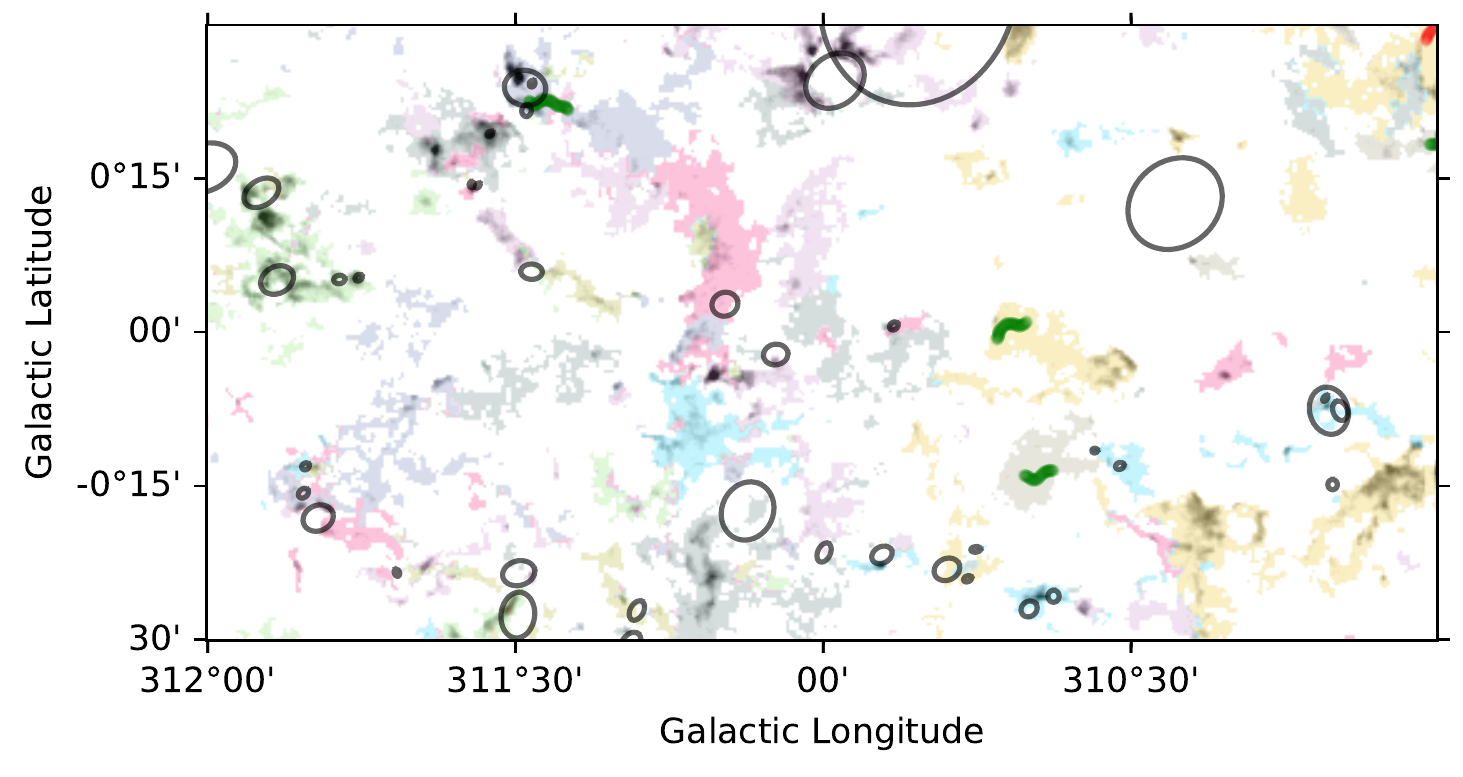}
    \caption{Elongated structures from ATLASGAL survey and bubbles from MWP survey overlaid on SEDIGISM clouds for $310 \degree \leq l \leq 312 \degree$.}
    \label{fig: SED overlap 310}
    \end{minipage}\hfill
\end{figure*}

\begin{figure*}[]
    \centering
    \begin{minipage}{\textwidth}
    \includegraphics[width = \textwidth, keepaspectratio]{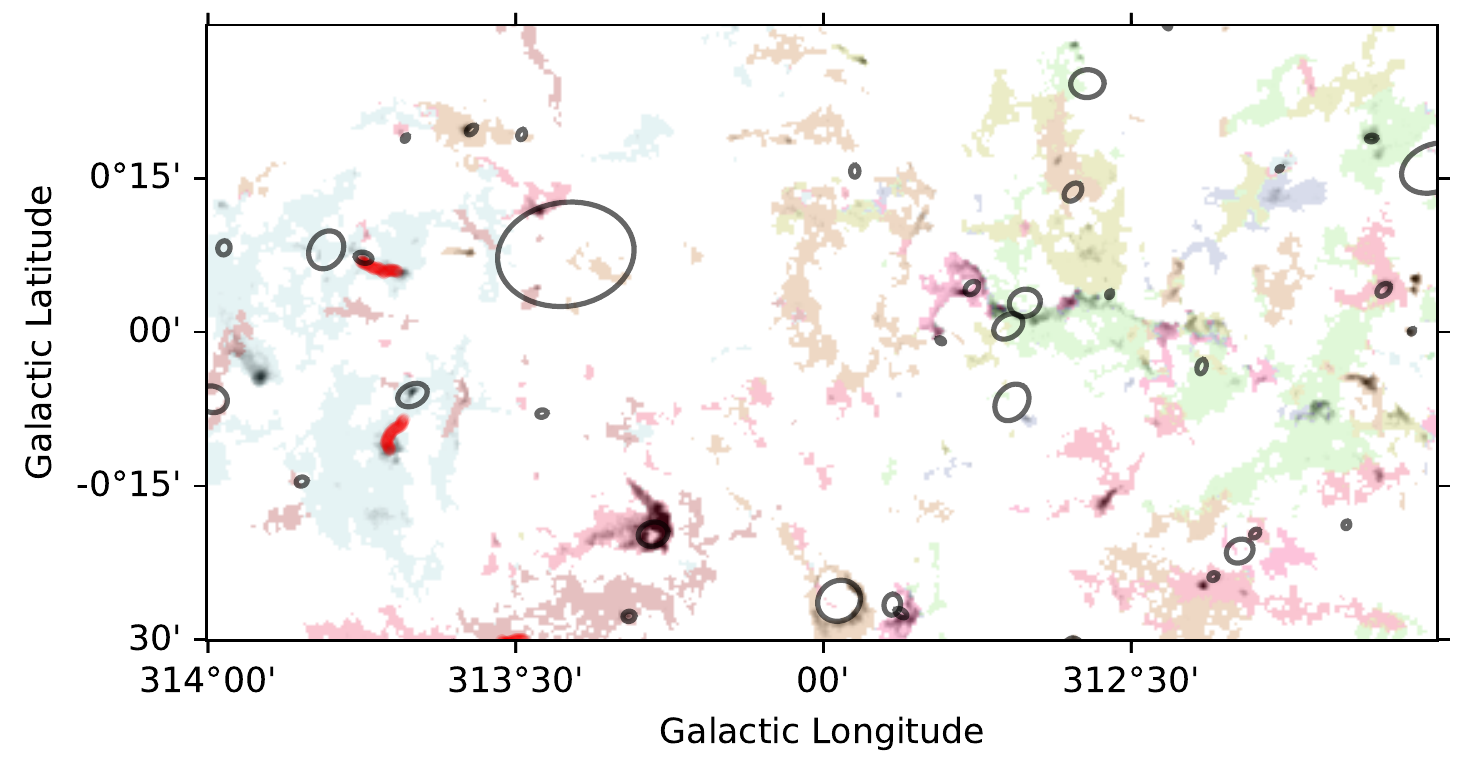}
    \caption{Elongated structures from ATLASGAL survey and bubbles from MWP survey overlaid on SEDIGISM clouds for $312 \degree \leq l \leq 314 \degree$.}
    \label{fig: SED overlap 312}
    \end{minipage}\hfill
\end{figure*}

\begin{figure*}[]
    \centering
    \begin{minipage}{\textwidth}
    \includegraphics[width = \textwidth, keepaspectratio]{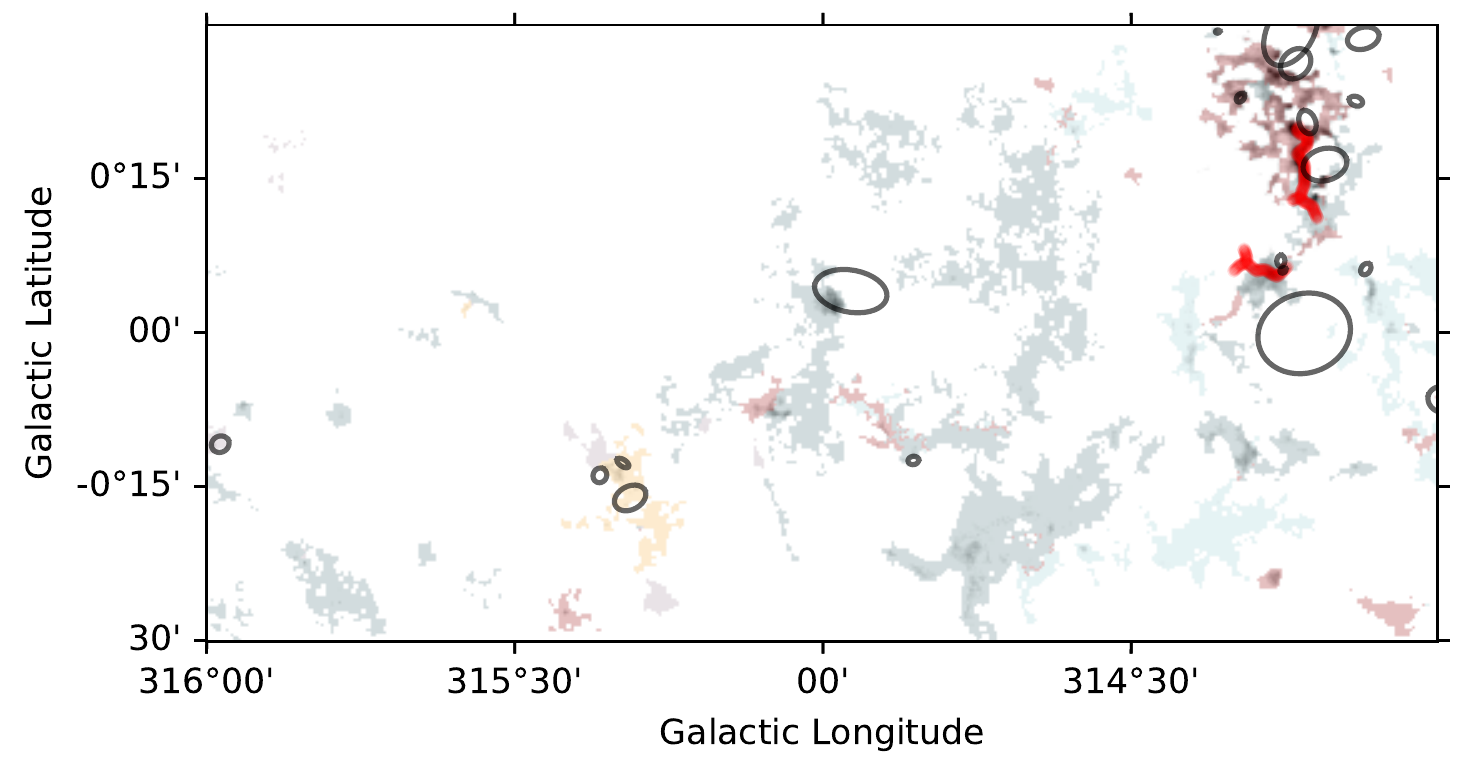}
    \caption{Elongated structures from ATLASGAL survey and bubbles from MWP survey overlaid on SEDIGISM clouds for $314 \degree \leq l \leq 316 \degree$.}
    \label{fig: SED overlap 314}
    \end{minipage}\hfill
\end{figure*}

\begin{figure*}[]
    \centering
    \begin{minipage}{\textwidth}
    \includegraphics[width = \textwidth, keepaspectratio]{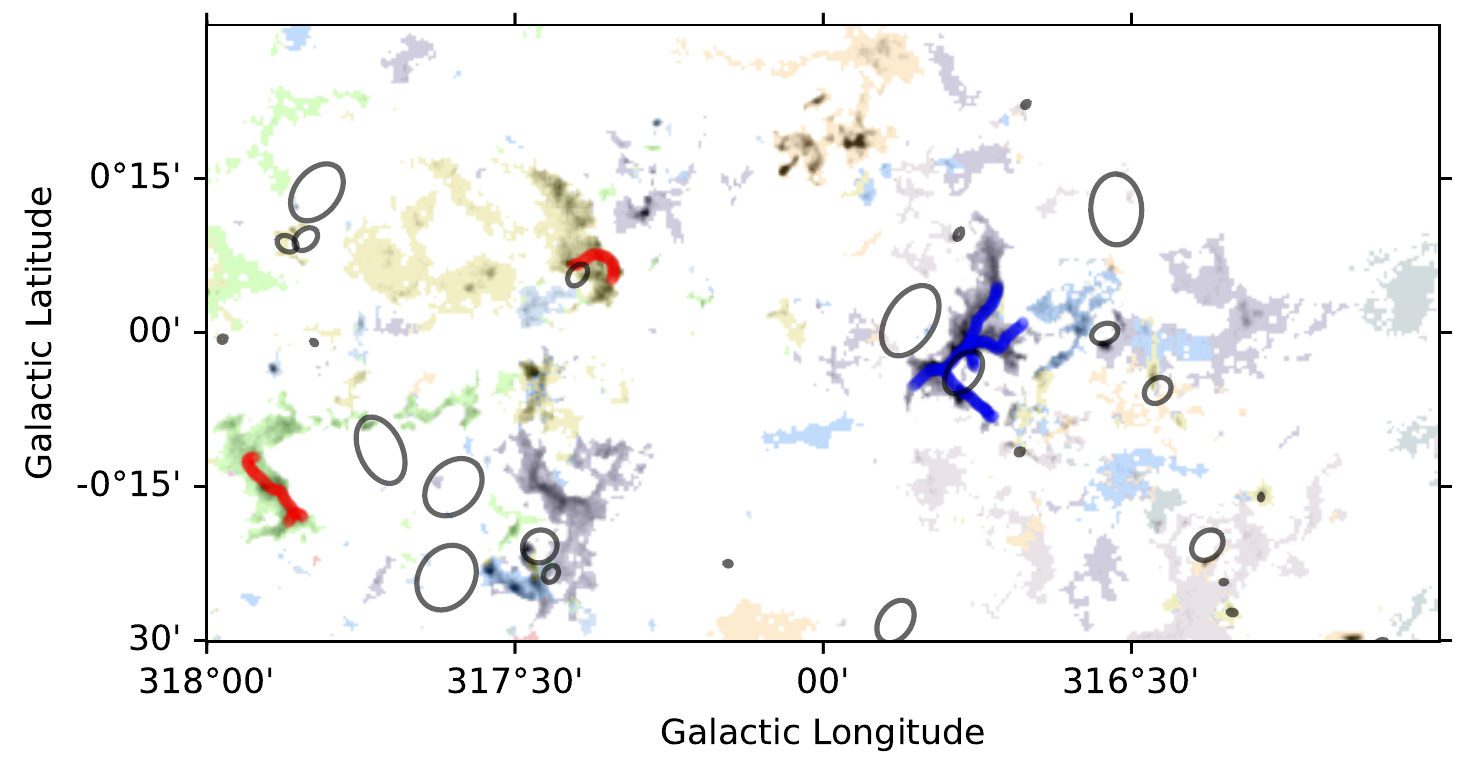}
    \caption{Elongated structures from ATLASGAL survey and bubbles from MWP survey overlaid on SEDIGISM clouds for $316 \degree \leq l \leq 318 \degree$.}
    \label{fig: SED overlap 316}
    \end{minipage}\hfill
\end{figure*}

\begin{figure*}[]
    \centering
    \begin{minipage}{\textwidth}
    \includegraphics[width = \textwidth, keepaspectratio]{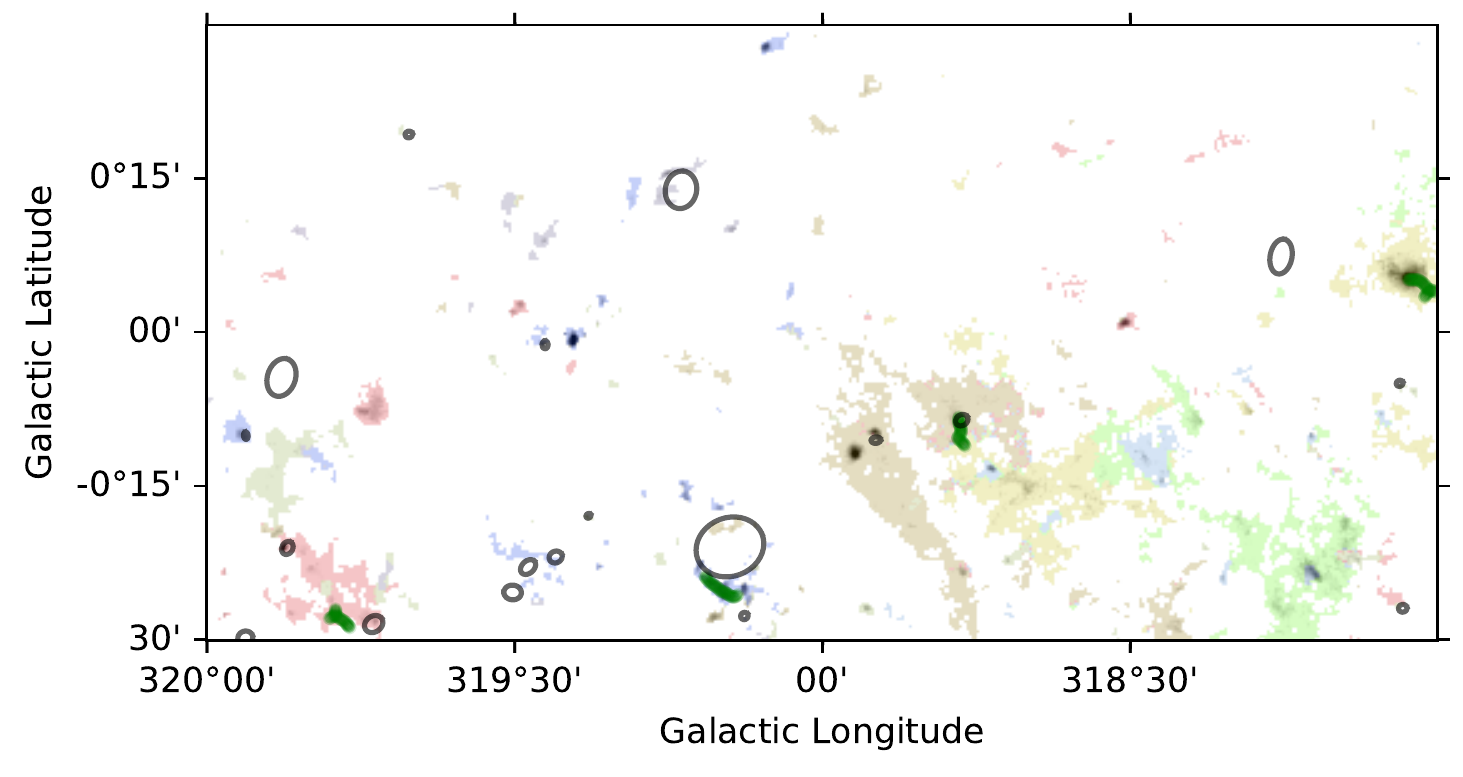}
    \caption{Elongated structures from ATLASGAL survey and bubbles from MWP survey overlaid on SEDIGISM clouds for $318 \degree \leq l \leq 320 \degree$.}
    \label{fig: SED overlap 318}
    \end{minipage}\hfill
\end{figure*}

\begin{figure*}[]
    \centering
    \begin{minipage}{\textwidth}
    \includegraphics[width = \textwidth, keepaspectratio]{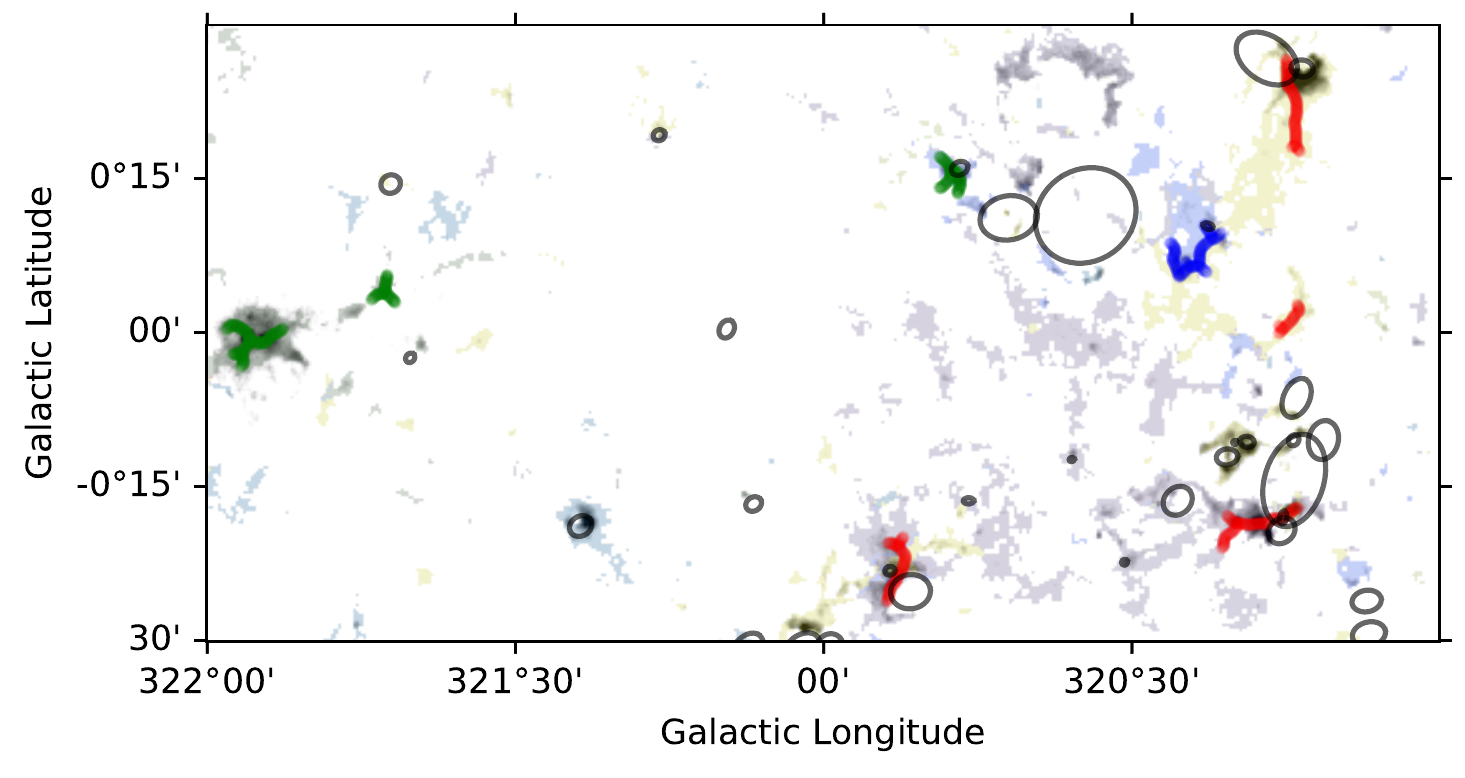}
    \caption{Elongated structures from ATLASGAL survey and bubbles from MWP survey overlaid on SEDIGISM clouds for $320 \degree \leq l \leq 322 \degree$.}
    \label{fig: SED overlap 320}
    \end{minipage}\hfill
\end{figure*}

\begin{figure*}[]
    \centering
    \begin{minipage}{\textwidth}
    \includegraphics[width = \textwidth, keepaspectratio]{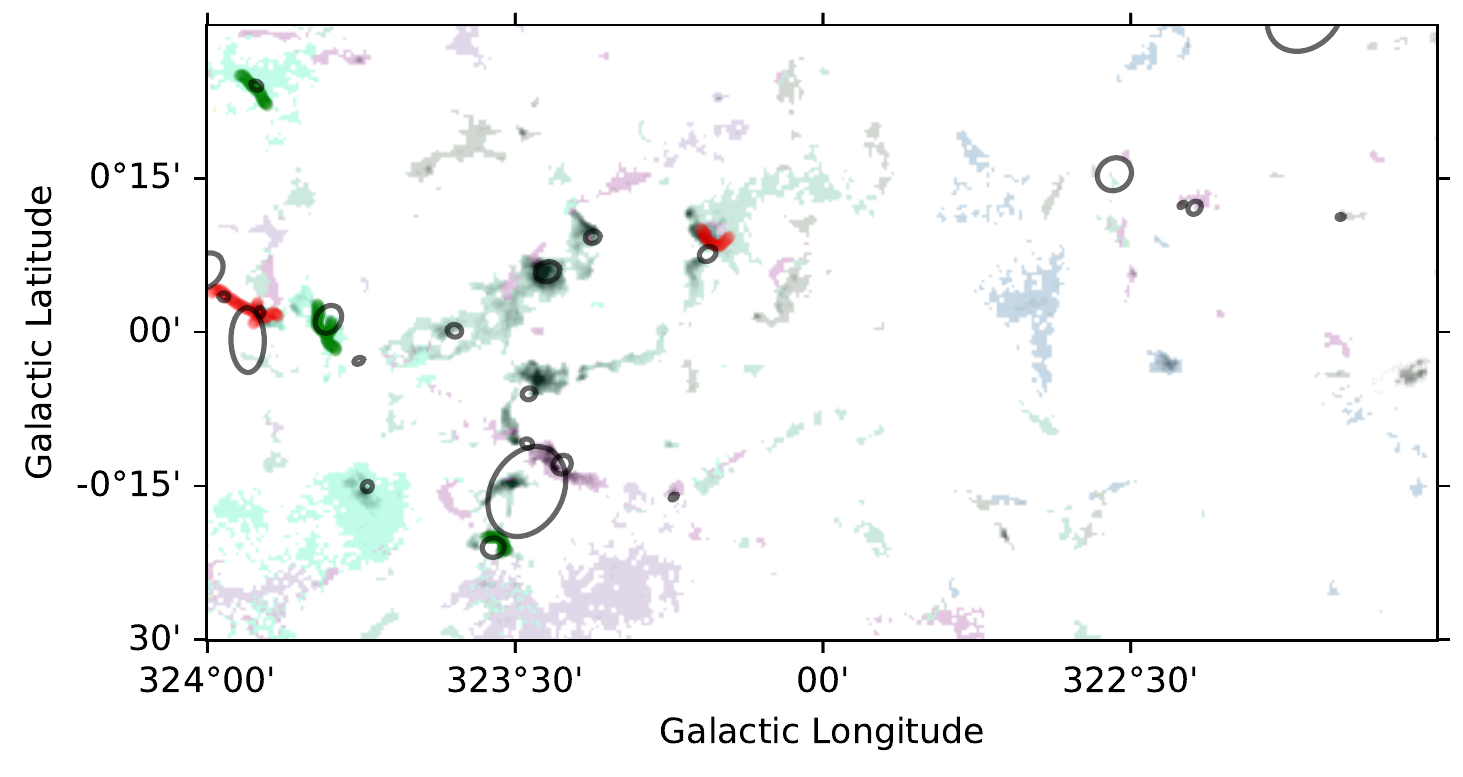}
    \caption{Elongated structures from ATLASGAL survey and bubbles from MWP survey overlaid on SEDIGISM clouds for $322 \degree \leq l \leq 324 \degree$.}
    \label{fig: SED overlap 322}
    \end{minipage}\hfill
\end{figure*}

\begin{figure*}[]
    \centering
    \begin{minipage}{\textwidth}
    \includegraphics[width = \textwidth, keepaspectratio]{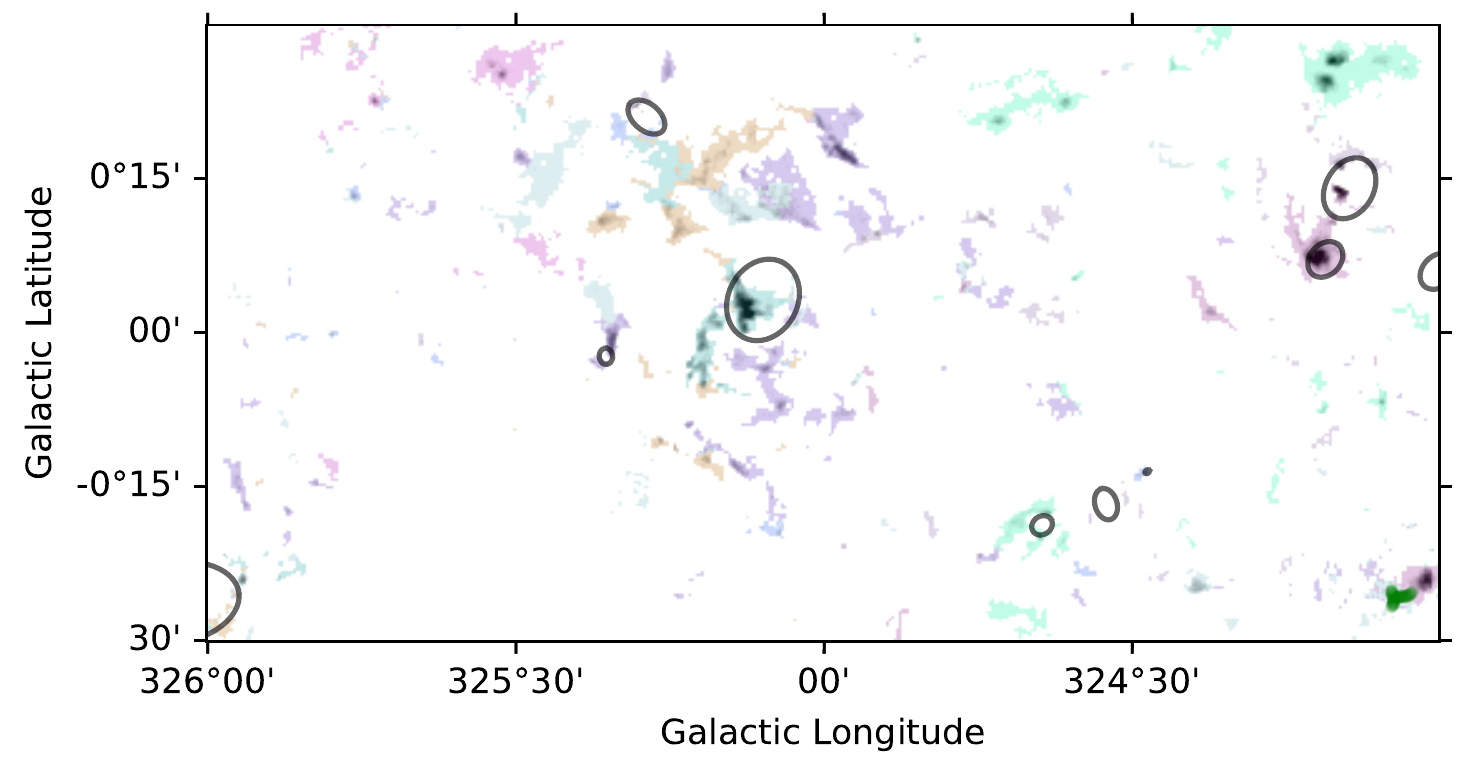}
    \caption{Elongated structures from ATLASGAL survey and bubbles from MWP survey overlaid on SEDIGISM clouds for $324 \degree \leq l \leq 326 \degree$.}
    \label{fig: SED overlap 324}
    \end{minipage}\hfill
\end{figure*}

\begin{figure*}[]
    \centering
    \begin{minipage}{\textwidth}
    \includegraphics[width = \textwidth, keepaspectratio]{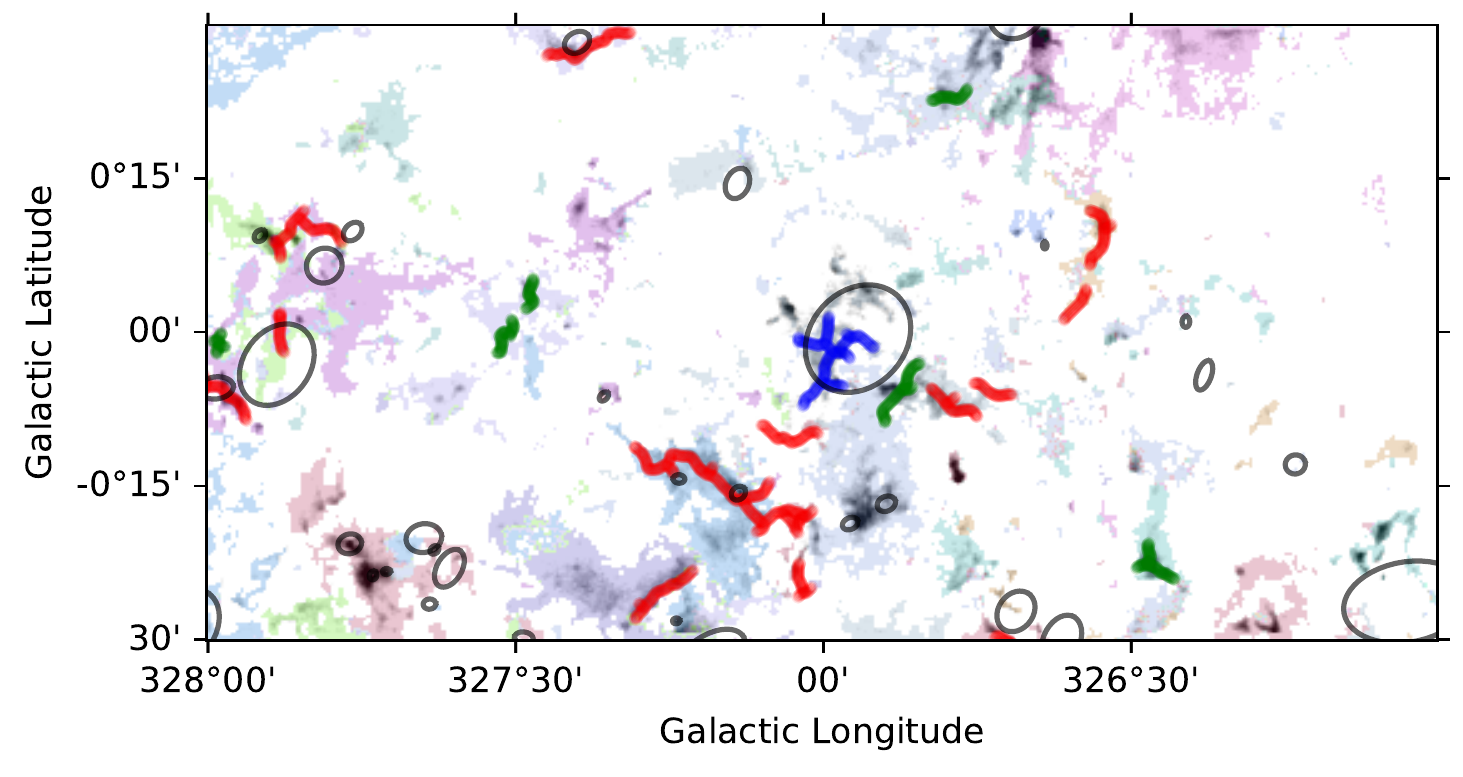}
    \caption{Elongated structures from ATLASGAL survey and bubbles from MWP survey overlaid on SEDIGISM clouds for $326 \degree \leq l \leq 328 \degree$.}
    \label{fig: SED overlap 326}
    \end{minipage}\hfill
\end{figure*}

\begin{figure*}[]
    \centering
    \begin{minipage}{\textwidth}
    \includegraphics[width = \textwidth, keepaspectratio]{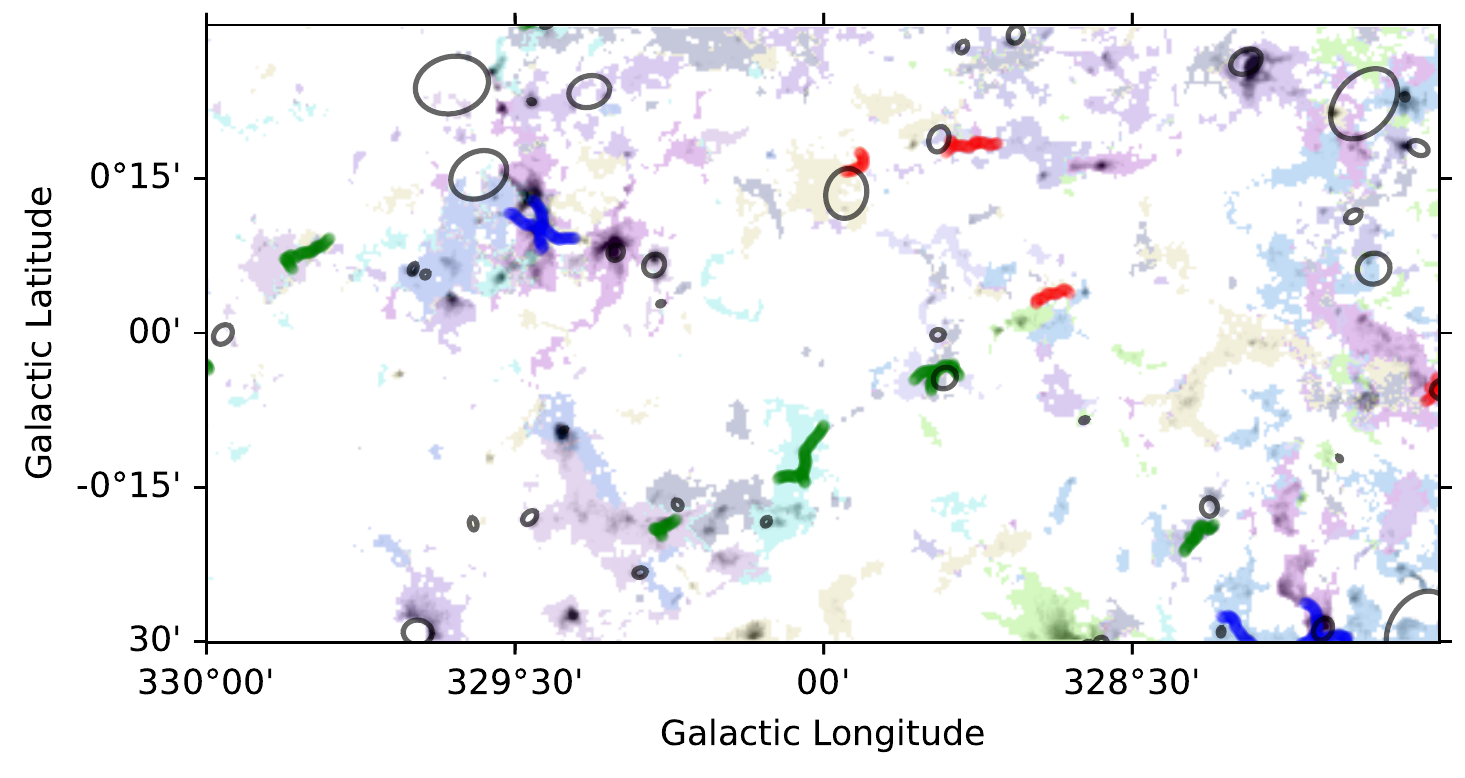}
    \caption{Elongated structures from ATLASGAL survey and bubbles from MWP survey overlaid on SEDIGISM clouds for $328 \degree \leq l \leq 330 \degree$.}
    \label{fig: SED overlap 328}
    \end{minipage}\hfill
\end{figure*}

\begin{figure*}[]
    \centering
    \begin{minipage}{\textwidth}
    \includegraphics[width = \textwidth, keepaspectratio]{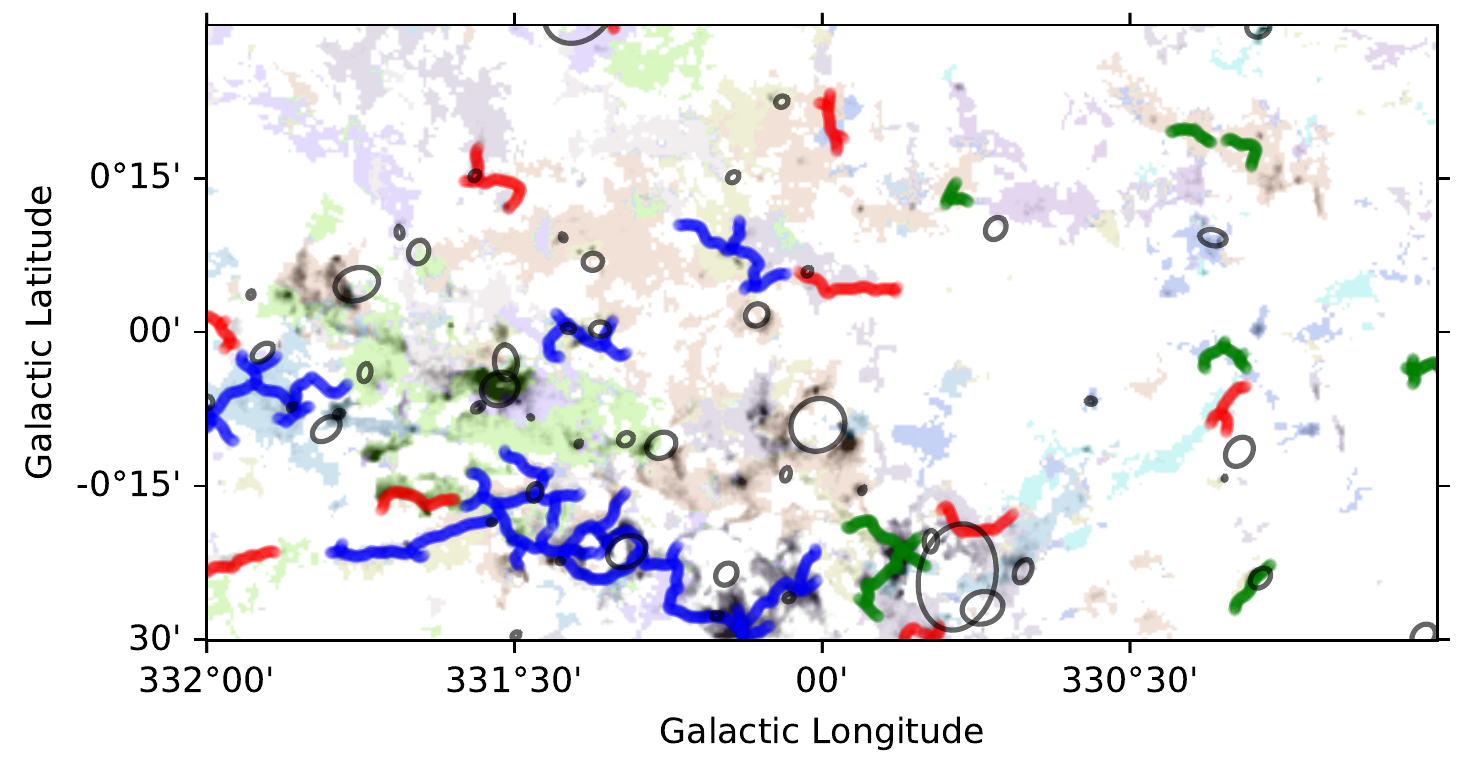}
    \caption{Elongated structures from ATLASGAL survey and bubbles from MWP survey overlaid on SEDIGISM clouds for $330 \degree \leq l \leq 332 \degree$.}
    \label{fig: SED overlap 330}
    \end{minipage}\hfill
\end{figure*}

\begin{figure*}[]
    \centering
    \begin{minipage}{\textwidth}
    \includegraphics[width = \textwidth, keepaspectratio]{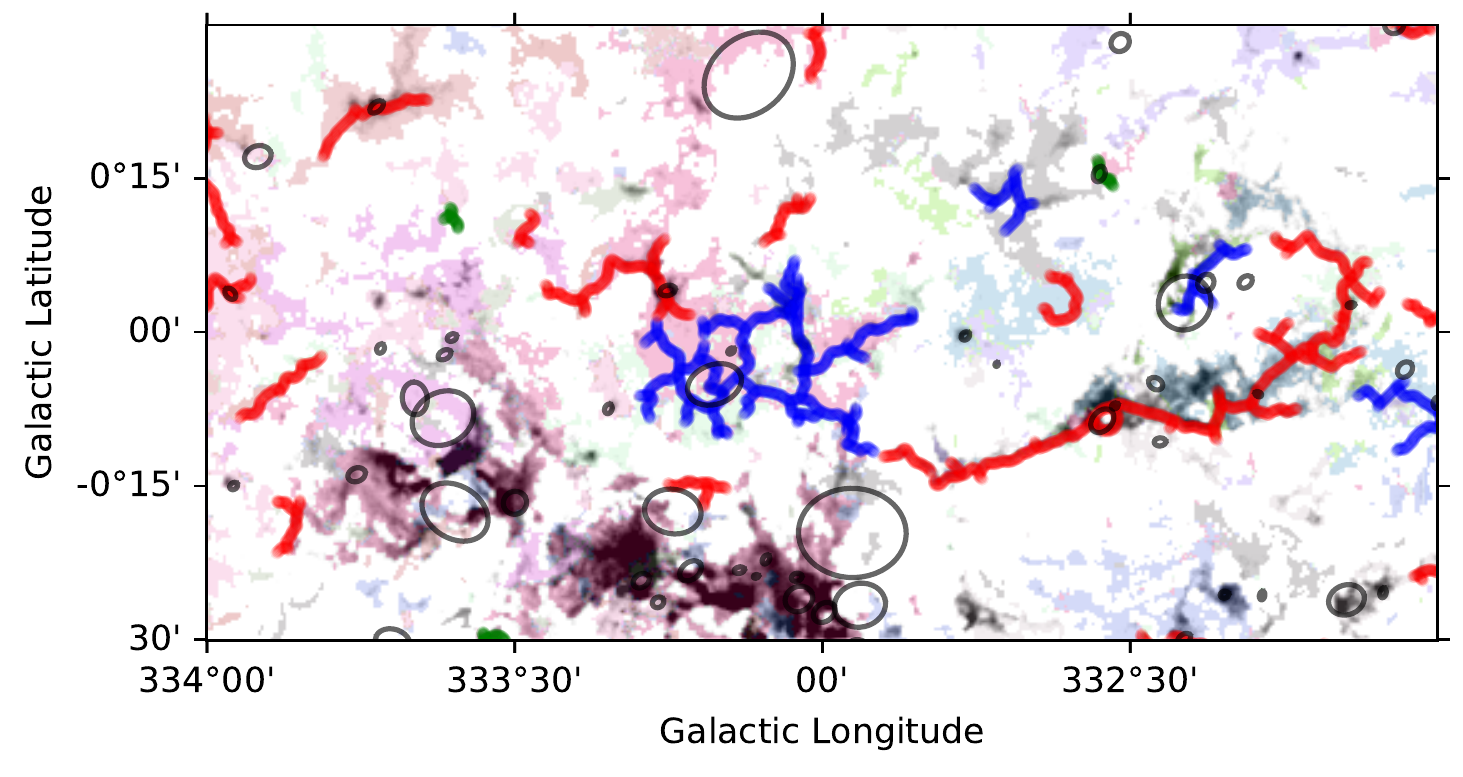}
    \caption{Elongated structures from ATLASGAL survey and bubbles from MWP survey overlaid on SEDIGISM clouds for $332 \degree \leq l \leq 334 \degree$.}
    \label{fig: SED overlap 332}
    \end{minipage}\hfill
\end{figure*}

\begin{figure*}[]
    \centering
    \begin{minipage}{\textwidth}
    \includegraphics[width = \textwidth, keepaspectratio]{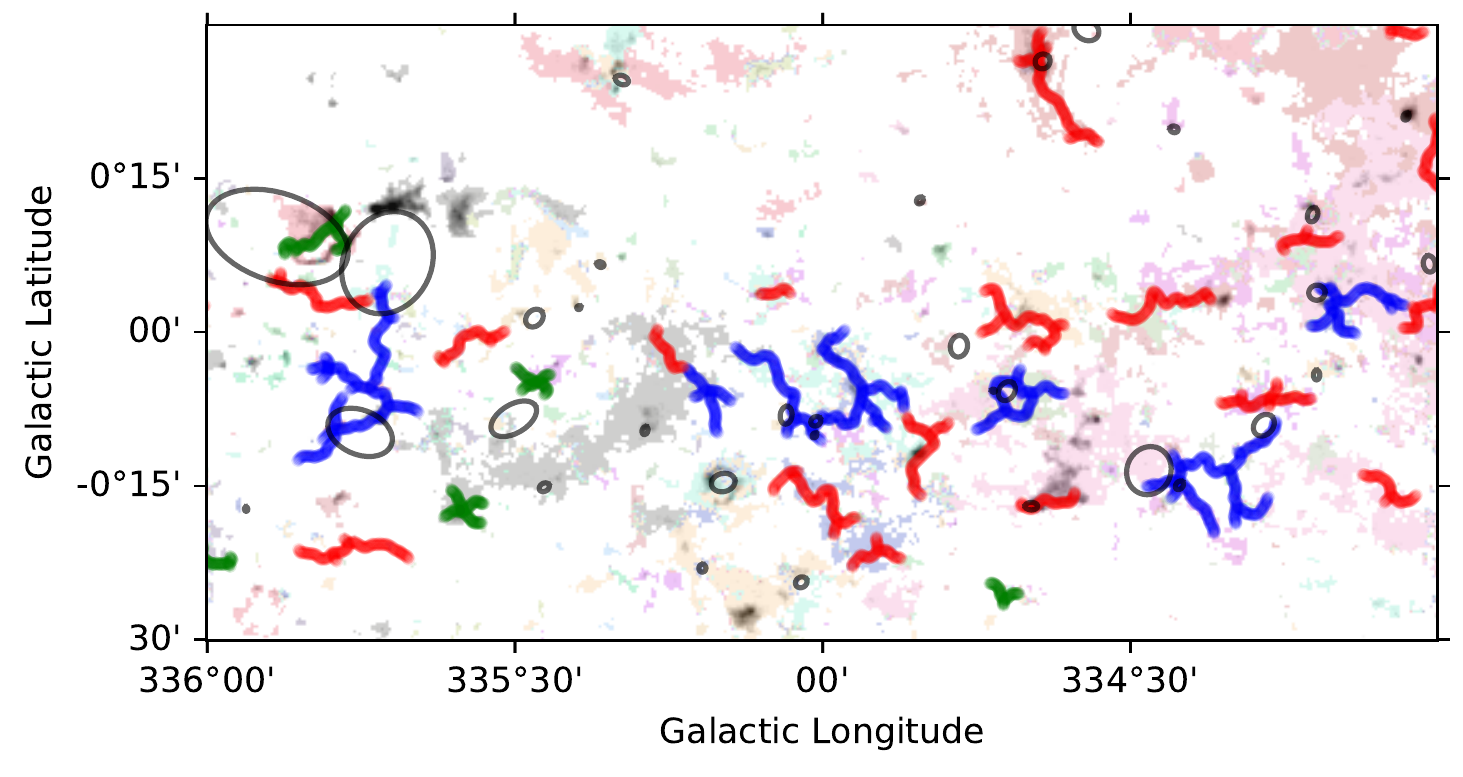}
    \caption{Elongated structures from ATLASGAL survey and bubbles from MWP survey overlaid on SEDIGISM clouds for $334 \degree \leq l \leq 336 \degree$.}
    \label{fig: SED overlap 334}
    \end{minipage}\hfill
\end{figure*}

\begin{figure*}[]
    \centering
    \begin{minipage}{\textwidth}
    \includegraphics[width = \textwidth, keepaspectratio]{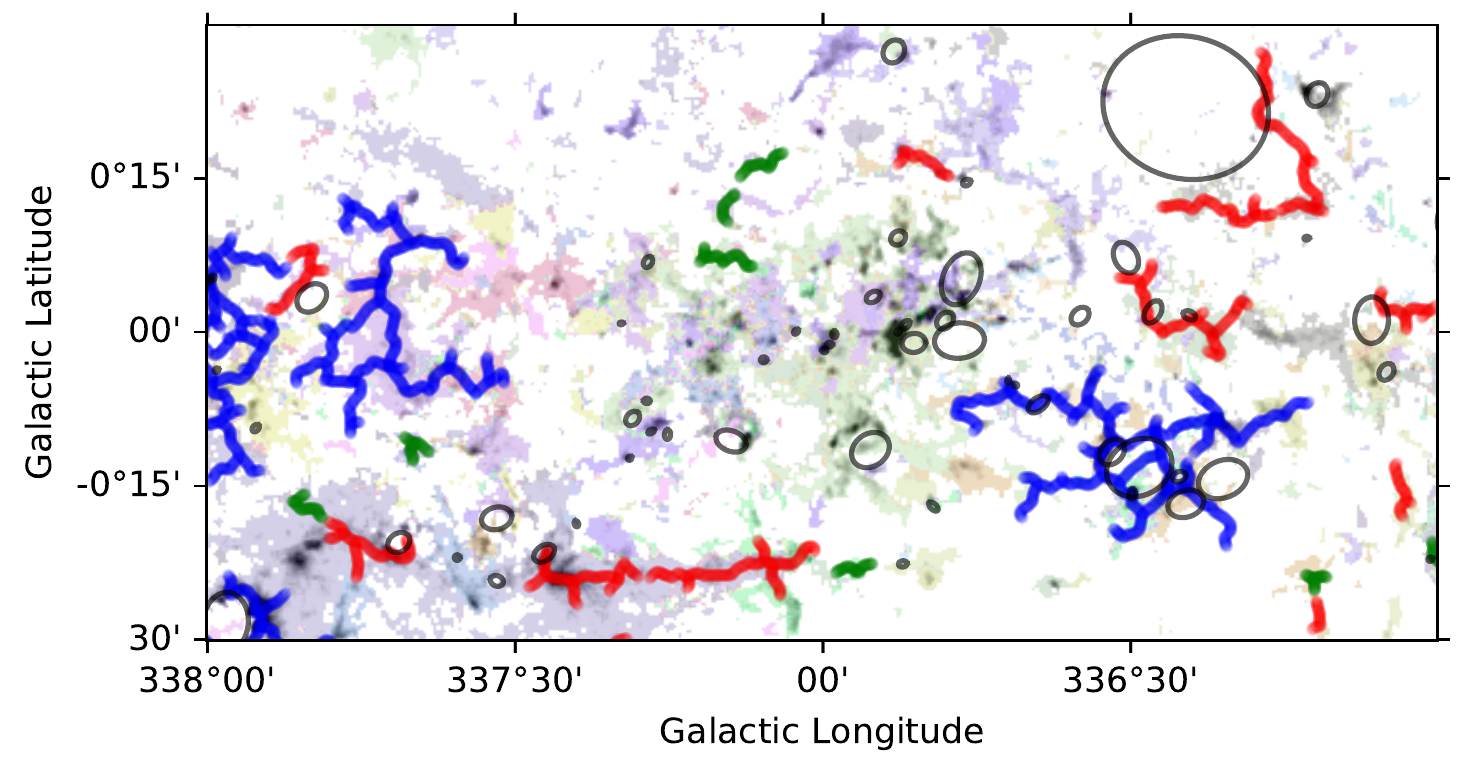}
    \caption{Elongated structures from ATLASGAL survey and bubbles from MWP survey overlaid on SEDIGISM clouds for $336 \degree \leq l \leq 338 \degree$.}
    \label{fig: SED overlap 336}
    \end{minipage}\hfill
\end{figure*}

\begin{figure*}[]
    \centering
    \begin{minipage}{\textwidth}
    \includegraphics[width = \textwidth, keepaspectratio]{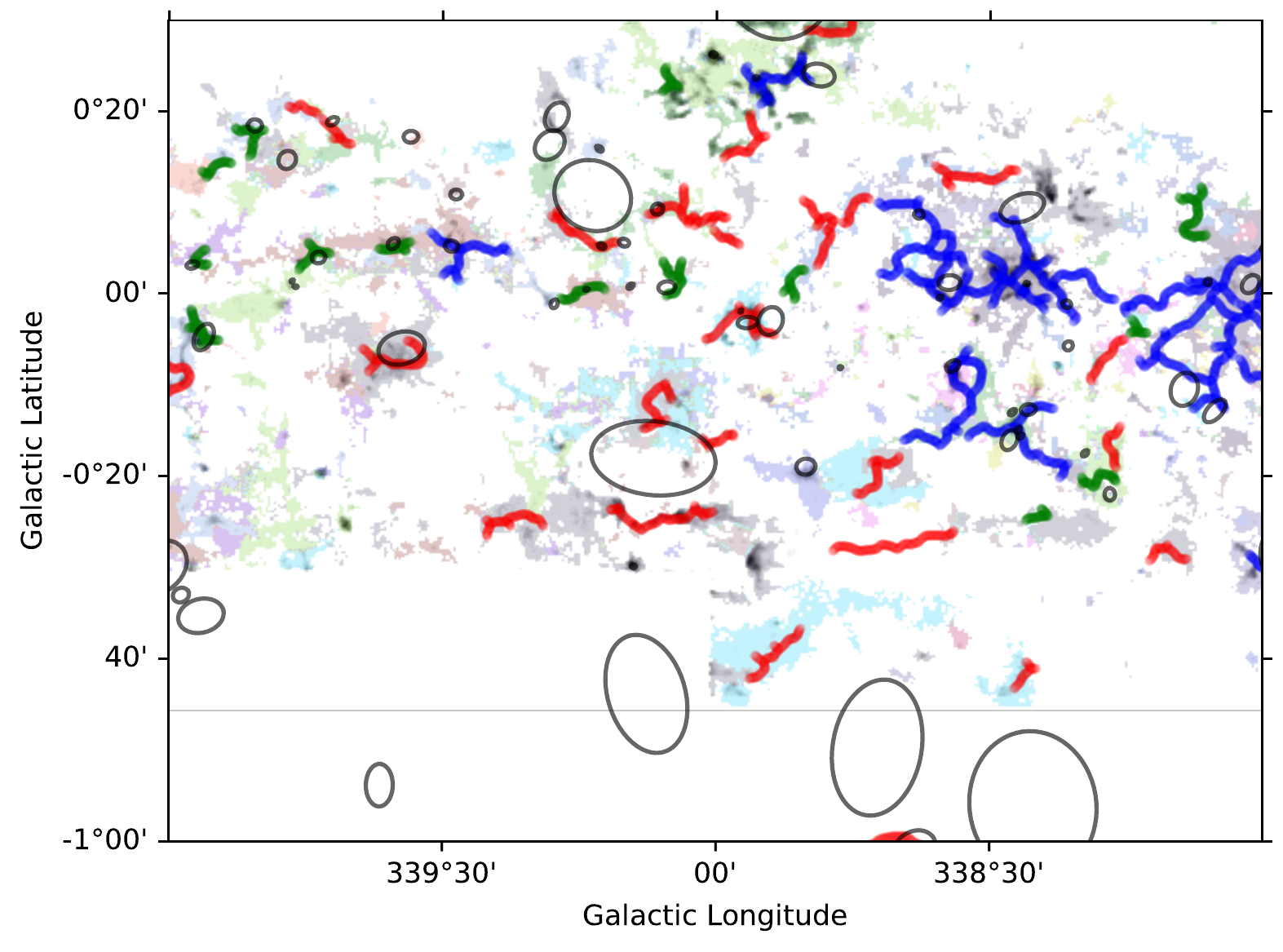}
    \caption{Elongated structures from ATLASGAL survey and bubbles from MWP survey overlaid on SEDIGISM clouds for $338 \degree \leq l \leq 340 \degree$.}
    \label{fig: SED overlap 338}
    \end{minipage}\hfill
\end{figure*}

\begin{figure*}[]
    \centering
    \begin{minipage}{\textwidth}
    \includegraphics[width = \textwidth, keepaspectratio]{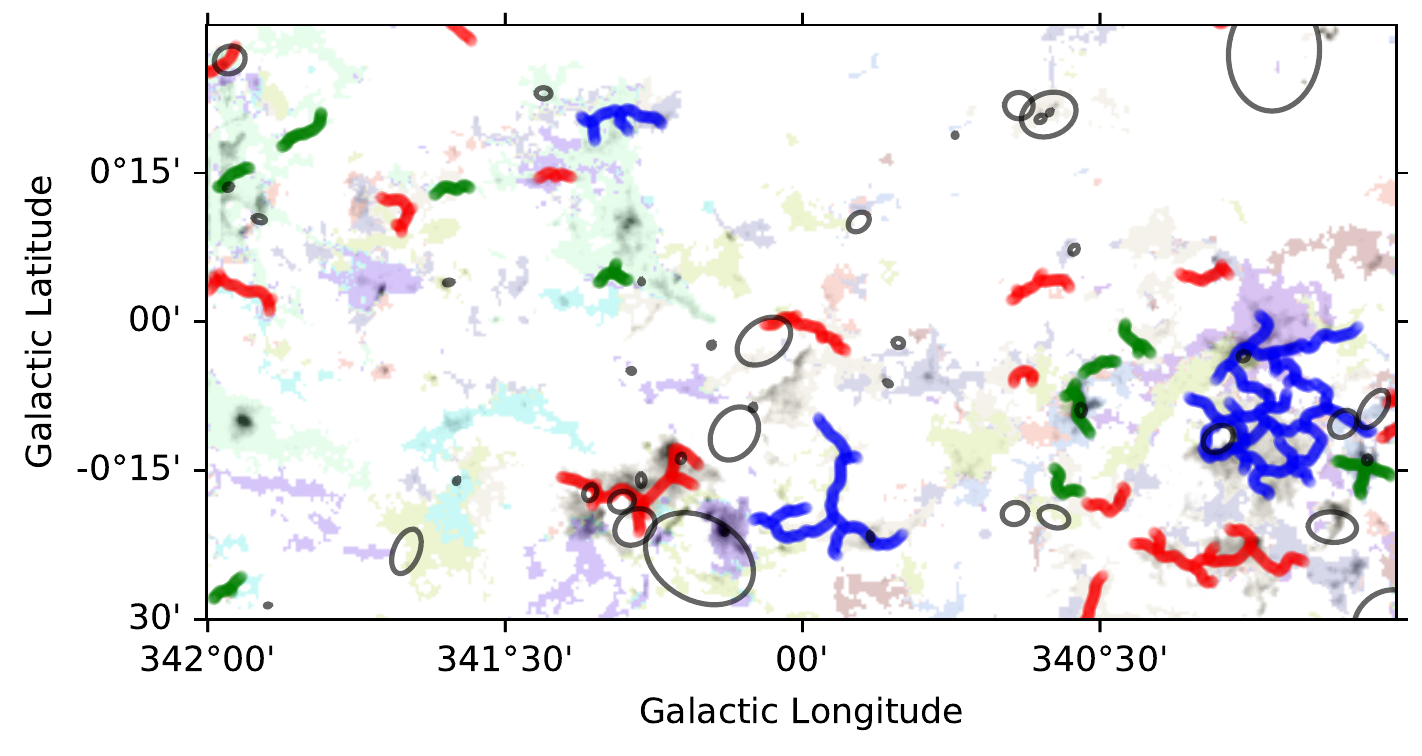}
    \caption{Elongated structures from ATLASGAL survey and bubbles from MWP survey overlaid on SEDIGISM clouds for $340 \degree \leq l \leq 342 \degree$.}
    \label{fig: SED overlap 340}
    \end{minipage}\hfill
\end{figure*}

\begin{figure*}[]
    \centering
    \begin{minipage}{\textwidth}
    \includegraphics[width = \textwidth, keepaspectratio]{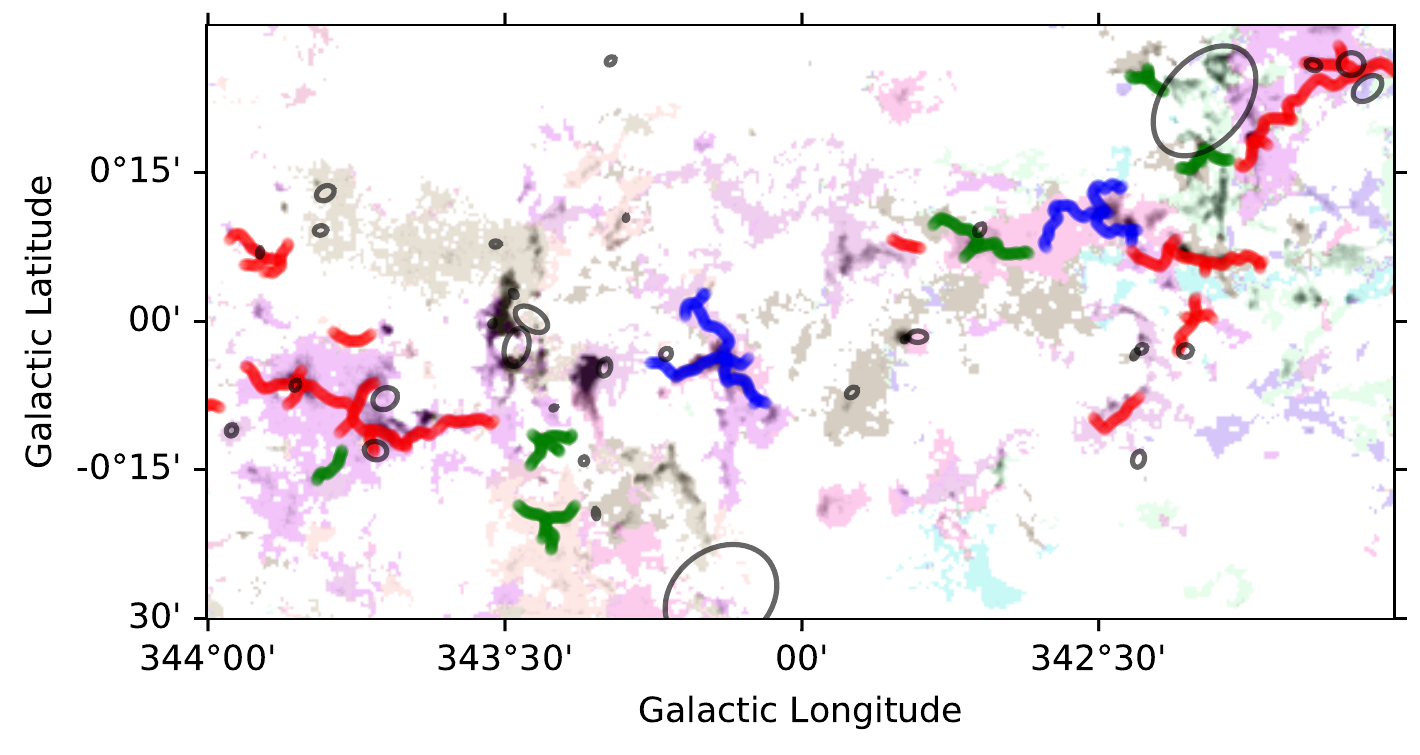}
    \caption{Elongated structures from ATLASGAL survey and bubbles from MWP survey overlaid on SEDIGISM clouds for $342 \degree \leq l \leq 344 \degree$.}
    \label{fig: SED overlap 342}
    \end{minipage}\hfill
\end{figure*}

\begin{figure*}[]
    \centering
    \begin{minipage}{\textwidth}
    \includegraphics[width = \textwidth, keepaspectratio]{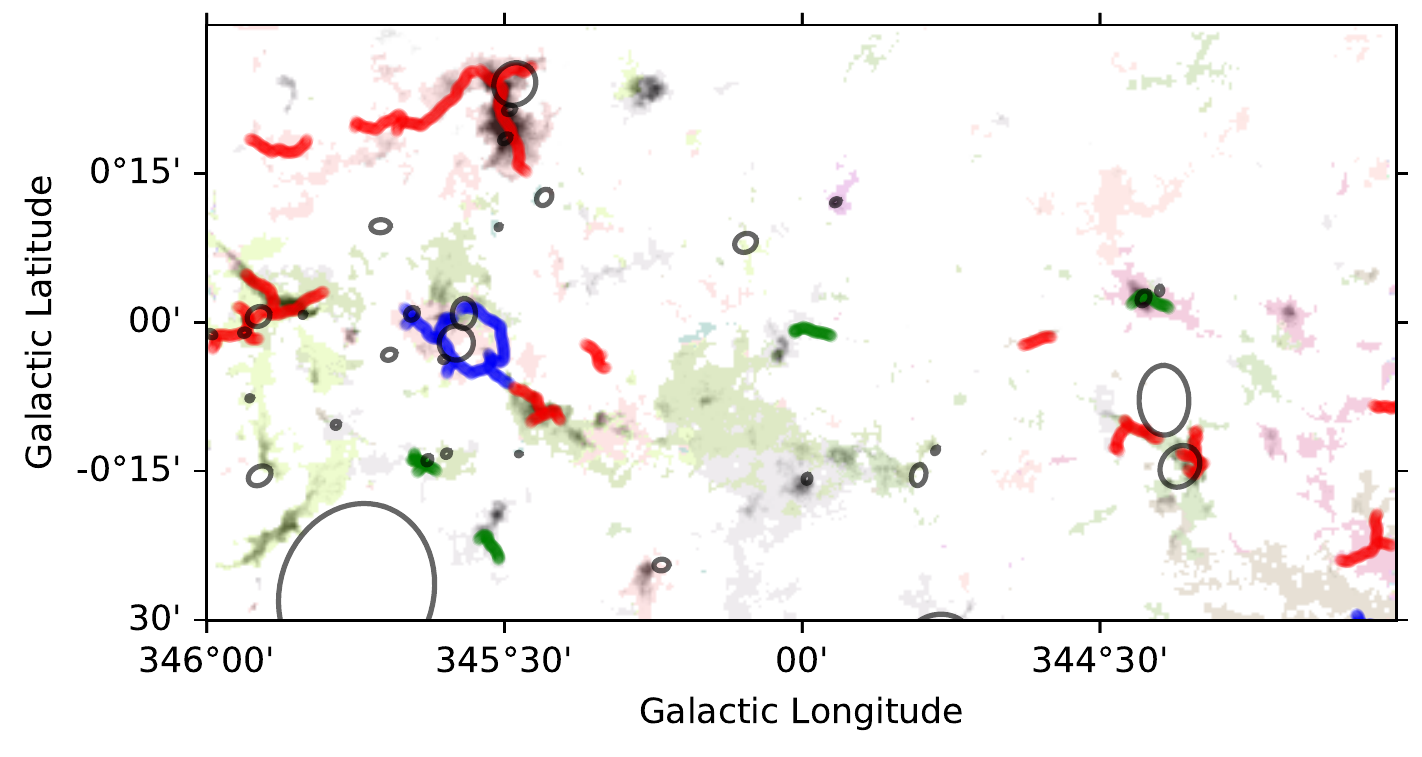}
    \caption{Elongated structures from ATLASGAL survey and bubbles from MWP survey overlaid on SEDIGISM clouds for $344 \degree \leq l \leq 346 \degree$.}
    \label{fig: SED overlap 344}
    \end{minipage}\hfill
\end{figure*}

\begin{figure*}[]
    \centering
    \begin{minipage}{\textwidth}
    \includegraphics[width = \textwidth, keepaspectratio]{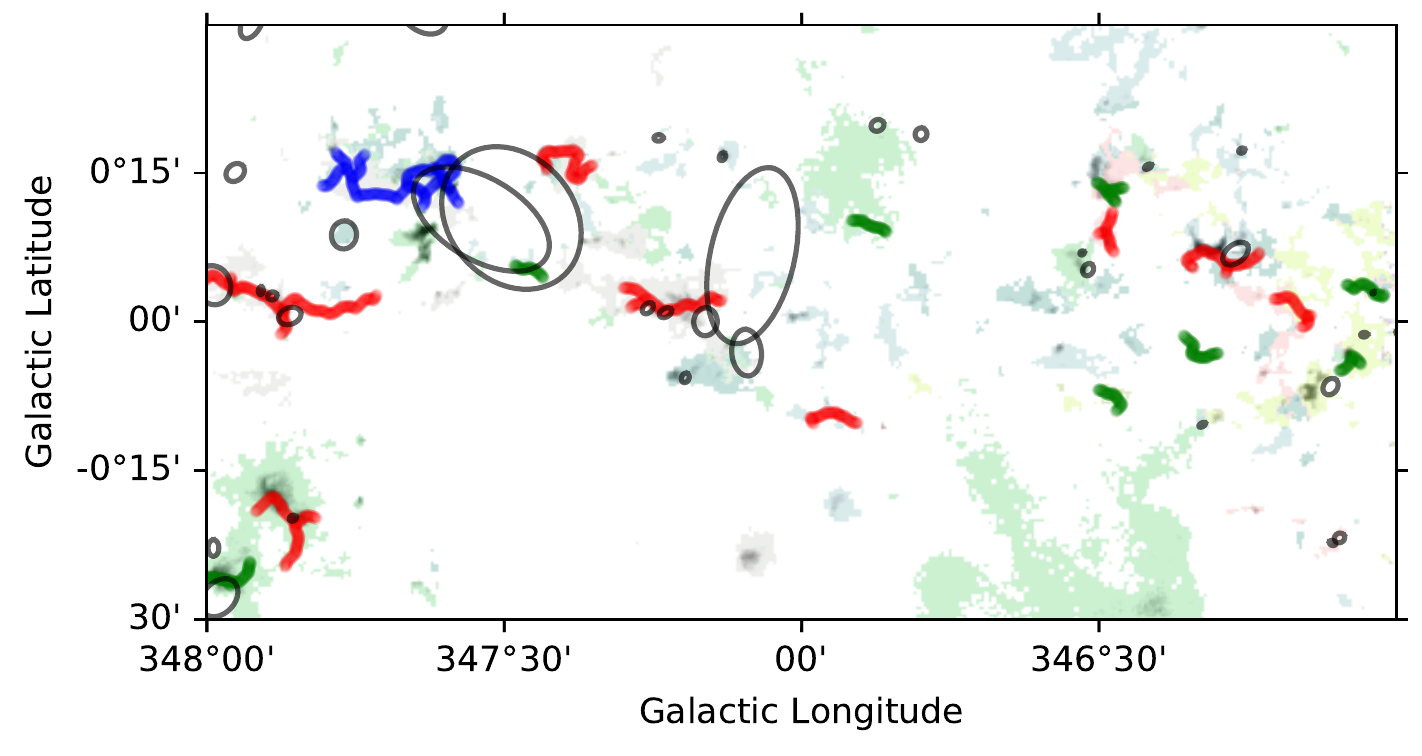}
    \caption{Elongated structures from ATLASGAL survey and bubbles from MWP survey overlaid on SEDIGISM clouds for $346 \degree \leq l \leq 348 \degree$.}
    \label{fig: SED overlap 346}
    \end{minipage}\hfill
\end{figure*}

\begin{figure*}[]
    \centering
    \begin{minipage}{\textwidth}
    \includegraphics[width = \textwidth, keepaspectratio]{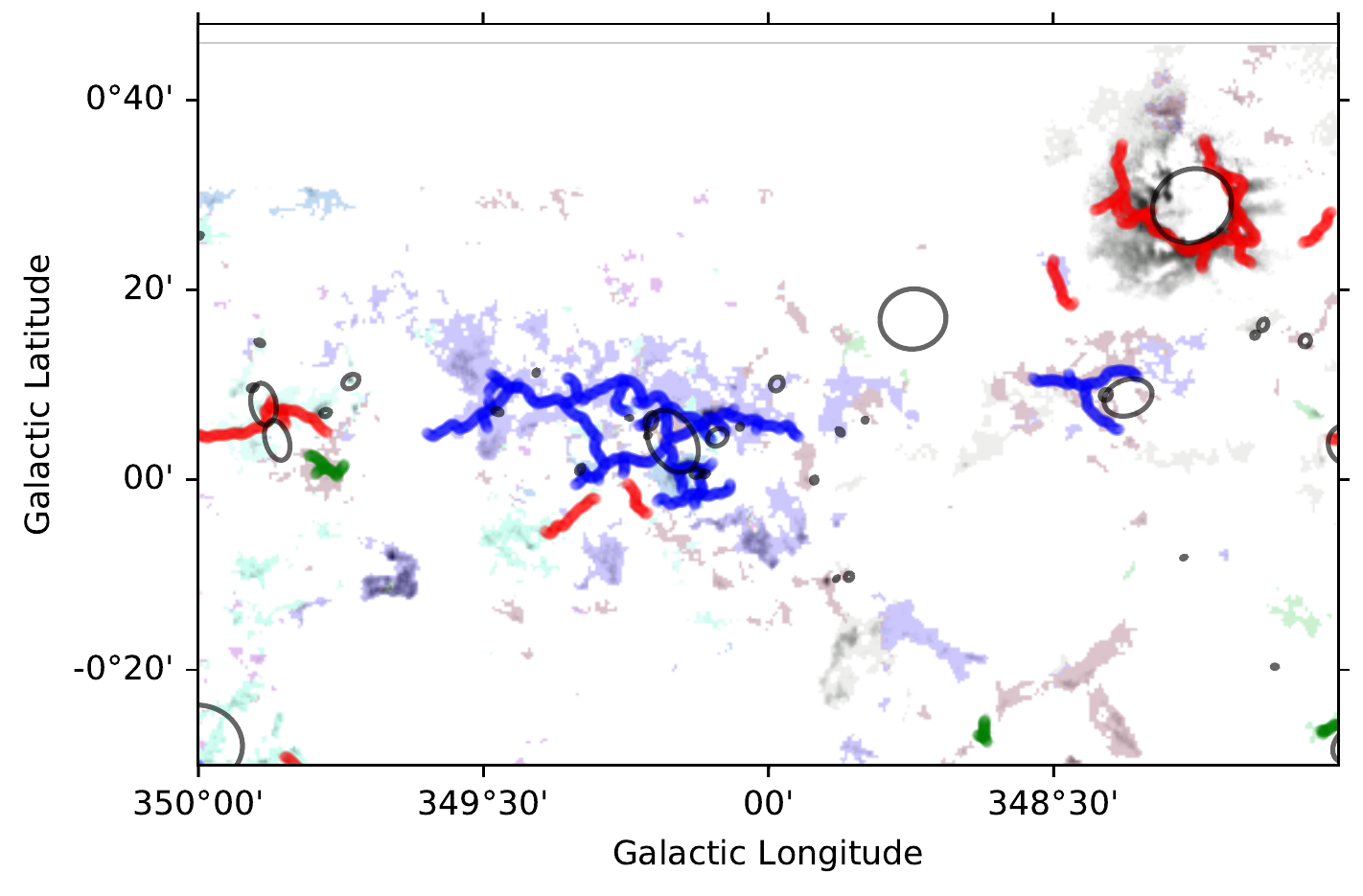}
    \caption{Elongated structures from ATLASGAL survey and bubbles from MWP survey overlaid on SEDIGISM clouds for $348 \degree \leq l \leq 350 \degree$.}
    \label{fig: SED overlap 348}
    \end{minipage}\hfill
\end{figure*}

\begin{figure*}[]
    \centering
    \begin{minipage}{\textwidth}
    \includegraphics[width = \textwidth, keepaspectratio]{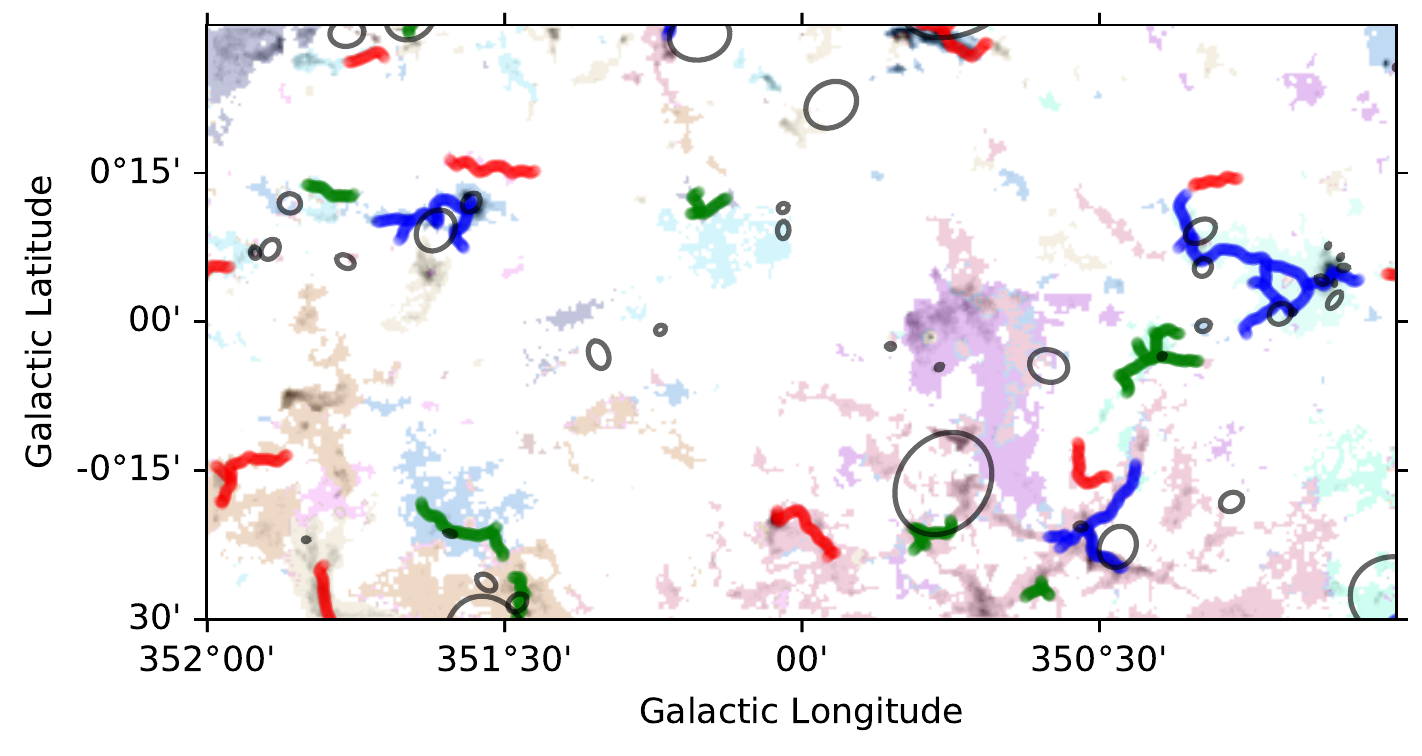}
    \caption{Elongated structures from ATLASGAL survey and bubbles from MWP survey overlaid on SEDIGISM clouds for $350 \degree \leq l \leq 352 \degree$.}
    \label{fig: SED overlap 350}
    \end{minipage}\hfill
\end{figure*}

\begin{figure*}[]
    \centering
    \begin{minipage}{\textwidth}
    \includegraphics[width = \textwidth, keepaspectratio]{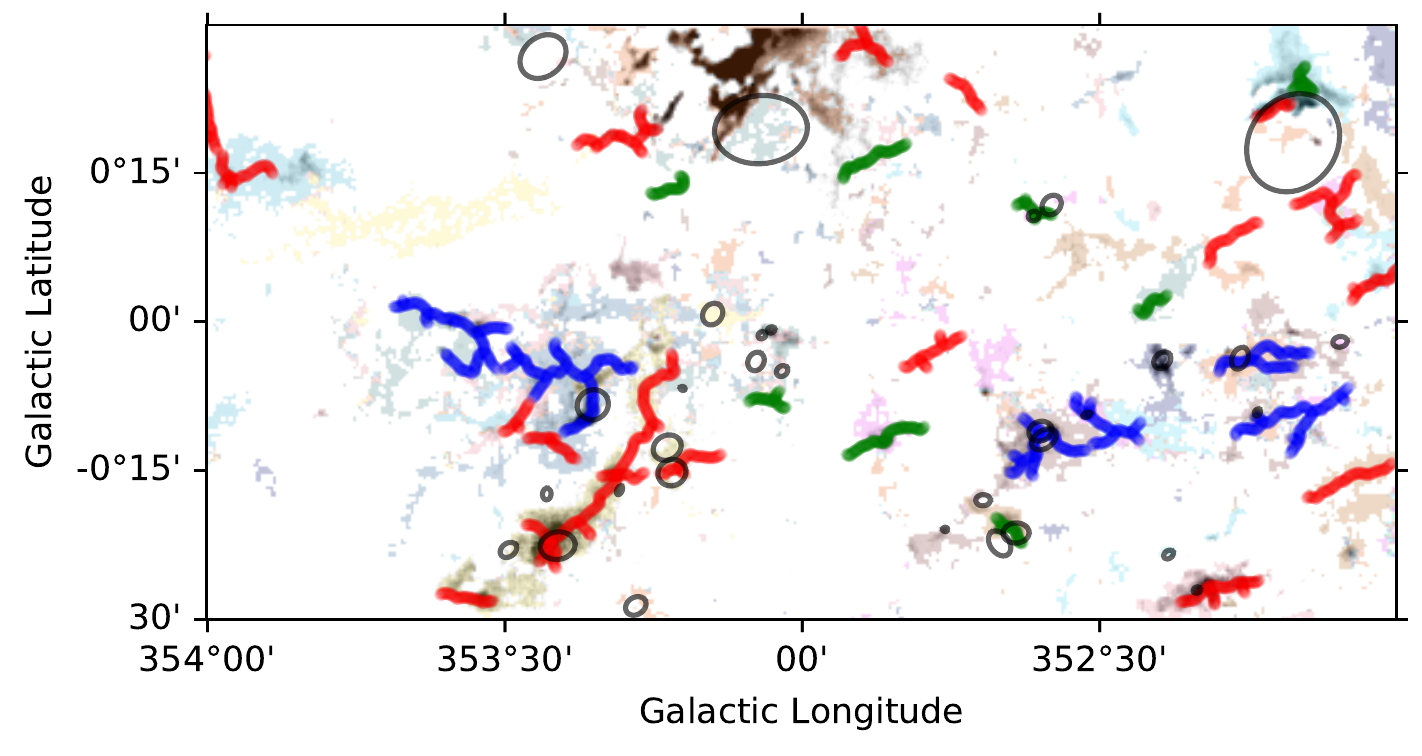}
    \caption{Elongated structures from ATLASGAL survey and bubbles from MWP survey overlaid on SEDIGISM clouds for $352 \degree \leq l \leq 354 \degree$.}
    \label{fig: SED overlap 352}
    \end{minipage}\hfill
\end{figure*}

\begin{figure*}[]
    \centering
    \begin{minipage}{\textwidth}
    \includegraphics[width = \textwidth, keepaspectratio]{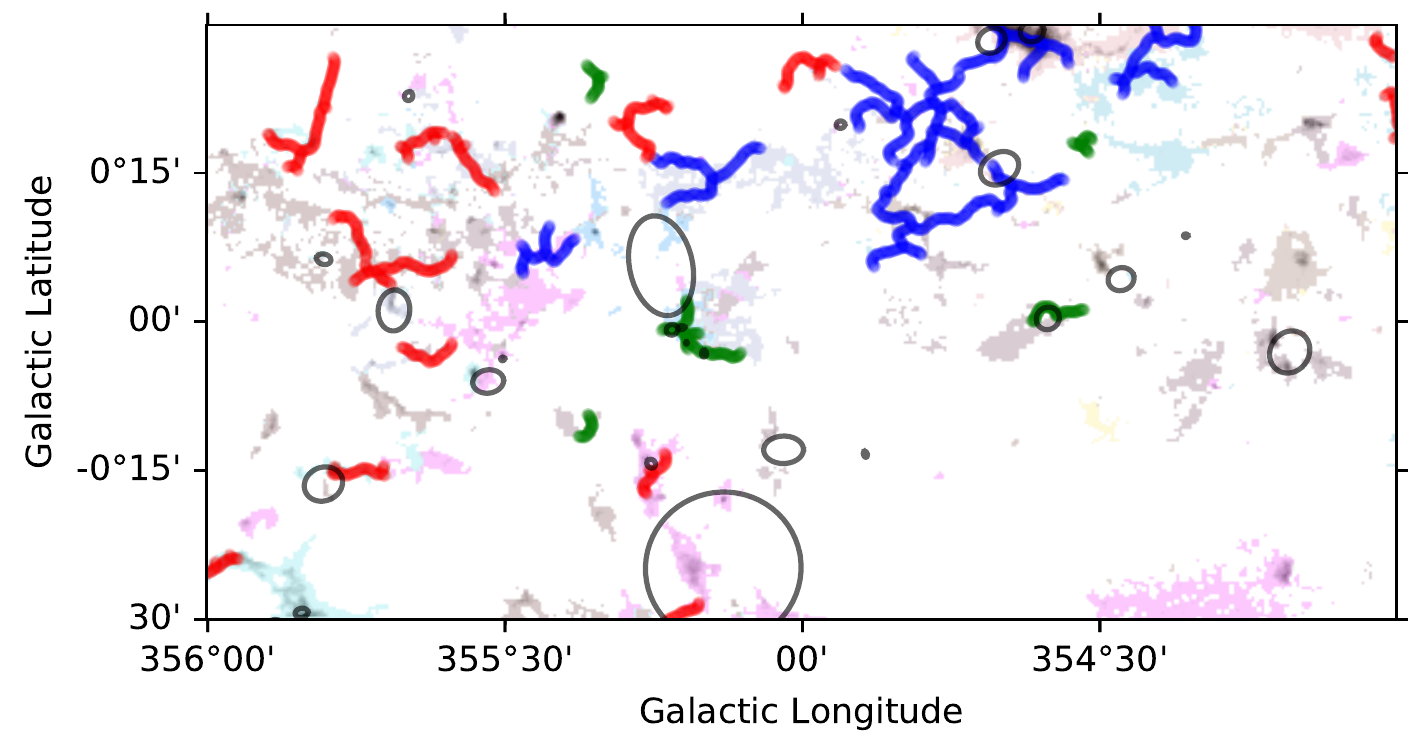}
    \caption{Elongated structures from ATLASGAL survey and bubbles from MWP survey overlaid on SEDIGISM clouds for $354 \degree \leq l \leq 356 \degree$.}
    \label{fig: SED overlap 354}
    \end{minipage}\hfill
\end{figure*}

\begin{figure*}[]
    \centering
    \begin{minipage}{\textwidth}
    \includegraphics[width = \textwidth, keepaspectratio]{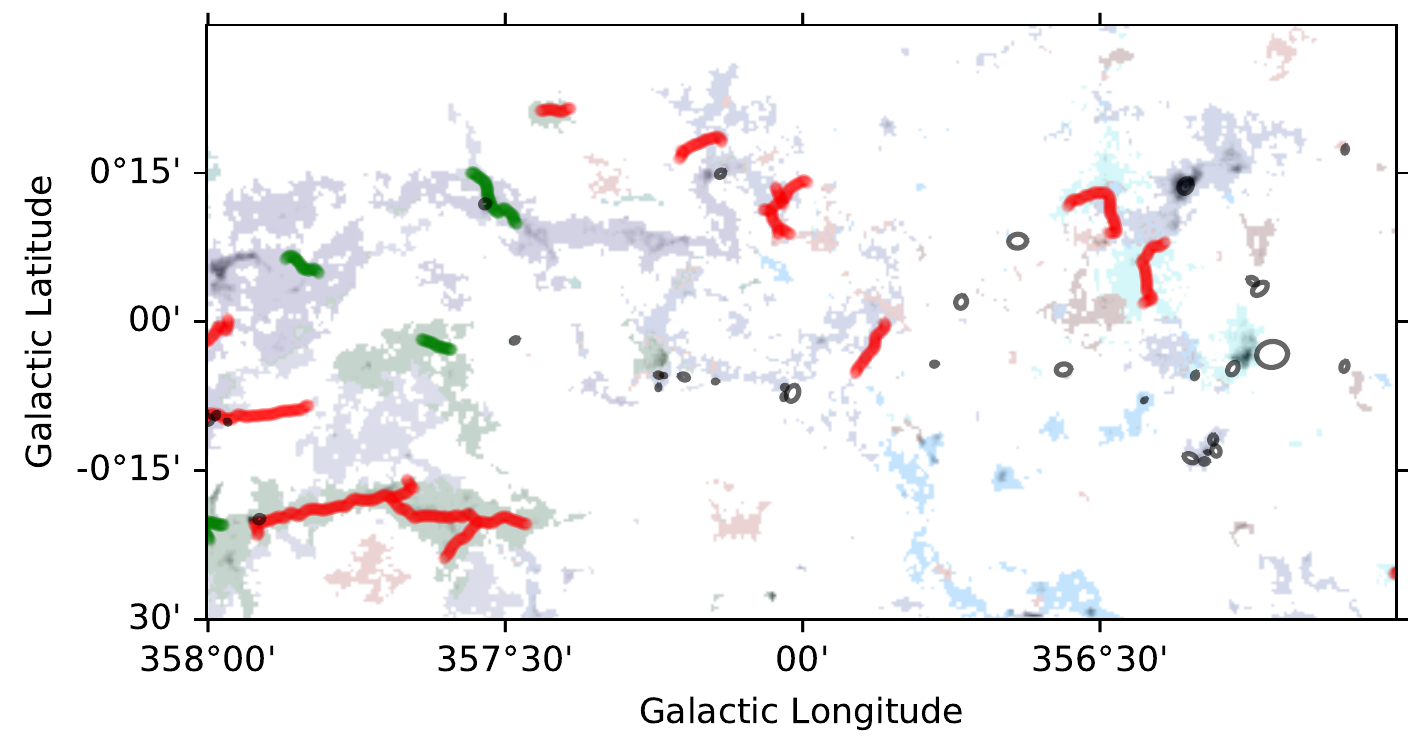}
    \caption{Elongated structures from ATLASGAL survey and bubbles from MWP survey overlaid on SEDIGISM clouds for $356 \degree \leq l \leq 358 \degree$.}
    \label{fig: SED overlap 356}
    \end{minipage}\hfill
\end{figure*}

\begin{figure*}[]
    \centering
    \begin{minipage}{\textwidth}
    \includegraphics[width = \textwidth, keepaspectratio]{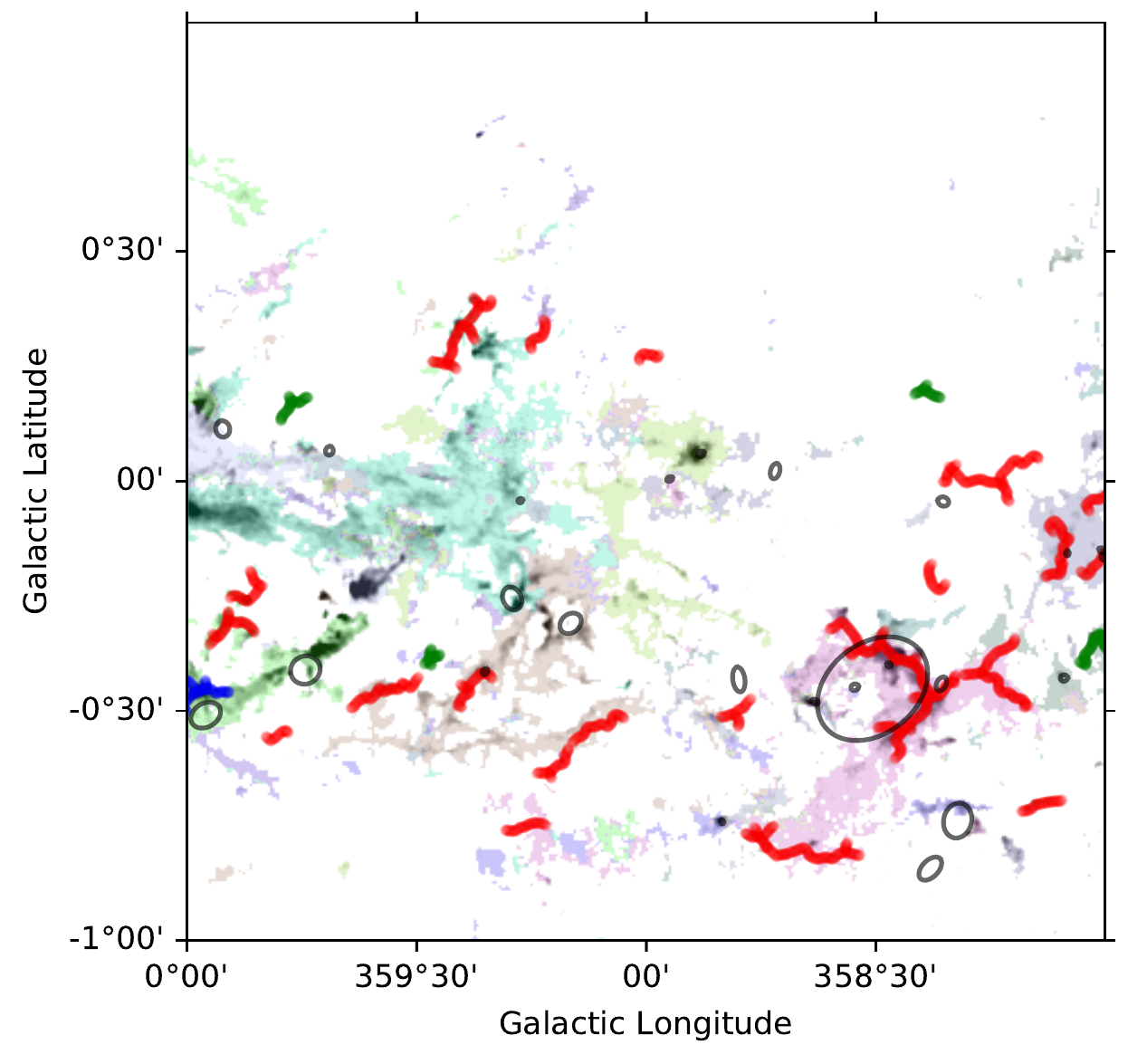}
    \caption{Elongated structures from ATLASGAL survey and bubbles from MWP survey overlaid on SEDIGISM clouds for $358 \degree \leq l \leq 0 \degree$.}
    \label{fig: SED overlap 358}
    \end{minipage}\hfill
\end{figure*}

\begin{figure*}[]
    \centering
    \begin{minipage}{\textwidth}
    \includegraphics[width = \textwidth, keepaspectratio]{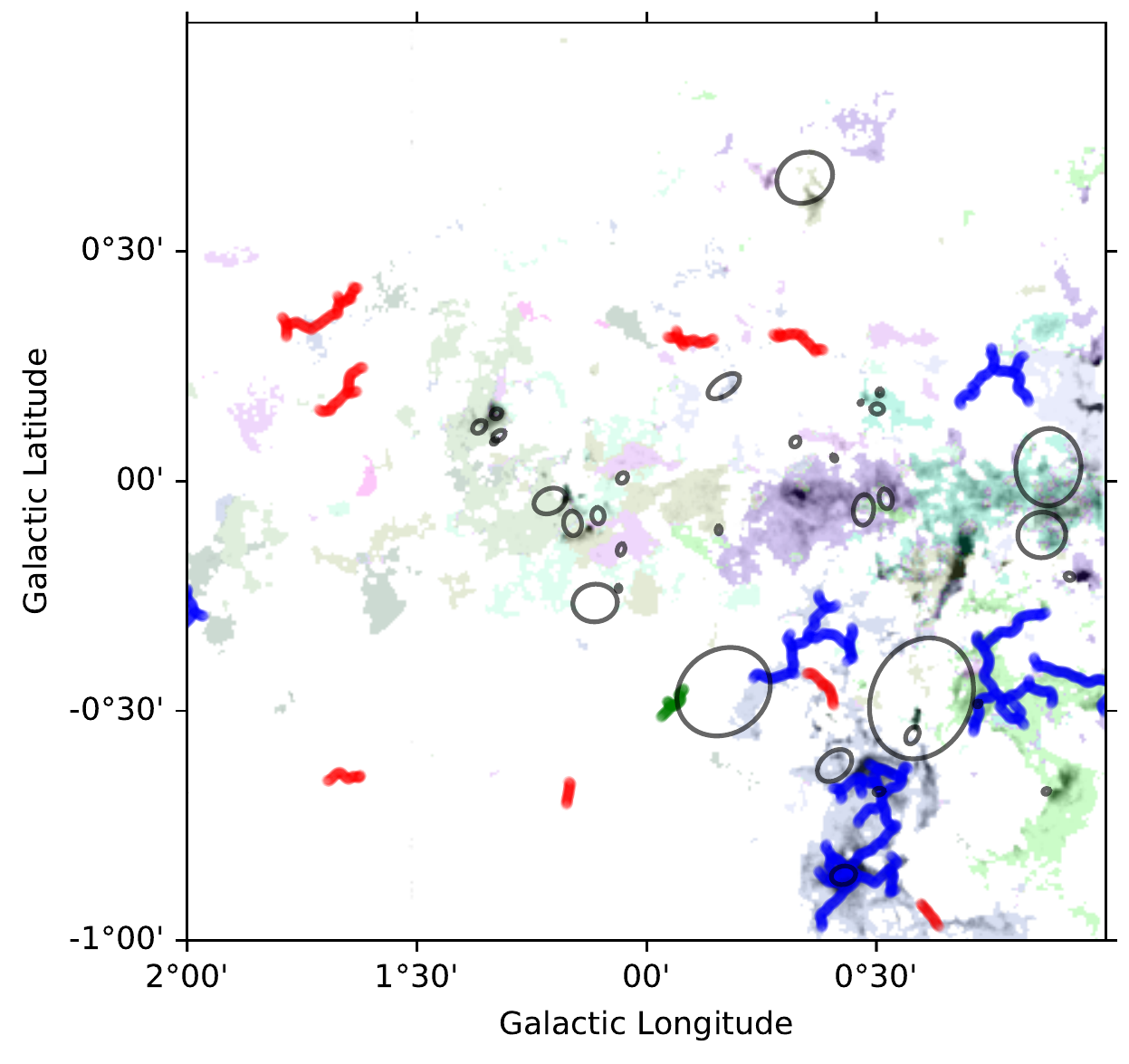}
    \caption{Elongated structures from ATLASGAL survey and bubbles from MWP survey overlaid on SEDIGISM clouds for $0 \degree \leq l \leq 2 \degree$.}
    \label{fig: SED overlap 0}
    \end{minipage}\hfill
\end{figure*}

\begin{figure*}[]
    \centering
    \begin{minipage}{\textwidth}
    \includegraphics[width = \textwidth, keepaspectratio]{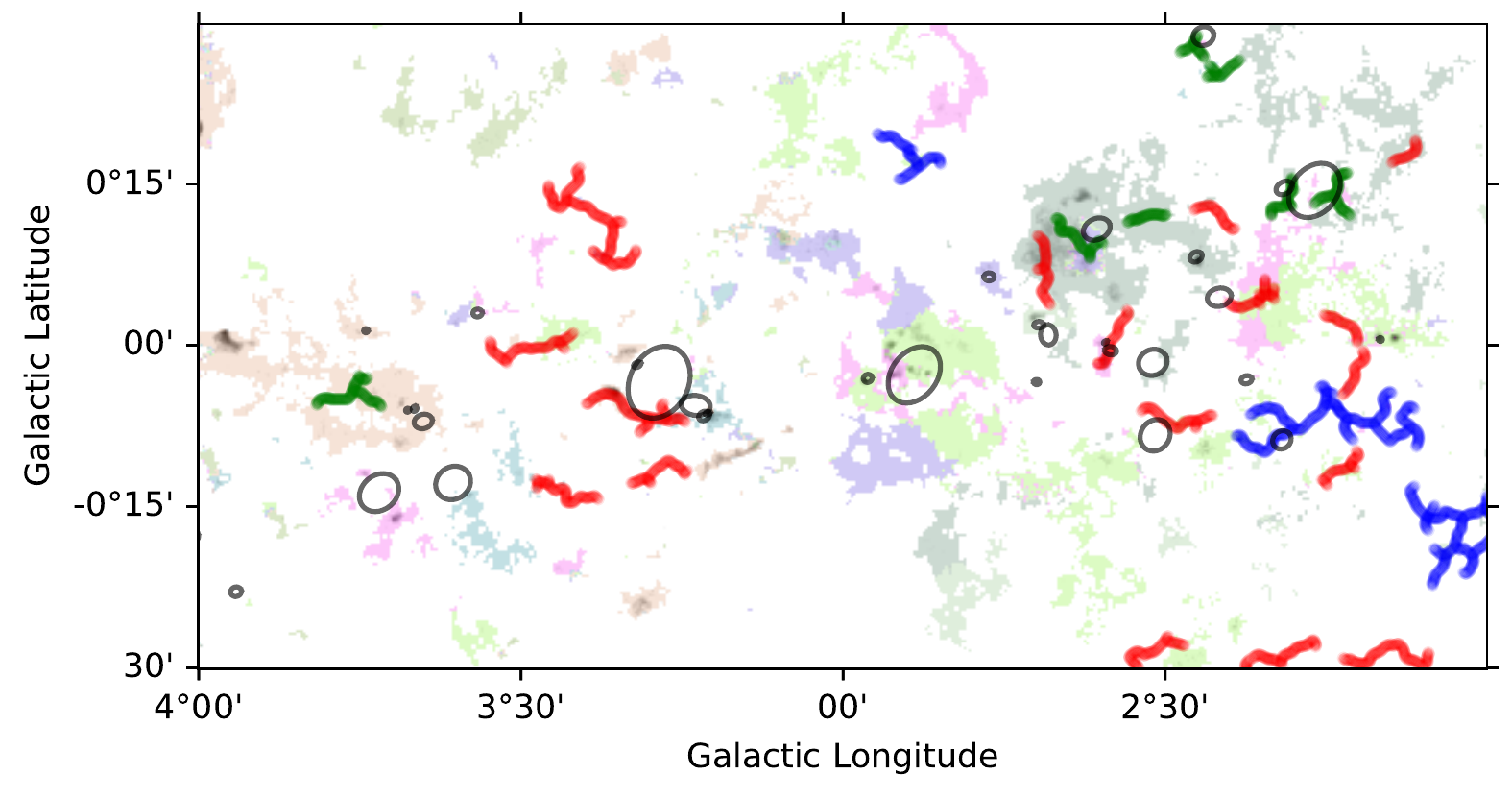}
    \caption{Elongated structures from ATLASGAL survey and bubbles from MWP survey overlaid on SEDIGISM clouds for $2 \degree \leq l \leq 4 \degree$.}
    \label{fig: SED overlap 2}
    \end{minipage}\hfill
\end{figure*}

\begin{figure*}[]
    \centering
    \begin{minipage}{\textwidth}
    \includegraphics[width = \textwidth, keepaspectratio]{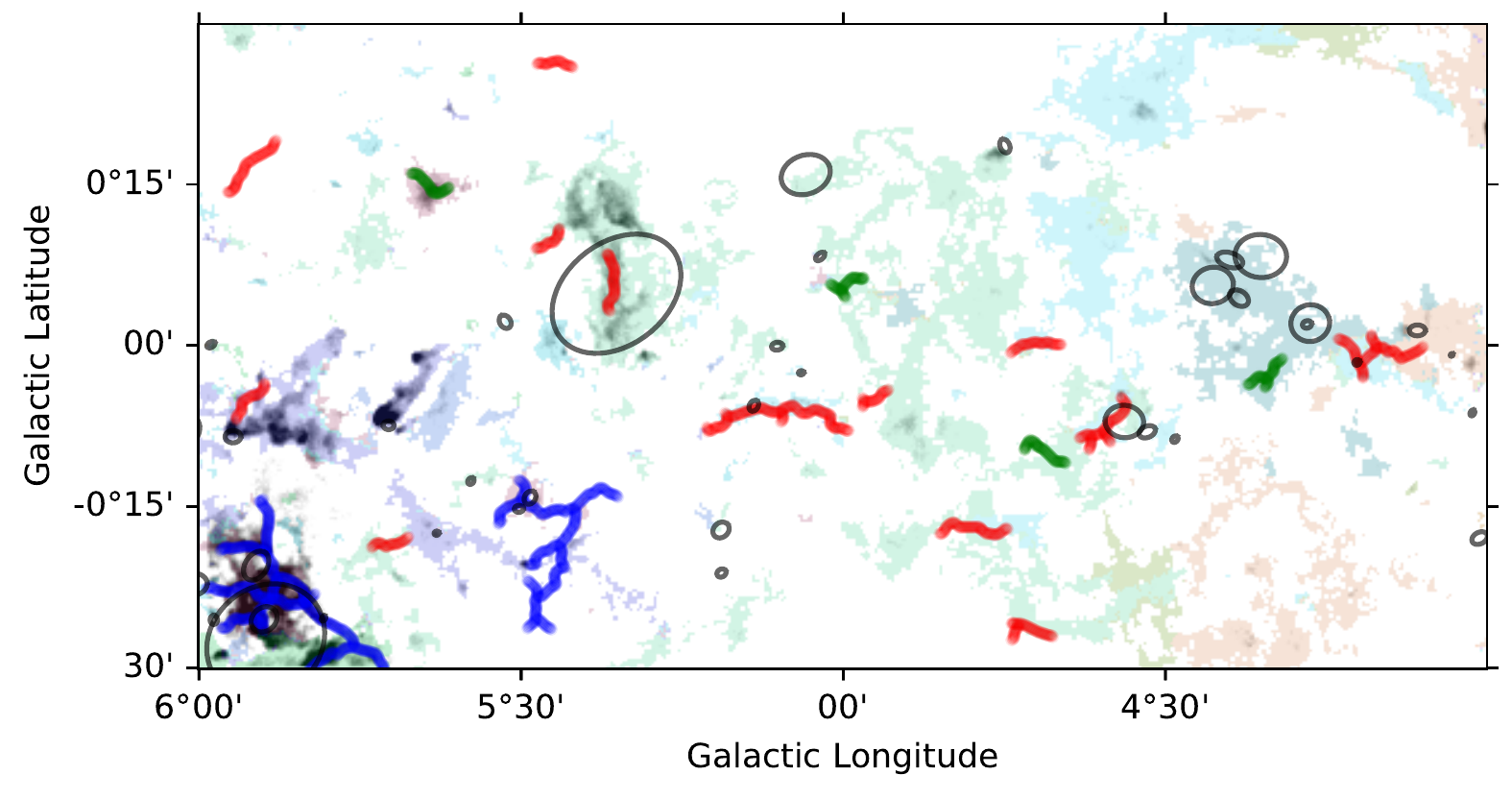}
    \caption{Elongated structures from ATLASGAL survey and bubbles from MWP survey overlaid on SEDIGISM clouds for $4 \degree \leq l \leq 6 \degree$.}
    \label{fig: SED overlap 4}
    \end{minipage}\hfill
\end{figure*}

\begin{figure*}[]
    \centering
    \begin{minipage}{\textwidth}
    \includegraphics[width = \textwidth, keepaspectratio]{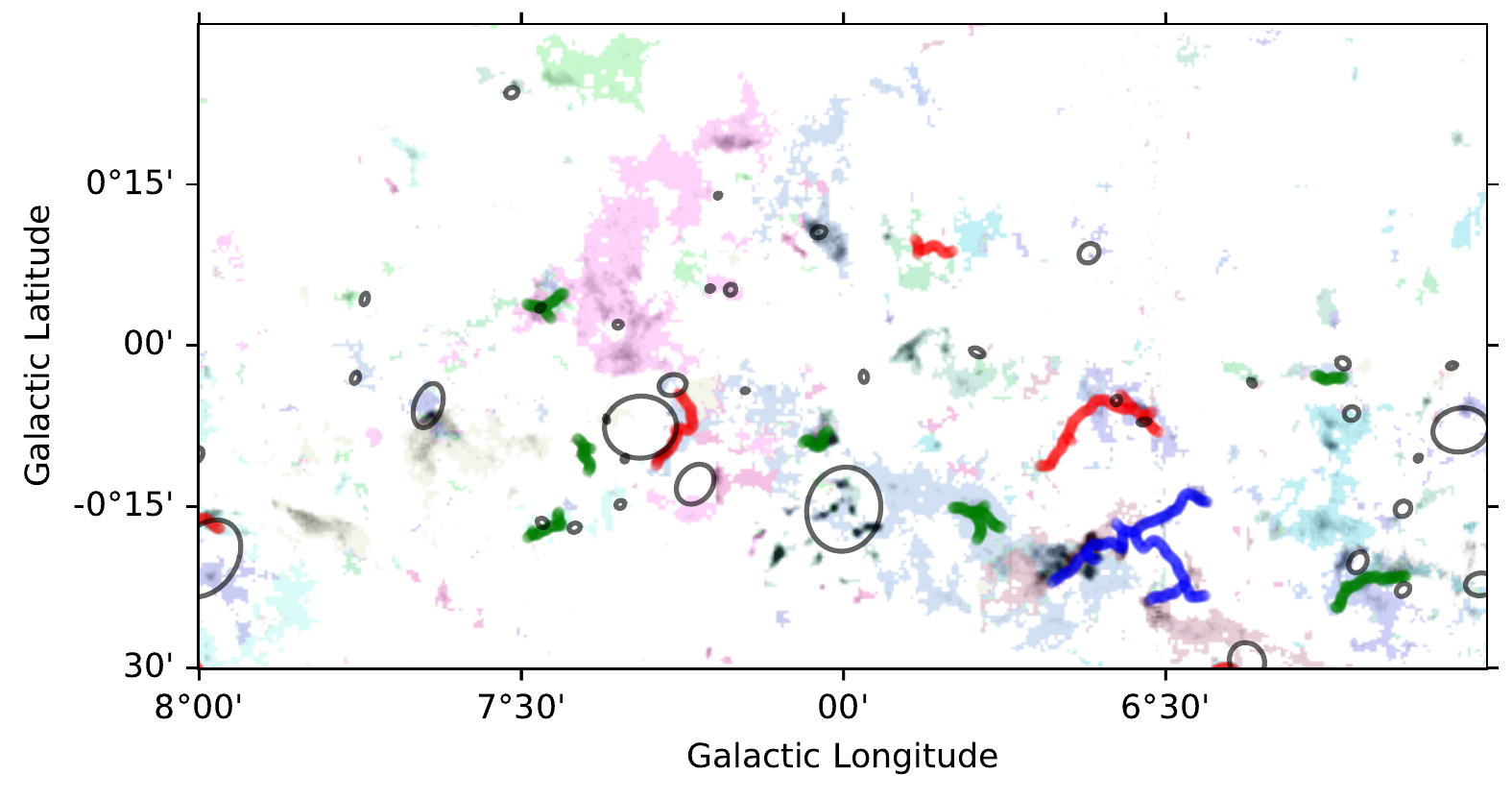}
    \caption{Elongated structures from ATLASGAL survey and bubbles from MWP survey overlaid on SEDIGISM clouds for $6 \degree \leq l \leq 8 \degree$.}
    \label{fig: SED overlap 6}
    \end{minipage}\hfill
\end{figure*}

\begin{figure*}[]
    \centering
    \begin{minipage}{\textwidth}
    \includegraphics[width = \textwidth, keepaspectratio]{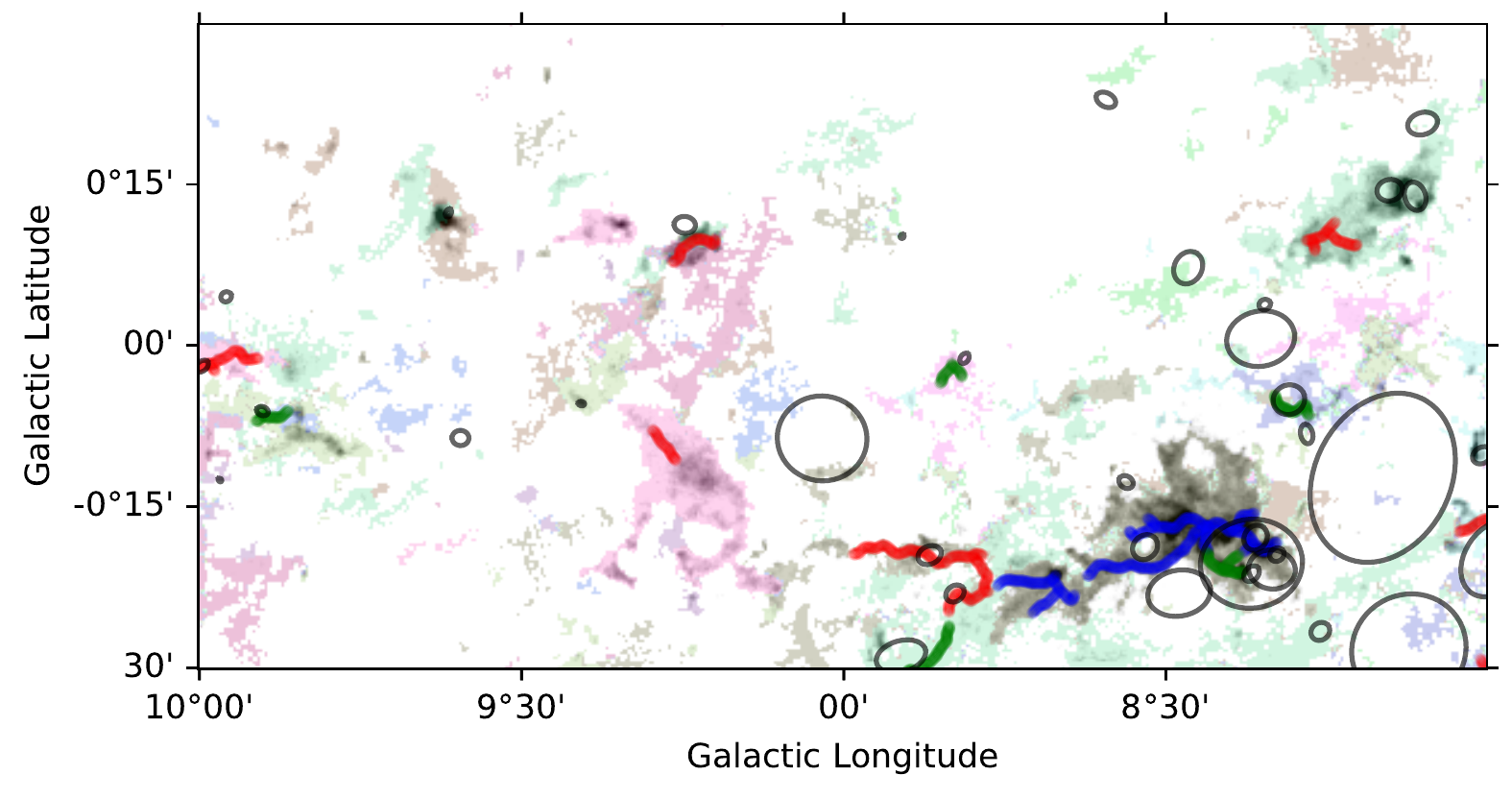}
    \caption{Elongated structures from ATLASGAL survey and bubbles from MWP survey overlaid on SEDIGISM clouds for $8 \degree \leq l \leq 10 \degree$.}
    \label{fig: SED overlap 8}
    \end{minipage}\hfill
\end{figure*}

\begin{figure*}[]
    \centering
    \begin{minipage}{\textwidth}
    \includegraphics[width = \textwidth, keepaspectratio]{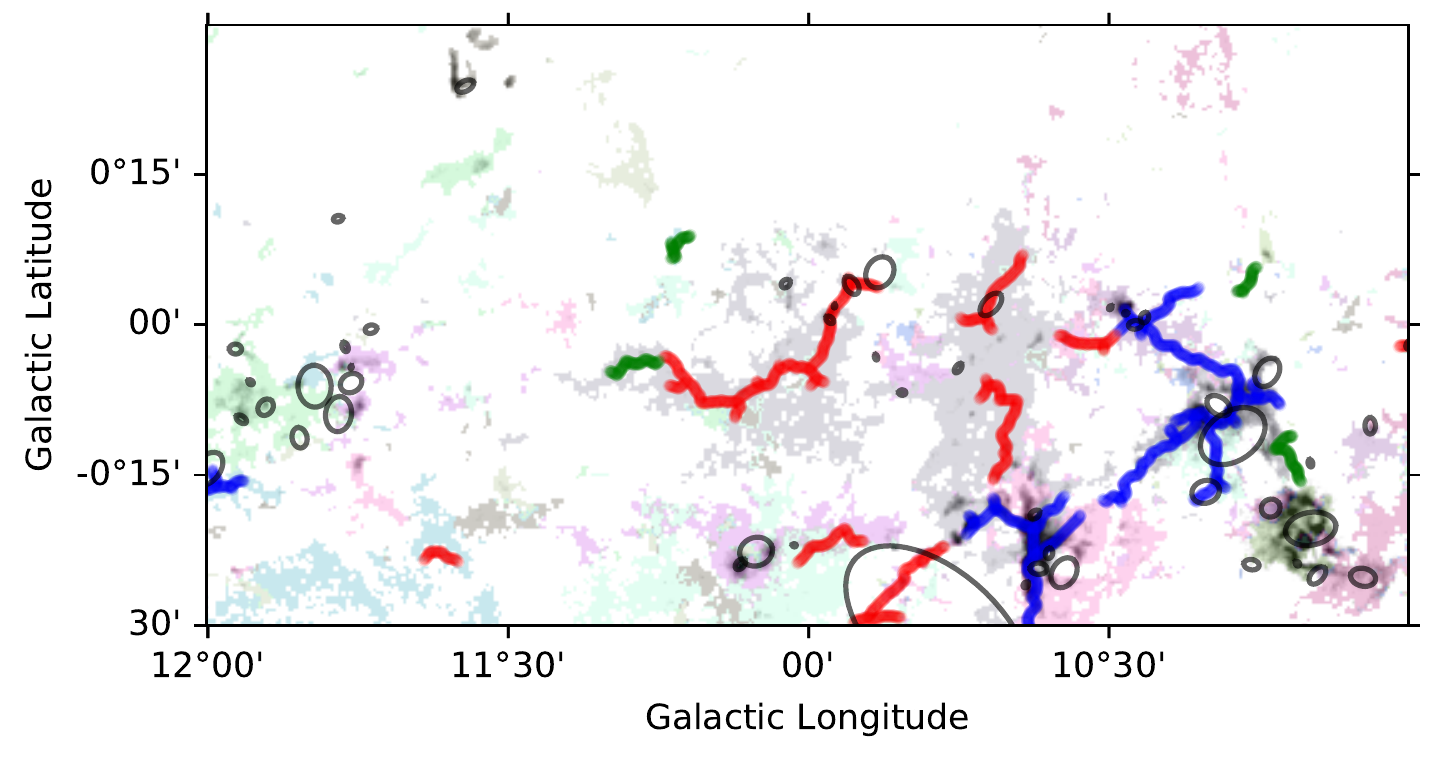}
    \caption{Elongated structures from ATLASGAL survey and bubbles from MWP survey overlaid on SEDIGISM clouds for $10 \degree \leq l \leq 12 \degree$.}
    \label{fig: SED overlap 10}
    \end{minipage}\hfill
\end{figure*}

\begin{figure*}[]
    \centering
    \begin{minipage}{\textwidth}
    \includegraphics[width = \textwidth, keepaspectratio]{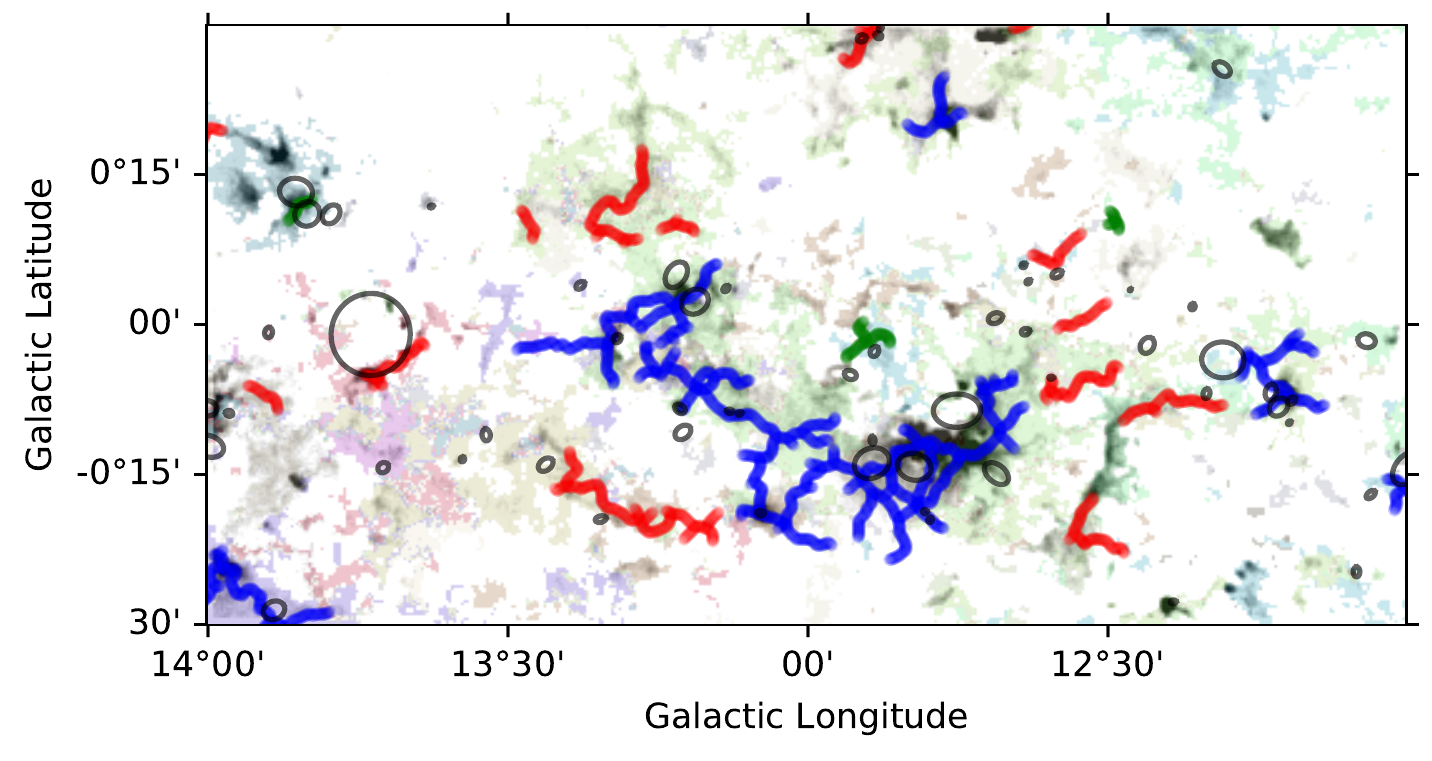}
    \caption{Elongated structures from ATLASGAL survey and bubbles from MWP survey overlaid on SEDIGISM clouds for $12 \degree \leq l \leq 14 \degree$.}
    \label{fig: SED overlap 12}
    \end{minipage}\hfill
\end{figure*}

\begin{figure*}[]
    \centering
    \begin{minipage}{\textwidth}
    \includegraphics[width = \textwidth, keepaspectratio]{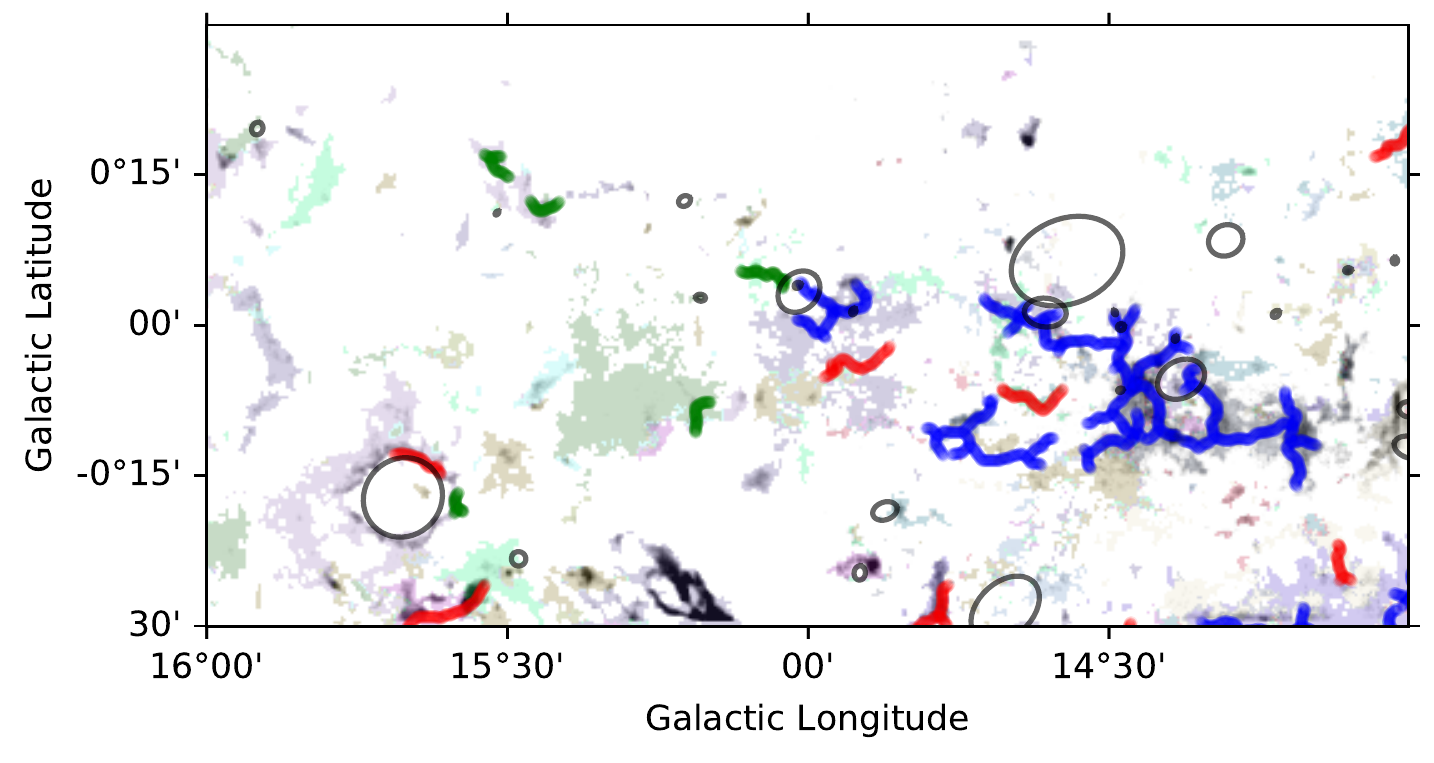}
    \caption{Elongated structures from ATLASGAL survey and bubbles from MWP survey overlaid on SEDIGISM clouds for $14 \degree \leq l \leq 16 \degree$.}
    \label{fig: SED overlap 14}
    \end{minipage}\hfill
\end{figure*}

\begin{figure*}[]
    \centering
    \begin{minipage}{\textwidth}
    \includegraphics[width = \textwidth, keepaspectratio]{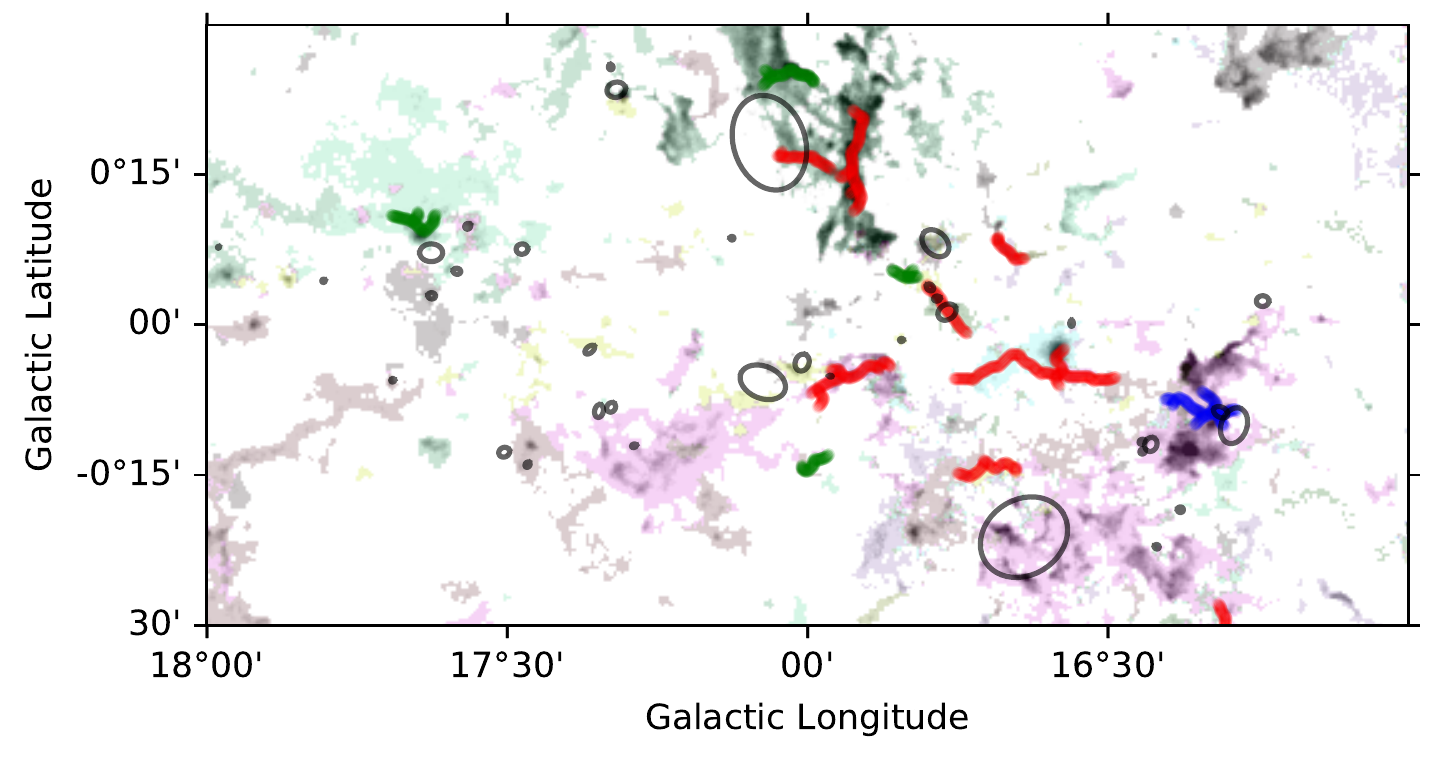}
    \caption{Elongated structures from ATLASGAL survey and bubbles from MWP survey overlaid on SEDIGISM clouds for $16 \degree \leq l \leq 18 \degree$.}
    \label{fig: SED overlap 16}
    \end{minipage}\hfill
\end{figure*}

We compare the SEDIGISM clouds with their counterparts from the ATLASGAL survey and MWP project. The figures \ref{fig: SED overlap 300} -- \ref{fig: SED overlap 16} show the overlap between the SEDIGISM clouds, AG-El and MWP bubbles. ATLASGAL elongated structures (AG-El) consist of `filaments', `networks of filaments' and `resolved elongated structures' \citep{ATLASGAL_filaments}. The filaments are coloured red, networks of filaments are coloured blue and resolved elongated structures are coloured green. The black ellipses on the figure represent the MWP bubbles. These structures are overlaid on the SEDIGISM cloud masks of various colours and $^{13}$CO peak intensity in reverse greyscale. The catalogue for AG-El contains the position for each pixel in the structure whereas MWP bubble catalogue only contains the positions of the centres for the ellipses (bubbles) and their major and minor axes. Thus our ellipses (Fig. \ref{fig: SED overlap 300} -- \ref{fig: SED overlap 16}) might not present the original structures of the actual bubbles, e.g. incomplete rings.

\section{\add{Original by-eye classification to morphological groups}}\label{App: ks test}

Our original by-eye classification tried to follow the indication of the $J$ plots algorithm -- filaments, rings and cores. However, we realised that clouds in SEDIGISM sample showed repetitive patterns that would be better suited to six cloud morphologies, which are described below:

\begin{enumerate}
    \item Bubble: This category consists of clouds with a structure similar to a ring or a bubble. They have negligible elongation and resemble complete gas bubbles.
    \item Filament: These are elongated structures resolved along lengths and widths, with the lengths distinctly larger than the widths. They also show similar intensity across the entire length.
    \item Core: The centrally concentrated clouds without obvious elongations are classified as cores. These are molecular clouds and have no direct relation to pre-stellar/proto-stellar cores.
    \item Elongated bubble: This category consists of stretched bubbles, bubbles with filamentary structures attached to them and clouds that resemble incomplete bubbles forming semi circles.
    \item Elongated core: Elongated clouds with centrally concentrated structures are classified in this category.
    \item Multiple Connected Clouds (MCC): Clouds containing multiple dense regions belong to this category. These may be elongated structures similar to globular filaments.
    \item Unclassified: Clouds which do not resemble any of the above categories.
\end{enumerate}

\begin{figure*}
    \centering
    \includegraphics[width = \textwidth, keepaspectratio]{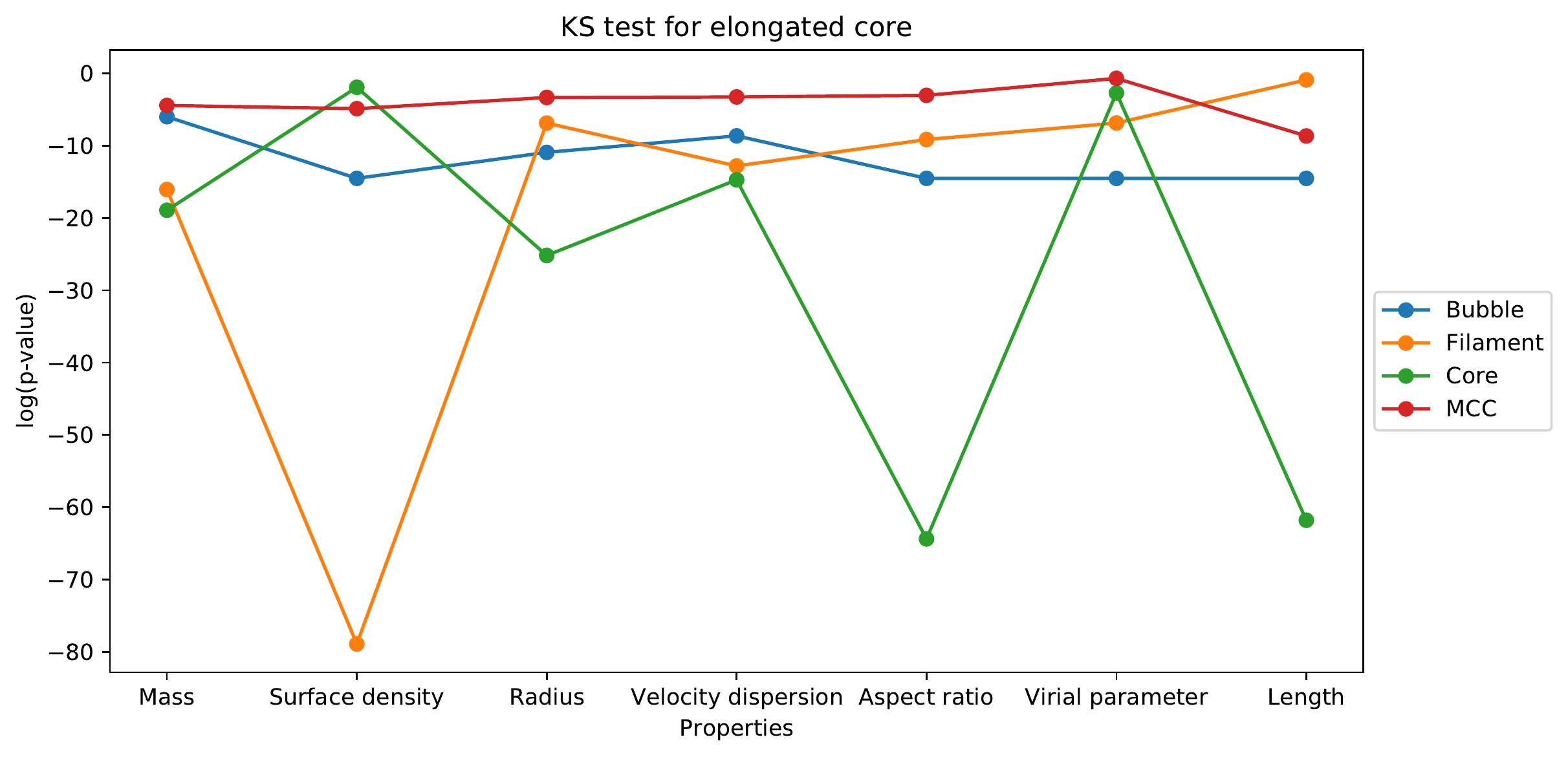}
    \caption{P-values obtained using KS test on elongated cores and the four morphological sub-classes -- bubbles, filaments, core and MCC -- for the various cloud properties. The different colours represent the morphologies that were compared with elongated cores for obtaining the p-value.}
    \label{fig: p-values ks ec}
\end{figure*}

\begin{figure*}
    \centering
    \includegraphics[width = \textwidth, keepaspectratio]{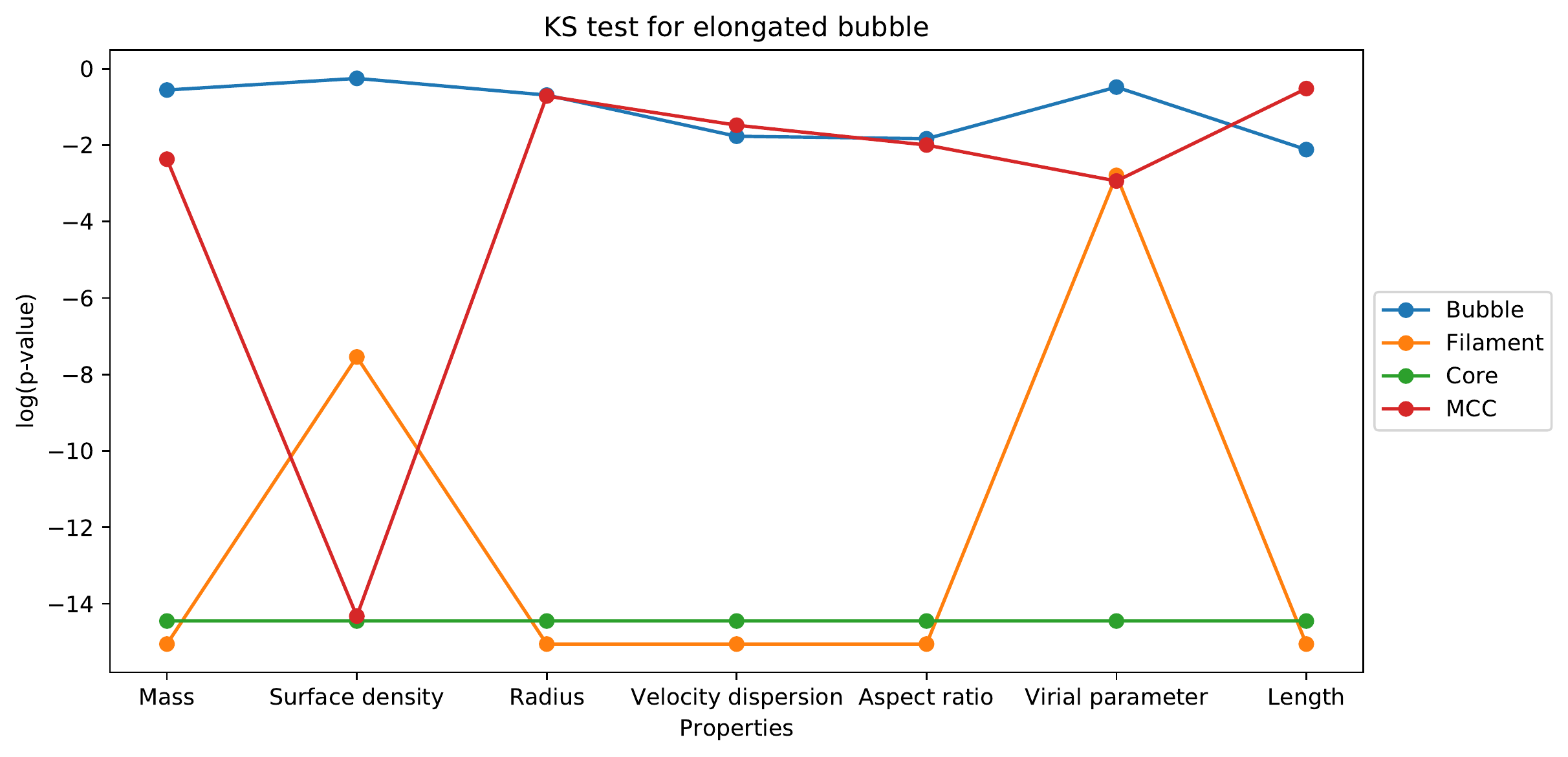}
    \caption{P-values obtained using KS test on elongated bubbles and the four morphological sub-classes -- bubbles, filaments, core and MCC -- for the various cloud properties. The different colours represent the morphologies that were compared with elongated bubbles for obtaining the p-value.}
    \label{fig: p-values ks eb}
\end{figure*}

These sub-classes were merged to get four major morphological groups (Sec. \ref{sec: by eye classification}). We first renamed the four sub-classes -- bubble, filament, core and MCC-- as ring-like, elongated, concentrated and clumpy clouds respectively. The remaining sub-classes were merged into the morphological groups using the two sample Kolmogorov–Smirnov (KS) test. The KS test was performed on the distributions of the seven properties i.e. mass, surface density, radius, velocity dispersion, aspect ratio, virial parameter and length. Based on the p-values from the KS test (Fig. \ref{fig: p-values ks ec} \& \ref{fig: p-values ks eb}), we classified the elongated bubbles as ring-like clouds and elongated cores as clumpy clouds. 

\begin{figure*}
    \centering
    \includegraphics[width = \textwidth, keepaspectratio]{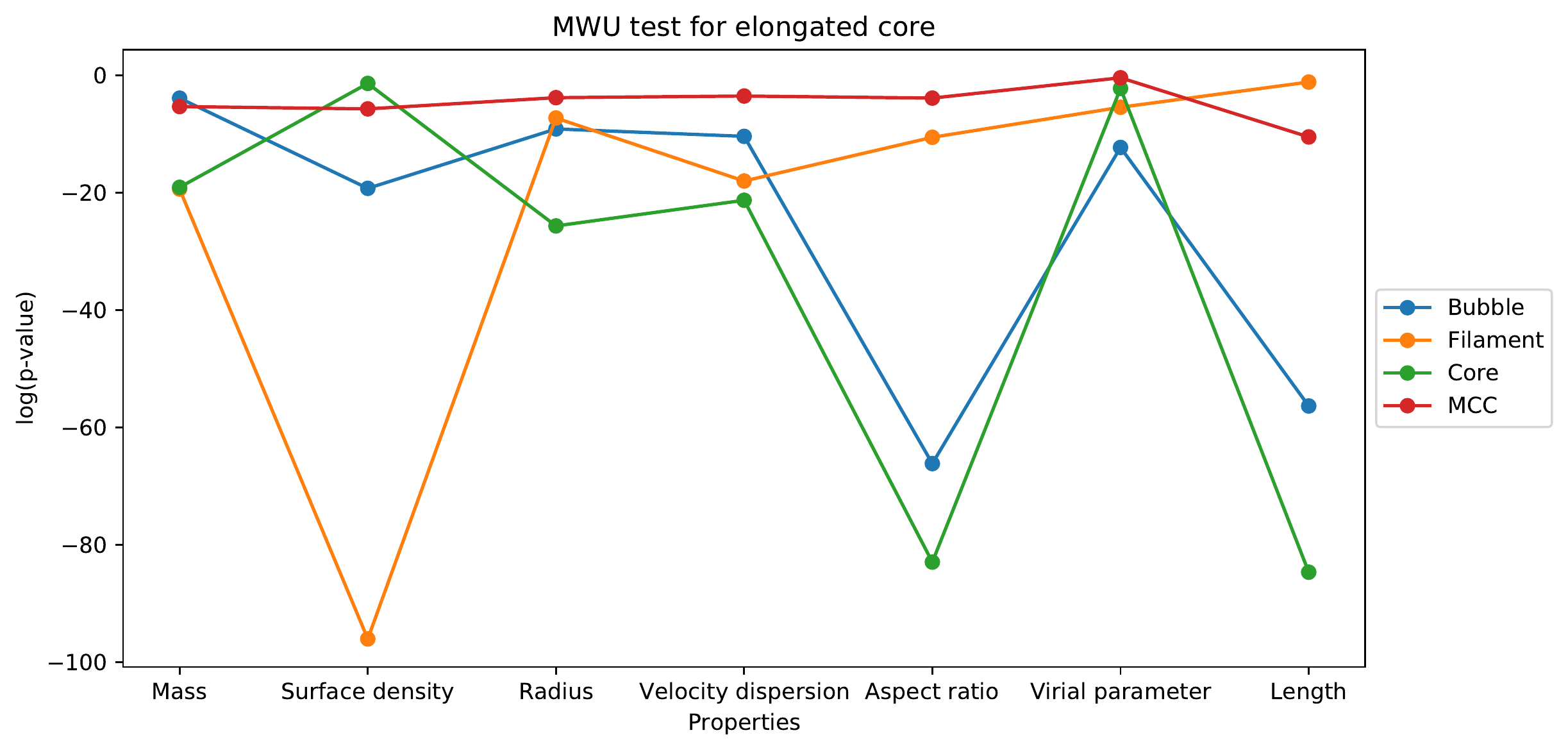}
    \caption{P-values obtained using MWU test on elongated cores and the four morphological sub-classes -- bubbles, filaments, core and MCC -- for the various cloud properties. The different colours represent the morphologies that were compared with elongated cores for obtaining the p-value.}
    \label{fig: p-values mwy ec}
\end{figure*}

\begin{figure*}
    \centering
    \includegraphics[width = \textwidth, keepaspectratio]{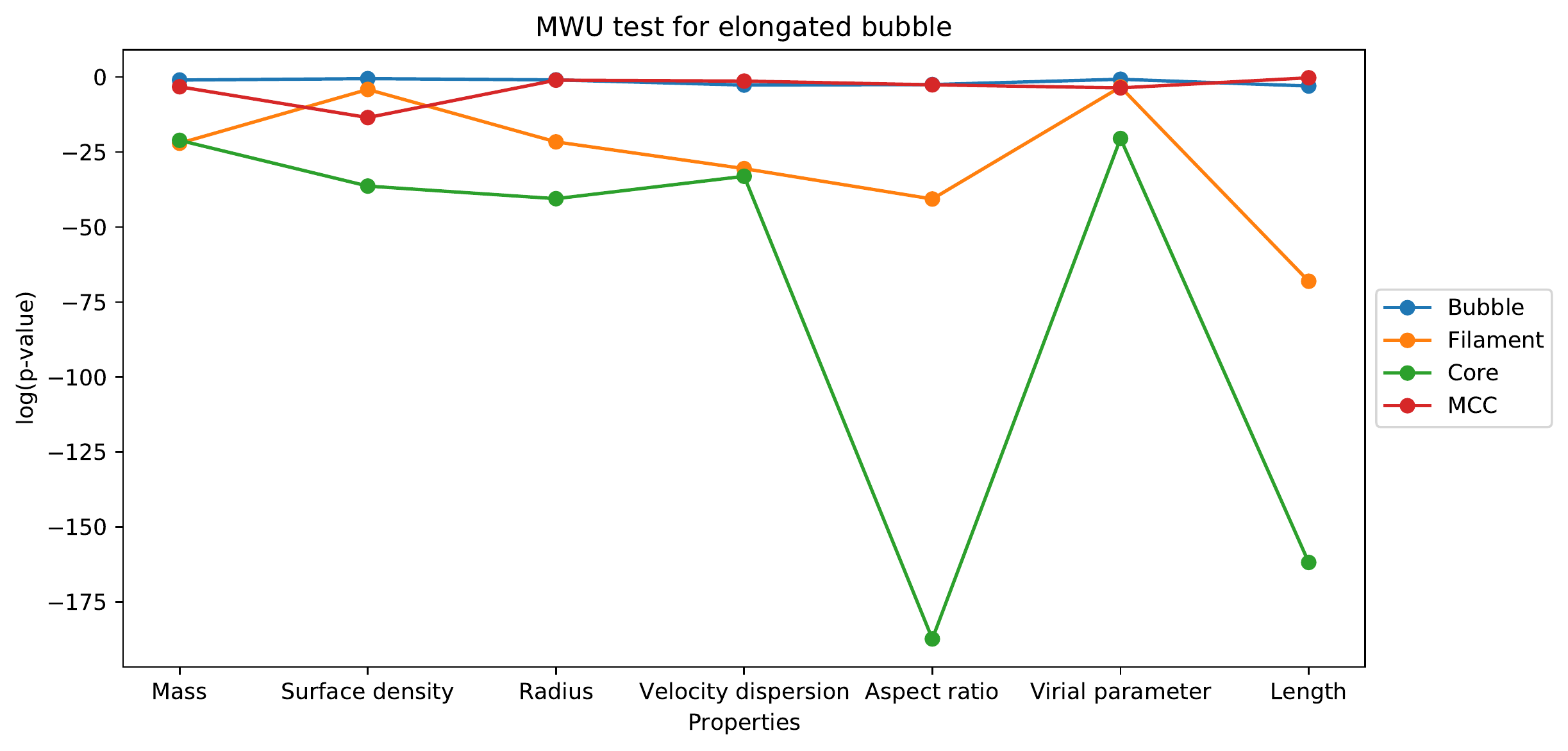}
    \caption{P-values obtained using MWU test on elongated bubbles and the four morphological sub-classes -- bubbles, filaments, core and MCC -- for the various cloud properties. The different colours represent the morphologies that were compared wth elongated bubbles for obtaining the p-value.}
    \label{fig: p-values mwy eb}
\end{figure*}

%As suggested by the referee, 
We also conducted the two sided Mann-Whitney U \citep[MWU;][]{fay_2010} test to compare the cloud properties distributions of elongated cores and elongated bubbles with other morphological sub-classes. The MWU test is a non-parametric test with a null hypothesis that neither distribution has stochastic dominance over other. It can be formally expressed as the probability of a variable drawn from a distribution 'X' having a greater value than a variable drawn from a distribution 'Y' being equal to the reverse, i.e. P(X > Y) = P(Y > X). A high p-value would suggest that the two distributions are similar. An advantage of MWU test over the KS test is that it is not affected due to different widths of the distributions. The p-values obtained from our analysis (Fig. \ref{fig: p-values mwy ec}) suggest classification of elongated cores as clumpy clouds. The p-values for elongated bubbles (Fig. \ref{fig: p-values mwy eb}) suggest that elongated bubbles have comparable distributions with both bubbles and MCC. However, a majority of the properties suggest that elongated bubbles have the closest distributions to bubbles. Thus, the morphological classes remain same as described by the KS test. 

\end{appendix}
\end{document}